\documentclass{pasj00}

\newcommand{\simgt}{\lower.5ex\hbox{$\; \buildrel > \over \sim \;$}}
\newcommand{\simlt}{\lower.5ex\hbox{$\; \buildrel < \over \sim \;$}}

\def\btheta{\mbox{\boldmath $\theta$}}

\begin{document}
\SetRunningHead{Okabe \& Umetsu}
{Subaru Weak Lensing Study of Seven Merging Clusters}
\Received{2007/02/24}
\Accepted{2007/10/20}

\title{Subaru Weak Lensing Study 
of Seven Merging Clusters: \\ Distributions of Mass and Baryons 
\thanks{Based on data collected at Subaru Telescope, which is
operated by the National Astronomical Observatory of Japan. }}

\author{Nobuhiro \textsc{Okabe}}
\affil{Astronomical Institute, Tohoku University,
Aramaki, Aoba-ku, Sendai, 980-8578, Japan}
\email{okabe@astr.tohoku.ac.jp}

\author{Keiichi \textsc{Umetsu}}
\affil{Institute of Astronomy and Astrophysics, Academia Sinica,\\
 P.O. Box 23--141,
 Taipei 106, Taiwan, Republic of China}
\email{keiichi@asiaa.sinica.edu.tw}

\KeyWords{gravitational lensing---X-rays: galaxies:
clusters---galaxies: clusters: individual 
(A520, A754, A1750, A1758, A1914, A2034, A2142) } 

\maketitle

\begin{abstract}
We present and compare projected distributions of mass, galaxies, and
the intracluster medium (ICM) for a sample of merging clusters of
 galaxies based on the joint weak-lensing, optical photometric, 
and X-ray analysis.
Our sample comprises seven nearby Abell clusters,
for which we have conducted systematic, deep imaging observations 
with Suprime-Cam on Subaru telescope.
Our seven target clusters, representing various 
merging stages and conditions, allow us to investigate in details the 
physical interplay between dark matter, ICM, and galaxies
associated with cluster formation and evolution.
A1750 and A1758 are binary systems consisting of two cluster-sized components, 
A520, A754, A1758N, A1758S, and A1914
are on-going cluster mergers, and
A2034 and A2142 are cold-front clusters.
In the binary clusters, the projected mass, optical light, and X-ray distributions 
are overall similar and regular without significant substructures.
On-going and cold-front merging clusters, on the other hand, reveal highly irregular
mass distributions. Overall the mass distribution appears to be similar
to the galaxy luminosity distribution, whereas their distributions are
quite different from the ICM distribution in a various ways.
We also measured for individual targets
the global cluster parameters such as the cluster mass,
 the mass-to-light ratio, and the ICM temperature.
A comparison of the ICM and virial temperatures of merging clusters from
 X-ray and weak-lensing analyses, respectively, shows that the ICM
 temperature of on-going and cold-front clusters is significantly higher
 than the cluster virial temperature by a factor of $\sim 2$. 
This temperature excess in the ICM could be explained by the effects of merger boosts.

\end{abstract}

\section{INTRODUCTION}

In hierarchical structure formation models
based on the Cold Dark Matter (CDM) scenario,
small structures form first, and they merge to form larger, more massive
objects.
Since clusters of galaxies are the largest gravitationally bound objects in
the universe,
they are still forming through mergers of sub-clusters and smaller
groups of galaxies along the filamentary structure within which
they are embedded.
The cluster merger is one of the most energetic events in the universe
because of its highest binding energy and huge energy release.
Dark matter, which is the dominant mass component of clusters,
governs the dynamics of cluster mergers,
and drives various phenomena
in the hot, diffuse intracluster medium (ICM). 
Colliding substructures with high collision velocities
trigger shocks or turbulences in the ICM. 
Merger shocks increase the temperature and entropy of the ICM,
resulting in increasing the X-ray luminosity of the cluster.
The merger-induced shocks and turbulences generate relativistic
particles, which then produce synchrotron radio emission and 
inverse-Compton emission on the Cosmic Microwave Background.
We however have not yet understood the cluster merger 
phenomena and the relationships
between the ICM and dark matter in detail.

Observationally, 
cluster mergers have been investigated based on
optical measures of cluster dynamics, namely 
line-of-sight velocity distributions of cluster member galaxies
as well as optical appearance of substructures.
Optical evidence of substructures
were reported by many authors
based on apparent concentrations of galaxies in projected space
(e.g., Abell, Neyman, \&  Scott 1964; 
       Baier \& Ziener 1977; Geller \& Beers 1982),
although a detection of subclustering of cluster galaxies is not 
direct evidence for the existence of underlying mass component. 
On the other hand, 
one can investigate the galaxy dynamics in clusters
by examining distribution of galaxies both in spatial and velocity space
(e.g. Zabludoff
et al. 1993; Ashman,  Bird \& Zepf 1994),

In recent years, X-ray satellites such as 
{\it ASCA}, {\it ROSAT}, {\it Chandra}, {\it XMM-Newton} and {\it Suzaku}
make it possible to study the physical processes 
in the ICM
involved in the formation and evolution of clusters.
In particular, precise spectro-imaging data of 
Chandra and XMM-Newton
allow us to derive the 
temperature, pressure, and entropy maps of the ICM,
which are useful tools to study the cluster merger process.
Recent Chandra and XMM-Newton observations have revealed 
a highly-complex X-ray morphology of merging clusters
(e.g. 
Govoni et al. 2004; 
Henry, Finoguenov, \&  Briel 2004;
Finoguenov, B$\ddot{{\rm o}}$hringer, \&  Zhang 2005).
Markevitch et al. (2000) discovered an unexpected ICM structure of
contact discontinuities, which are called ``cold fronts''. 
However, 
although such X-ray observations of the ICM physics
provide us with detailed information of the cluster merger,
X-ray study alone is not sufficient for
understanding of 
complex physical processes acting during the cluster merger.
This is because
cluster mergers are dynamically governed by dark matter 
which mainly 
accounts for $\sim 80\%$ of the total mass of clusters. 
Further, 
the ICM in merging clusters is not in, or close to, 
hydrostatic equilibrium,
so that the X-ray determined total mass of merging clusters, 
under the assumption of hydrostatic equilibrium,  will be 
different from the gravitational mass of the clusters.

Gravitational lensing effects on background galaxies, on the other hand,
are powerful, unique tools to study the mass distributions 
in clusters regardless of the physical/dynamical state of matter in the
systems. Weak lensing image distortions of background galaxy images can be used
to map the distribution of matter in clusters on scales
of $\sim 100$ kpc  to $1$--$2$ Mpc scales 
in a model-independent way (see, e.g.,
Bartelmann \& Schneider 2001; 
Umetsu, Tada, \& Futamase 1999).
Therefore weak lensing enables the direct study of mass in clusters
even when the clusters are in the process of pre/mid/post merging,
where the assumption of hydrostatic equilibrium or isothermality are no
more valid.

In this way, 
weak lensing, X-ray, and optical photometric observations 
provide complementary
information of the physical properties of clusters.
Hence, a joint analysis of weak lensing, X-ray, 
and optical photometric observations
will yield a comprehensive picture of the 
physical state of ICM, dark matter and member galaxies.

In this paper
we present weak-lensing 
mass distributions of a sample of seven nearby
merging clusters, reconstructed using weak shear data
taken with {\it Suprime-Cam} on the 8.2m {\it Subaru telescope},
and make a quantitative comparison of the mass maps with 
X-ray and optical galaxy distributions.
The Subaru/Suprime-Cam is the most ideal, working instrument for 
weak lensing shape measurements of background galaxies because of 
its wide field-of-view of $34'\times 27'$, photon collecting power
of the $8.2$m mirror, and excellent image quality with stable, small 
PSF anisotropy (e.g., Miyazaki et al. 2002; Hamana et al. 2003; Sato et
al. 2003; Broadhurst et al. 2005).
We selected for the Subaru weak lensing observations
seven nearby merging clusters ($0.05 \simlt z \simlt 0.28$)
of different merging stages and properties,
A520, A754, A1750, A1758, A1914, A2034 and A2142,
for which archival X-ray data of Chandra and/or XMM-Newton
are available.
Detailed analysis results  of individual clusters will be presented
elsewhere.
Details of 
our X-ray, optical and weak-lensing analyses are presented
 in \S\ref{sec:data}. 
In \S \ref{sec:maps}, we compare the resulting projected
distributions of mass, optical luminosity density,
and X-ray emission for our sample of merging clusters.
In \S \ref{sec:glob} we derive global cluster parameters
such as the cluster mass, the cluster mass-to-light ratio, and the
ICM temperature.
Finally, in \S\ref{sec:dis}, we summarize the projected offset
distributions of cluster galaxies and ICM with respect to the mass,
and
discusss the implications of
our results for understanding of cluster merger phenomena.

Throughout the paper
we adopt the concordance cosmology of 
$\Omega_{m0}=0.3$, $\Omega_{\Lambda}=0.7$ and
$H_0=70 \ {\rm km s}^{-1} {\rm Mpc}^{-1}$,
and use the AB magnitude system.

\section{DATA ANALYSIS}\label{sec:data}

\subsection{Data Preparation}

\label{subsubsec:subarudata}
\subsubsection{Subaru Data Analysis}

We observed all of the target clusters, except A520 for which
archival data are available (SMOKA),
with Subaru/Suprime-Cam (ID: S05A-159, PI: N. Okabe). 
For each target cluster 
we chose $R_{\rm c}$ or $i'$ band filters for the weak lensing
shape analysis, and $g'$ or $V$ for separating member and background
galaxies based on the color-magnitude diagram.
All of our observations were taken using the 
AG (acquisition and guide) probe for guide stars,
while the archival $i'$-band data of A520 were taken 
with and without guide probe on
17th November 2001
and
19th October 2001, respectively.
We retrieved from SMOKA
a total of seven $i'$-band images taken under good seeing conditions
($\sim 0\farcs 6$),
and analyzed separately the following three imaging data sets:
(A)  all of seven $i'$ images taken both
with and without guide probe ($7\times 240$s exposure),
(B) four $i'$ images taken without guide probe ($4\times 240$s exposure),
and
(C) three $i'$ images taken with guide probe ($3\times 240$s exposure).
Our main analysis results are based on the deepest data set (A)
(see \S \ref{subsec:a520}),
which was also used by Mahdavi et al. (2007) for their multi-telescope,
multi-bandpass weak lensing analysis of the cluster. 
We compare in Appendix the resulting mass maps
derived from the thee different imaging data sets.
We note that the A520 data were taken with large dithering offsets
of $\approx 2\farcm 3$,
whereas the other cluster data were taken with 
a dithering offset of $1'$.
A summary of observation parameters and conditions is given in Table
\ref{tab:subaru_data}.

Each Subaru data set was reduced  using the software SDFRED (Yagi et
al. 2002; Ouchi et  al. 2004). 
For each CCD frame we first estimated and subtracted 
the bias level by the median value in the overscan region.
Next we carried out flat-fielding for all of the CCD frames
to correct for the pixel-to-pixel variation in sensitivity. 
We then combined mosaic CCD images and 
corrected for the field distortion due to the camera optics.
In order to obtain the local sky background map 
we first measure the median sky level in each mesh of a grid, 
and then compute a high-resolution background map via the 2D bilinear
interpolation. Here we took the mesh size to be $32\times 32$ pixels
with $0\farcs 202$ pixel$^{-1}$.
We subtract the sky background measured 
from our images.
We matched the
point spread function (PSF) between the frames
and masked out areas and objects
vignetted by the AG probe and bad pixels, such
as satellite trails.
Finally we generate a median combined image from these individual
frames.
Here stacking parameters such 
as the coordinate shifts, rotations, and
scalings 
were determined by stellar objects common to all exposures.
A summary of the seeing FWHMs of final reduced images
for the weak lensing analysis 
is shown in Table \ref{tab:subaru_data}. 

In order to make an accurate comparison between
optical and X-ray images,
we performed astrometry using the IRAF 2.11 tasks
(CCMAP and CCSETWCS)
with our object catalogs generated using
SExtractor (Bertin \& Arnouts 1996) and the USNO-A2.0 catalog as our
reference.
After the astrometry correction
typical rms residuals are about 1 pixel ($=0\farcs 202$).

\subsubsection{Chandra Data Analysis}

We used archival Chandra ACIS data for the X-ray analyses of
A520, A754, A1914 and A2142. 
The X-ray observations are identified
by their ID numbers (Obs. ID) in 
Table \ref{tab:subaru_data}.
The data were reduced
using the CIAO 
(version 3.3 for A520, A754 and A2142, 
 and ver 3.2.2 for A1914) 
for the X-ray data processing. 
Standard screening was applied to
the photon list. 
Bad pixels and columns were removed with ASCA grades
0, 2, 3, 4 and 6. 
The data were then cleaned of periods of anomalous background rates
which are more than $3\sigma$ significance from the mean value.
We found an aspect offset 
in the reduced data
of A754 (${\rm Obs.ID}=577$),
which was then corrected using the aspect calculator.
We applied the CTI correction to our data
with focal plane temperature of $-120^\circ{\rm C}$,
because of the charge loss in 
ACIS-I CCD chips during transfer to the readout node.
Background was derived from blank sky data.
We normalized the background data 
so that the count rates in the $10-12$ keV band
matched our observed count rates in the outer region of 
Chandra field-of-views.

The raw X-ray image was binned by a factor of 8 in 
each dimension (DETX and DETY).
Point sources detected by 
WAVDETECT in the CIAO package was excluded from 
the images.
X-ray images were adaptively smoothed, and background and exposure images
were smoothed using the same kernel. Smoothed X-ray images were obtained
by substituting the smoothed backgrounds and dividing them by smoothed
exposure maps.

\subsubsection{XMM-Newton Data Analysis}

We used archival XMM-Newton data for
A1750, A1758 and A2034 
from three EPIC cameras (MOS1, MOS2 and PN).
X-ray observations are identified
by their ID numbers (Obs. ID) in Table \ref{tab:subaru_data}.
We created calibrated event files using SAS version 6.5.0. 
The XMM-Newton
data during high background flares were removed. 
To identify the high background periods,
we generated light curves in the $>10$ keV energy band.
The light curves were binned in 100 s intervals.
We obtained good time intervals (GTIs) 
by applying $2\sigma$ clipping. 
We confirmed that
GTIs are consistent with previous works done by 
Belsole et al. (2004) and David \& Kempner (2004).
The data were then filtered to 
leave only events
with ${\rm PATTERN} \leq 12$ and \#XMMEA\_EM for MOS and ${\rm FLAG}=0$ and
${\rm PATTERN}\leq4$ for PN.

We followed the double-subtraction method (Arnaud et al. 2002) for
background modeling.
Background files were produced by taking into account three different
background components: 
particle-induced background, 
cosmic X-ray background (CXB) 
and residual soft photon contamination.
The particle background was estimated from the filter wheel closed (FWC)
data released from the EPIC Background Working Group.
The particle background was re-normalized 
in the $0.3-10$ keV energy band
by the
ratio of the observation counts over the FWC particle counts 
outside the XMM field-of-view.
For a CXB background estimation, we used blank-sky background data 
of the same instrument, filter and mode as those used in each
observation. 
The blank-sky background counts were normalized to match the 
observed event counts in a source-free annulus at large radius.
We obtained soft X-ray excess images 
by matching in the low energy band
the background count rates to 
our observed count rates in the outer region of the XMM field-of-view.
Finally these normalized background images were
combined to form the total X-ray background.

The raw X-ray image was then smoothed by a factor of 64 in each
dimension
(DETX and DETY).
Finally we obtained for each target adaptively smoothed,
background-subtracted and exposure-corrected mosaic of the MOS1, 
MOS2 and PN images,
in the same manner as done for the Chandra data.

\subsection{Cluster Galaxy Selection}\label{subsec:gal}

\begin{figure}
\begin{center}
\FigureFile(90mm,90mm){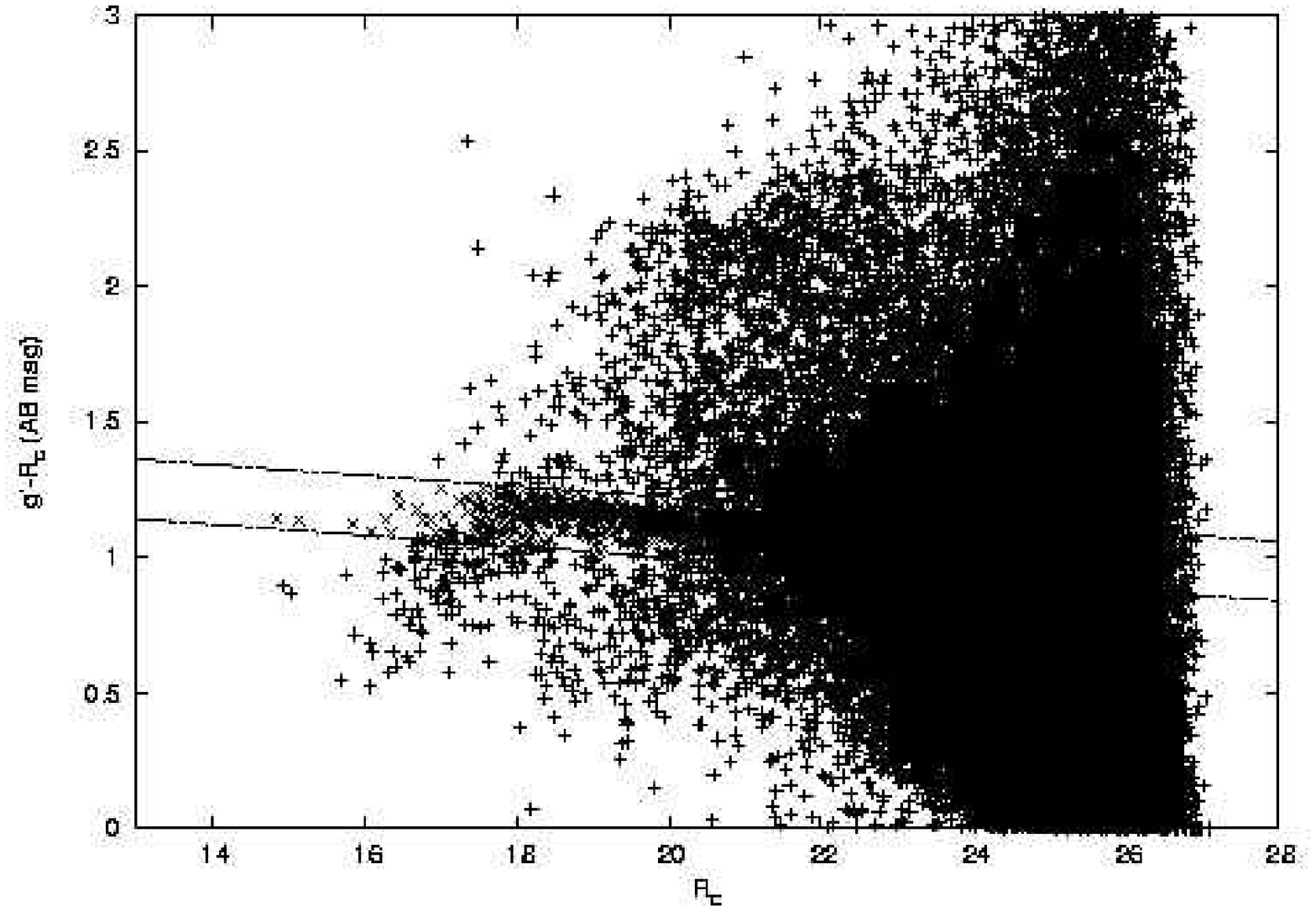}
\end{center}
\caption{The color-magnitude diagram of A2142: The early-type galaxies
 in clusters are well
 separated by their colors thanks to the redshifted $4000$\AA~ break. }\label{fig:CMR}
\end{figure}

We performed object detection and 
aperture photometry using SExtractor 
(Bertin \& Arnauts 1996)
in
dual-image mode with deep $i'$ or $R_{\rm c}$ 
as the detection image,
yielding a common object catalog for the two filters,
$(g',R_{\rm c})$ or $(V, i')$.
We extracted all objects with isophotal area larger than
10 pixels above $3\sigma$ ${\rm pixel}^{-1}$ of the local sky level. 
We used MAG\_AUTO and MAG\_APER of SExtractor output
as object's total magnitude and aperture magnitude, respectively. 
For aperture photometry 
the aperture size is set to 10 pixels ($2\farcs 02$),
 except 20 pixels ($4\farcs 04$) for A754 at a very low redshift of
$z=0.0542$.
We define for each target cluster a sample of cluster member galaxies
according to their color ($g'-R_{\rm c}$ or $V-i'$) and magnitude
($R_{\rm c}$ or $i'$).
Figure \ref{fig:CMR} shows 
the color magnitude diagram (CMD) for one of our target clusters,
A2142, after the removal of stellar objects.
The CMD in Figure \ref{fig:CMR} exhibits a tight color-magnitude (CM)
sequence of early-type cluster galaxies,
since the two filters bracket the rest-frame $4000$\AA~break.
We quantify the CM sequence of galaxies with a linear relation
in color-magnitude space.
We set the maximum magnitude of the linear CM relation
to $R_{\rm c},i'=22$ ABmag.
It can be seen 
that the luminous end of CM sequence extends to the high-luminosity
regime of $R_{\rm c} \approx 16$ ABmag.
For individual galaxies,
we convert from apparent to absolute magnitudes by 
using the k-correction for early-type galaxies.
Here we assume that all of the member galaxies 
in each target cluster are located at a single
redshift summarized in Table \ref{tab:sample}.

We measure the optical luminosity of an individual cluster target
by summing all the cluster galaxies within a given radius from the
cluster center.
 Here we define the cluster center to be either 
at the brightest cluster galaxy (hereafter BCG).
To do this we use a sample of bright CM-sequence galaxies
with magnitudes $R_{\rm c}, i' < 22$ ABmag.
We note that BCGs tend to deviate from the linear CM relation.
Hence we visually checked if such brightest galaxies were
properly included in our cluster member sample, and
included them if they were missing.
A field correction is applied to the measured luminosity 
density of cluster galaxies 
to account for contamination by field galaxies:
We estimate the background luminosity density from an annular region outside of the target region, and then subtract this
background contribution from the observed luminosity density of the
cluster.

In order to measure the total luminosity of  cluster galaxies,
we also need to correct for incompleteness of the sample
due to the magnitude limit.
We assume that cluster galaxies follow
a Schechter luminosity function (Schechter 1976) of the form:
\begin{eqnarray}
\phi(L)=\frac{dN}{dL}
=\frac{\phi^*}{L^*}\left(\frac{L}{L^*}\right)^{-p} \exp(-L/L^*).
\end{eqnarray}
By integrating $\phi(L)$ down to the luminosity cutoff
corresponding to the magnitude limit $L_{\rm min}$ of our cluster member
sample, 
we obtain the following relation between 
the total luminosity $L_{\rm tot}$ and the observed luminosity
$L_{\rm obs}$ of the cluster:
\begin{eqnarray}
L_{{\rm tot}}=L_{{\rm obs}}
\frac{\Gamma(2-p)}{\Gamma(2-p, L_{{\rm lim}}/L^*)},
\end{eqnarray}
where $\Gamma(a,x)$ is the upper incomplete gamma function. 
We adopt the following parameters for the cluster luminosity function:
$p=1.03, M^*_{R_c}=-21.89+5\log(h)$ 
in $R_{\rm c}$-band and 
$p=0.70, M^*_{i'}=-22.31+5\log(h_{70})$ 
in $i'$-band (Goto et al. 2002), 
where $M^*$ is the absolute magnitude corresponding to 
the characteristic luminosity $L^*$. 
The resulting luminosity correction factors,
$\Gamma(2-p)/\Gamma(2-p, L_{\rm lim}/L^*)$, 
are of order of unity as shown in Table \ref{tab:cmr}.

\subsection{Weak Lensing}

In the weak lensing analysis we aim to reconstruct 
from image distortions of background galaxies
the dimensionless surface mass density 
\begin{equation}
\label{eq:kappa}
\kappa(\btheta)=\Sigma^{-1}_{\rm cr}\,\Sigma(\btheta)
\end{equation}
of clusters in units of the critical surface mass
density for gravitational lensing,
\begin{eqnarray}
\label{eq:sigmacrit}
  \Sigma_{\rm cr}=
\frac{c^2}{4\pi G}\frac{D_s}{D_d D_{ds}} \label{eq:sigma_cr},
\end{eqnarray}
where $D_s, D_d$ and $D_{ds}$ are the angular diameter distances 
from the observer to the sources, from the observer to the deflecting
lens, 
and from the lens to the sources.
In the following we closely follow the standard notation 
of Bartelmann \& Schneider (2001).

\subsubsection{Weak Lensing Distortion Analysis}\label{subsubsec:wlana}

We use for our weak lensing analysis
the IMCAT package developed by N. Kaiser
\footnote{http://www.ifa.hawaii/kaiser/IMCAT},
following the formalism developed by Kaiser, Squires, \& Broadhurst (1995).
We have modified the method somewhat following the procedures
described in Erben et al. (2001). We used the same analysis pipeline
as in 
    Broadhurst, Takada, Umetsu et al. (2005),    
    Medezinski, Broadhurst, Umetsu et al. (2007),
    and 
    Umetsu \& Broadhurst (2007).
Note that our weak lensing analysis on A520 is based on 
the deepest data set in Table \ref{tab:sample}, and
a comparison of the mass reconstructions between different
imaging data sets is given in Appendix, demonstrating the effects
of different observation conditions and strategies
 on the weak lensing analysis.

We measure the image ellipticity 
$e_{\alpha} = \left\{Q_{11}-Q_{22}, Q_{12} \right\}/(Q_{11}+Q_{22})$ 
from the weighted quadrupole moments of the surface brightness of
individual galaxies,
\begin{equation}
Q_{\alpha\beta} = \int\!d^2\theta\,
 W({\theta})\theta_{\alpha}\theta_{\beta} 
I({\btheta})
\ \ \ (\alpha,\beta=1,2)
\end{equation} 
where $I(\btheta)$ is the surface brightness distribution of an object,
$W(\theta)$ is a Gaussian window function matched to the size of the
object.

Firstly the PSF anisotropy needs to be corrected using the star images
as references:
\begin{equation} 
e'_{\alpha} = e_{\alpha} - P_{sm}^{\alpha \beta} q^*_{\beta}
\label{eq:qstar}
\end{equation}
where $P_{sm}$ is the {\it smear polarisability} tensor 
being close to diagonal, and
$q^*_{\alpha} = (P_{*sm})^{-1}_{\alpha \beta}e_*^{\beta}$ 
is the stellar anisotropy kernel.
We select bright, unsaturated foreground 
stars identified in a branch of the 
circularized half-light radius ($r_h$) vs.
magnitude ($i'$) diagram 
to measure $q^*_{\alpha}$. 
In Table \ref{tab:sample} we summarize 
basic information and statistics of the 
stellar samples used for PSF corrections in our weak lensing analysis.
To obtain a smooth map of $q^*_{\alpha}$ used in equation (\ref{eq:qstar}), 
we divided the co-added mosaic
image, whose size is about $\sim 11{\rm K}\times 9{\rm K}$ pixels,
into $5\times 4$ chunks.
The chunk length is determined based on the typical coherent scale of
PSF anisotropy patterns.
By this, PSF anisotropy in individual chunks can be well described
by fairly low-order polynomials.
The typical number of stars per chunk is $20$--$50$, depending
on the field (see $N^*$ of Table \ref{tab:sample}).
We fitted the $q^*$ in each chunk independently
with second-order bi-polynomials, $q_*^{\alpha}(\btheta)$,
in conjunction with iterative $\sigma$-clipping rejection on each
component of the residual:
$\delta e_{\alpha}^* =
e^*_{\alpha}-P_{*sm}^{\alpha\beta}q^*_{\beta}(\btheta)$.
As summarized in Table \ref{tab:estar}, uncorrected ellipticity
components of stellar objects have on average a mean 
of $1-2 \%$ with a few $\%$
of rms, or variation of PSF across the data field.
On the other hand, the residual $\delta e^*_{\alpha}$ after correction 
is reduced to $|\overline{\delta e}_{\alpha}^*| \simlt 10^{-4}$.
After the anisotropic PSF correction the rms value of stellar ellipticities,
$\sigma(\delta e^*)\equiv \sqrt{\left<|\delta e^*|^2\right>}$, 
is reduced from a few $\%$ to $(4-8)\times 10^{-3}$.
We show in Figure \ref{fig:e+eres} 
distributions of stellar ellipticity components
before and after the PSF anisotropy correction
for our target clusters.
Figure \ref{fig:emap} shows the distortion fields of
stellar ellipticities before and after the PSF anisotropy correction.
From the rest of the object catalog, we select objects with
$\overline{r_h}^* \lesssim r_h \lesssim 15$ pixels
as 
a weak lensing galaxy sample,
where $\overline{r_h}^*$ is the median value of stellar half-light
radii (see Table \ref{tab:estar}), corresponding to the half median width of
circularized PSF over the data field.
An apparent magnitude cut off is also made 
to remove from the weak
lensing galaxy sample bright foreground/cluster galaxies and very faint
galaxies with noisy shape measurements. 
Table \ref{tab:fbg} summarizes 
the magnitude range and 
the mean surface number density $n_g$ of 
background galaxies 
for our sample of target clusters. 
Without color selection the $n_g$ is ranging from 
$n_g \approx 37 \,  {\rm arcmin}^{-2}$  (A520) to 
$n_g \approx 72 \, {\rm arcmin}^{-2}$ (A2142),
depending on the depth, 
the seeing condition of the observations, and the degree of
contamination of cluster/foreground galaxies.

Second, we need to correct the isotropic smearing effect on 
image ellipticities
caused by seeing and the Gaussian window function used for the shape
measurements. 
The reduced shear $g_\alpha = \gamma_{\alpha}/(1-\kappa)$ 
can be estimated from 
\begin{equation}
\label{eq:raw_g}
g_{\alpha} =(P_g^{-1})_{\alpha\beta} e'_{\beta}
\end{equation}
with the {\it pre-seeing shear polarizability} tensor
$P^g_{\alpha\beta}$.
In the weak lensing limit where $|\kappa|, |\gamma_{\alpha}| \ll 1$,
the image ellipticity is linearly proportional to the gravitational
shear $\gamma_{\alpha}$:
$g_{\alpha} \approx \gamma_{\alpha} \approx (P_g^{-1})_{\alpha\beta} 
e'_{\beta}$.
We follow the procedure described in Erben et al. (2001)
to measure $P_g$ for an individual galaxy (see also 
\S~3.4 of Hetterscheidt et al. 2007).
We adopt the scalar correction scheme, namely,
\begin{equation}
(P_{g})_{\alpha\beta}=\frac{1}{2}{\rm tr}[P_g]\delta_{\alpha\beta}\equiv
P_g^{\rm s}\delta_{\alpha\beta}
\end{equation}
(Hudson et al. 1998; Hoekstra et al. 1998; Erben et al. 2001;
Hettersheidt et al. 2007).
However, the $P_{g}^{\rm s}$ measured for individual galaxies
are still noisy
especially for small and faint galaxies.  
We thus adopt a smoothing scheme in object parameter space proposed by 
Van Waerbeke et al. (2000; see also Erben et al. 2001; Hamana et al. 2003).
In this scheme
we first identify $N$ neighbors for each object in
$r_g$-${\rm mag}$ parameter space.
We then calculate over the local ensemble
the median value $\langle P_g^{\rm s}\rangle$
of $P_g^{\rm s}$ 
and the variance $\sigma^2_{g}$ 
of $g=g_1+ig_2$ using equation (\ref{eq:raw_g}).
The dispersion $\sigma_g$ is used as an rms error of the shear estimate
for individual galaxies.
We adopt $N=30$.
For each cluster field we compute the mean variance $\bar{\sigma}_g^2$
over the background galaxy sample.
In Table \ref{tab:fbg} we listed the mean rms
$\bar{\sigma}_g\equiv \sqrt{\bar{\sigma}_g^2}$ per galaxy, which is of
the order of $\bar{\sigma}_g\approx 0.4$.
Finally we use the following estimator for the reduced shear:
\begin{equation}
g_{\alpha} = e'_{\alpha}/\left< P_g^{\rm s}\right>.
\end{equation}

\subsubsection{Weak Lensing Mass Reconstruction}\label{subsubsec:massmap}

Having obtained the shear estimates for a sample of
background galaxies, we then pixelize 
the distortion data 
into a regular grid of pixels using a Gaussian 
$w_g(\theta)\propto \exp[-\theta^2/\theta_g^2]$ with $\theta_g = {\rm
FWHM}/\sqrt{4\ln{2}}$. 
Further we incorporate in the pixelization
a statistical weight $u_g$ for an individual galaxy, so that the
smoothed estimate of the reduced shear field at an angular position
$\btheta$ is written as
\begin{equation}
\label{eq:smshear}
\bar{g}_{\alpha}(\btheta) = \frac{\sum_i w_g(\btheta-\btheta_i) u_{g,i} g_{\alpha,i}}{\sum_i
w_g(\btheta-\btheta_i) u_{g,i}}
\end{equation}
where $g_{\alpha,i}$ is 
the reduced shear estimate of the $i$th galaxy
at angular position $\btheta_i$, 
and $u_{g,i}$ is the statistical weight of $i$th galaxy
which is taken as the inverse variance,
$u_{g,i}=1/(\sigma_{g,i}^2+\alpha^2)$, 
with $\sigma_{g,i}$ being the rms error for
the shear estimate of $i$th galaxy (see \S~\ref{subsubsec:wlana})
and $\alpha^2$ being the softening constant variance (Hamana et
al. 2003). 
We choose $\alpha=0.4$, which is a typical value of 
the mean rms $\bar{\sigma}_g$ over the background sample.
The case with $\alpha=0$ corresponds to an inverse-variance weighting.
On the other hand,
the limit  $\alpha\gg \sigma_{g,i}$ yields a uniform weighting.  
We have confirmed that our results are insensitive to the 
choice of $\alpha$ 
(i.e., inverse-variance or uniform weighting)
under the adopted smoothing parameters.

The error variance for the smoothed shear $\bar{g}=\bar{g}_1+i\bar{g}_2$ 
(\ref{eq:smshear}) is
then given as
\begin{equation}
\label{eq:smshearvar}
\sigma^2_{\bar{g}}(\btheta) = 
\frac{\sum_i w_{g,i}^2 u_{g,i}^2 \sigma^2_{g,i}}
{ \left( \sum_i w_{g,i} u_{g,i} \right)^2}
\end{equation} 
where $w_{g.i}=w_g(\btheta-\btheta_i)$ and 
we have used $\langle g_{\alpha,i}\, g_{\beta,j}\rangle = 
(1/2)\sigma_{g,i}^2\delta^{\rm K}_{\alpha\beta}\delta^{\rm K}_{ij}$
with $\delta^{\rm K}_{\alpha\beta}$ and $\delta_{ij}^{\rm K}$ being the
Kronecker's delta.

We then invert the pixelized reduced-shear field (\ref{eq:smshear}) to
obtain the lensing convergence field. In the map-making we assume the
linear shearing in the weak-lensing limit, that is,
$g_{\alpha}=\gamma_{\alpha}/(1-\kappa) \approx \gamma_{\alpha}$.
We adopt the two inversion methods, namely,
the Kaiser \& Squires inversion method (Kaiser \& Squires 1993) 
and the noise-filtering 
inversion method outlined 
in Seitz \& Schneider (2001). The first method
makes use of the 2D Green function in an infinite space, while the
latter is based on the finite-field solution of the inversion problem
for the reconstruction kernel. The finite-field method must be used for
a nearby cluster where the data field is dominated by positive, biased
density field with $\langle \kappa \rangle >0$.
In the linear map-making process, the pixelized shear field is weighted
by the inverse of the variance (\ref{eq:smshearvar}).
Note that this
weighting scheme corresponds to using only the diagonal part of the
noise covariance matrix,
$N(\btheta_i,\btheta_j) = \langle
\overline{\Delta g}(\btheta_i)\overline{\Delta g}(\btheta_j)\rangle$,
which is only an approximation of the actual inverse noise weighting in
the presence of pixel-to-pixel correlation due to non-local Gaussian 
smoothing. 
In Table \ref{tab:fbg} we summarize the Gaussian FWHM used in
the pixelization and the rms noise level in the reconstructed $\kappa$
field for our sample of target clusters. 

The smoothing scale is chosen so as to optimize the weak lensing
detection of target mass structures, depending both on 
the size of the structure
and
the strength of noise power ($\propto \bar{\sigma}_g^2/n_g$).
Under the adopted smoothing parameters,
the typical rms reconstruction error in the $\kappa$ map is 
$\sim 0.02$
except for A1914,
in which a subarcmin-scale (${\rm FWHM}=0\farcm 75$), 
high resolution reconstruction 
was adopted to resolve substructures (see \S \ref{subsec:a1914}).

\subsubsection{Red Background Galaxy sample}\label{sussubsesc:red}

As demonstrated by Broadhurst et al. (2005) 
and Medezinski et al. (2007), it is
crucial to make a secure selection of {\it background} galaxies in order
to minimize the dilution of the lensing signal by cluster/foreground
galaxies and to make an accurate determination of the cluster mass.
To do this,
we define a sample of {\it red background galaxies}
whose colors are redder 
than the CM sequence of cluster member galaxies
due to large $k$-corrections.
These red background galaxies are largely composed of early to mid-type
galaxies at moderate redshifts (Medezinski et al. 2007). Cluster member
galaxies are not expected to extend to these colors in any significant
numbers because the intrinsically reddest class of cluster galaxies,
i.e. E/S0 galaxies, are defined by the CM sequence and lie blueward of
chosen sample limit, so that even large photometric errors will not 
carry them into our red sample.

In Table \ref{tab:red} we summarize our selection criteria for the red
background sample and the resulting mean surface density of the red
galaxies.  The larger the cluster redshift, the redder the CM sequence
of cluster galaxies due to larger $k$-corrections. Hence, the number of
red background galaxies, selected in this way, will decrease with
cluster redshift. In particular, A1758 is at a moderately high redshift
of $z=0.279$, and there are very few such red galaxies remained. We
therefore relaxed the selection criteria of the red background sample
for the case of A1758 (see Table \ref{tab:red}).

\subsubsection{Tangential Distortion and Cluster Mass  Profile}\label{subsubsec:fitting}
 
For an individual cluster,
we derive an azimuthally-averaged
shear profile as a function of projected
radius from the fiducial cluster center,
which is chosen as the optical center
 (see \S \ref{subsec:gal}).

The tangential component $g_{+}$ is used to obtain the azimuthally averaged
distortion due to lensing, and computed from the distortion
coefficients $(g_{1},g_{2})$ of each object:
\begin{equation}
g_{+}=-( g_{1}\cos2\phi +  g_{2}\sin2\phi),
\end{equation}
where $\phi$ is 
the position angle of an object with respect to the cluster center,
and 
the uncertainty in the $g_+$ measurement
is $\sigma \equiv \sigma_g/\sqrt{2}$ in terms of the rms error $\sigma_g$
for the reduced shear measurement.

The estimation of $g_{+}$ only has significance 
when evaluated statistically over a large number
of background galaxies because of the intrinsic spread in shapes and
orientations as well as the measurement errors in the shape measurement.
In radial bins we
calculate the weighted average of the $g_+$s and its weighted error:
\begin{eqnarray}
\label{eq:mean_gt}
\langle{}g_{+}(\theta_n)\rangle{}
&=& \frac{\sum_i u_{g,i} g_{+,i} } 
         {\sum_i u_{g,i} }, \\ 
\label{eq:err_gt}
\sigma_+(\theta_n)
&=&
\sqrt{\frac{\sum_i u_{g,i}^2 \sigma^2_{i}}
{ \left( \sum_i u_{g,i} \right)^2}},
\end{eqnarray} 
where the index $i$ runs over all of the objects located within the
$n$th annulus with a median radius of $\theta_n$,
and $u_{g,i}=1/(\sigma_{g,i}^2+\alpha^2)$ is the inverse variance weight
softened with $\alpha=0.4$ (see \S \ref{subsubsec:massmap}).

For a parameter-free estimation of the cluster mass profile
we use 
the aperture-densitometry, or so-called the $\zeta_c$-statistic 
(Fahlman et al. 1994; Clowe et al. 2000) of the form:
\begin{eqnarray}
\label{eq:zetac}
 \zeta_c(\theta; \theta_{\rm inn},\theta_{\rm out})
&=&\bar{\kappa}(<\theta) - 
   \bar{\kappa}(\theta_{\rm inn}< \theta <\theta_{\rm out}) \\
                     &= & 2\int^{\theta_{\rm inn}}_{\theta} d \ln \theta'
                      \langle \gamma_+ (\theta)
		      \rangle \nonumber \\ 
                     & + &  \frac{2}{1-\theta_{\rm inn}^2/\theta_{\rm out}^2} 
                      \int^{\theta_{\rm out}}_{\theta_{\rm inn}}
		      d \ln \theta' \langle \gamma_+(\theta)\rangle  \nonumber 
\end{eqnarray}
where $\theta_{\rm inn}$ and $\theta_{\rm out}$ 
are the inner and outer radii of 
the annular background region 
in which the mean background contribution,
$\bar{\kappa}(\theta_{\rm inn}<\theta<\theta_{\rm out})$, is estimated;
The $\langle \gamma_+ \rangle$ is an azimuthal average of the
tangential component of the gravitational shear, which we take 
$\langle \gamma_+(\theta)\rangle \approx \langle g_+(\theta)\rangle$ 
in the weak lensing limit. 
Then
the projected mass of the cluster 
inside the projected radius $\theta$ is given as
\begin{eqnarray}
 M_\zeta(<\theta)=\pi \theta^2 \Sigma_{\rm cr}  \zeta_c(\theta;
  \theta_{\rm inn}, \theta_{\rm out}).
\end{eqnarray}
$M_{\zeta}(<\theta)$ is regarded as a lower bound to the true enclosed
mass $M(<\theta)$, 
because of the subtraction of
the mean $\kappa$ within the outer region,
$\theta_{\rm inn}< \theta < \theta_{\rm out}$.
Errors on $\zeta_c$ are calculated by propagating 
the rms errors $\sigma_+(\theta_n)$ (equation [\ref{eq:err_gt}]) 
for the tangential shear measurement.

The lensing properties such as $\kappa$ and $\gamma_{\alpha}$ depend on
the source redshift and the background cosmology through the critical
lensing surface mass density, $\Sigma_{\rm cr}$, defined in equation
(\ref{eq:sigmacrit}). 
For all of the cluster targets,
we assume the mean redshift of the red background galaxies is 
$\langle z_s \rangle=1$ (e.g., Broadhurst et al. 2005).
Since
our target clusters are located at low redshifts ($z_d\simlt 0.2$ except
$z_d=0.279$ for A1758) 
$\Sigma_{\rm cr}$ depends weakly on the source redshift 
as shown in Figure \ref{fig:lensS},
so that a precise knowledge of
the redshift distribution of background galaxies is not crucial (see,
e.g., Bartelmann \& Schneider 2001).

In order to quantify and constrain the cluster mass properties
we fit the following two different mass models to $\zeta_c$-statistic
measurements: singular
isothermal sphere (SIS) and universal density profile proposed by
Navarro, Frenk, \& White (1996, hereafter NFW).
The details of the halo models are given in Appendix.
We parametrize our halo models using the cluster virial properties,
such as the cluster virial mass, $M_{\rm vir}$, the cluster virial
epoch, $z_{\rm vir}$, and the cluster virial radius, $r_{\rm vir}$:
\begin{equation}
M_{\rm vir} = \frac{4\pi}{3}\bar{\rho}(z_{\rm vir})\Delta_{\rm
vir}r_{\rm vir}^3, 
\end{equation}
where 
 $\Delta_{\rm vir}$
is the overdensity with respect to the mean cosmic density
$\bar{\rho}(z_{\rm vir})$ at the cluster virial redshift,
predicted by the dissipationless spherical
tophat collapse model (Peeebles 1980; Eke, Cole, \& Frenk 1996; Bullock
et al. 2001).
We assume the cluster redshift $z_d$ is
 equal to the cluster virial redshift $z_{\rm vir}$.

The SIS model has a one-parameter functional form described by the
one-dimensional velocity dispersion $\sigma_v$ of the cluster,
which is related with the cluster virial redshift $z_{\rm vir}$ and the
cluster virial mass $M_{\rm vir}$ as
\begin{equation}
\sigma_v(M_{\rm vir},z_{\rm vir}) = \frac{1}{2} r_{\rm vir} H_0 
\sqrt{\Omega_{m0} \Delta_{\rm vir} (1+z_{\rm vir})^3}
\end{equation}
Thus, $\sigma_{v}\propto M_{\rm vir}^{1/3}$.
Alternatively we can introduce the virial temperature of an SIS halo:
\begin{eqnarray}
\label{eq:TSIS}
\mu m_p \sigma_v^2 = k_B T_{{\rm SIS}}
\end{eqnarray}
where $\mu=0.62$ is the mean molecular weight and $m_p$ is the proton
mass.

On the other hand, the NFW model has a two-parameter functional form,
and we take the virial mass, $M_{\rm vir}$, and the concentration
parameter, $c_{\rm vir}=r_{\rm vir}/r_s$, with $r_s$ being the inner
characteristic radius of the NFW profile.

\section{DISTRIBUTIONS OF MASS AND BARYONS IN MERGING CLUSTERS}\label{sec:maps}

Here we present and compare 
in Figures \ref{fig:a754}-\ref{fig:a520}
the resulting two dimensional maps of 
the convergence $\kappa$, 
optical luminosity density,
and  X-ray emission for our sample of merging clusters.
Table \ref{tab:fbg} lists the properties of 
our background galaxy samples and the parameters relevant for weak
lensing mass reconstructions.
The cluster targets are selected to be on the various merging stages
based on previous detailed X-ray studies, for which
the physical relationship between mass and baryons though the merging
processes is yet unknown.
X-ray properties of the target clusters are summarized in Table
\ref{tab:sample}. 
A1750 and A1758 are binary clusters 
which are presumably in a pre-merger phase.
A2034 and A2142 are cold front clusters. A754, A1914
and A520 are on-going mergers, which 
have irregular temperature distributions and 
radio halos (Govoni et al. 2004).

\subsection{A754} \label{subsec:a754}

A754 is an on-going merger at a redshift of $z=0.054$. 
For such a low redshift cluster ($z<0.1$), 
the expected lensing signal is very low (Figure \ref{fig:lensS}) and a weak lensing analysis is challenging. 
As seen from the Chandra X-ray contours in the top-right panel of Figure
\ref{fig:a754}, there are two major gas components 
in the east and west sides of the data field. 
The west gas clump (XC in Figure \ref{fig:a754})
corresponds  to the main cluster having a BCG.
From this configuration,
the east X-ray gas clump 
(XE in Figure \ref{fig:a754})
seems to have been running
through the cluster center from the west.
The XMM-Newton image with a wider field-of-view has shown a 
moving feature of the east clump consistent with the Chandra X-ray
observation 
(Henry, Finoguenov, \& Briel. 2004). 
The X-ray contours of the east gas clump are compressed and elongated 
towards the north-east direction, 
indicating that
the east gas clump is currently moving towards the northeast. 
Its averaged temperature within $9$ arcmin radius
is $10.0\pm0.3$ keV 
at the $90\%$ confidence level
(Markevitch et al. 2003; see
also Table \ref{tab:sample}).
The X-ray temperature maps of A754 were
derived by Henry \& Briel (1995) with ROSAT data, 
Henriksen \& Markevitch (1996) with ASCA data, 
Markevitch et al. (2003) with Chandra data,
and Henry,  Finoguenov \& Briel (2004) with XMM-Newton data. 
Henry \& Markevitch (1996) obtained an X-ray temperature of 
$8.5$--$9.0$ keV from the ASCA data.
Henry et al. (2004) derived an X-ray temperature
within $12'$ to be $8.6\pm 0.1\pm 0.6$ where the first and the second
errors are statistical and systematic uncertainties at $1\sigma$ confidence, 
respectively.
The ASCA and XMM-Newton temperatures of A754 are somewhat lower than
the Chandra temperature. However, the ASCA and Chandra measurements
covered larger volumes of this on-going merger.
The Chandra temperature map shows a strongly irregular feature,
where the cool gas is offset 
from the X-ray brightness peak
towards the northeast by $\sim 200$ kpc.

Since this nearby cluster extends outside the field-of-view of
Subaru/Suprime-Cam, we used a finite-field method for reconstructing the
mass distribution of A754 as shown in Figure \ref{fig:a754}. 
We also compared the $\kappa$ map derived 
using the Kaiser \& Squires method with the finite-field based $\kappa$ map.
The $\kappa$ maps derived with the different methods are qualitatively
similar, and the main features and quantitative 
properties of the mass peaks are
consistent with each other within $1\sigma$ reconstruction error, 
which means that the mass reconstruction is
insensitive to the boundary conditions.
The Gaussian FWHM used for the mass reconstruction
is $1\farcm 67$.
The reconstructed $\kappa$ map shows
two mass clumps in the west and east sides of the data field.
The west clump is located in the X-ray cluster central region (Henry,
Finoguenov \& Briel 2004), and hence we refer to this as the central
mass clump 
(C in Figure \ref{fig:a754}).
We found moderate signal-to-noise ratios of 5.8 and 5.1 
in the $\kappa$ map
for the central and east mass 
(E in Figure \ref{fig:a754})
clumps, respectively,
thanks to the superb image quality of Subaru/Suprime-Cam.

The peak locations in the luminosity map of 
cluster sequence galaxies are in good agreement with these mass
substructures (bottom-left panel).
The central mass peak coincides with the BCG (top-left panel). 
The east mass concentration contains a luminous, large elliptical
galaxy.

The east mass clump is offset 
from the east X-ray clump towards the southwest,
which is opposite to the moving direction of this gas clump
(bottom-right panel).
This kinematic feature is different from the case of 
the bullet cluster,
1E0657-56
(Clowe,Gonzale \& Markevitch 2004).
The cool gas region (see Markevitch et al. 2003) is 
located outside the region enclosed by the 
$1\sigma$ mass contour around the east mass clump.

A possible scenario explaining the observed
 merger geometry of the east mass and X-ray clumps,
which cannot be simply explained by the ram-pressure stripping, 
is the following:
the east mass substructure
just reaches its apocenter of the merger orbit and falls back towards the
center for its second impact.
In the rest frame of the mass clump, the X-ray core, which is
initially bounded in the mass clump, feels the force in the opposite
direction to the acceleration (and moving) direction of the mass clump.
This explains the observed configuration of the
mass and gas clumps associated with the east substructure.
As a result, the gas clump escapes away from the potential well
of the mass substructure and would cool adiabatically as it expands.

\subsection{A1750} \label{subsec:a1750}

A1750 is a binary cluster at $z=0.086$, well studied by Einstein, ROSAT,
ASCA and XMM-Newton X-ray satellites (Forman et al. 1981; Novicki, Jones, \& Donnelly 1998; Donnelly et al. 2001; Belsole et al. 2004).
The XMM-Newton observation (Belsole et al. 2004) has shown 
that the southern cluster, A1750C, has a
higher X-ray luminosity than the northern cluster, A1750N. 
The projected distance between the two X-ray peaks is 
about $900$ kpc. 
The X-ray temperatures of A1750C and A1750N are
$3.87\pm0.10$ keV and $2.84\pm0.12$ keV at 90\% confidence, 
respectively, 
(Belsole et al. 2004).
The projected separation 
is therefore shorter than the sum the two virial radii predicted by the 
$M-T_X$ relation.
Hence, it is likely that 
the two clusters just started to interact with each other.
Indeed, the temperature in the middle region of the two clusters
is higher than the cluster temperatures:
$T_X=5.12^{+0.77}_{-0.69}$ keV.
Belsole et al. (2004) estimated a Mach number of $1.64$
for the merger shocks
by applying the Rankine-Hugoniot jump conditions 
under the assumption that the 
pre-shock temperature is approximated by 
the averaged temperature in the central regions of A1750C and A1750N
and
that the post-shock temperature is given by the temperature in the middle
region of the two clusters. 
Belsole et al. (2004) also concluded that A1750C
is an unrelaxed cluster for the following four reasons: 
(1) the discontinuity of the gas density
profile in the southeast region (region 3 in their notation),
(2) the shift of X-ray centroid from the position of 
the BCG,
(3) the lack of evidence for a cooling flow, and 
(4) excess entropy in
the central region, compared with other relaxed clusters.

No significant offset among distributions of the galaxy luminosity,
mass
and ICM
is found in the binary cluster A1750, as shown in Figure \ref{fig:a1750}.

The X-ray surface brightness peak in A1750C coincide with the
position of the BCG
(top-right panel) but is offset from the
centroid of the X-ray emission,
which is consistent with the X-ray analysis by
Belsole et al. (2004).
The position of the main mass peak ($7.7\sigma$) in A1750C coincides
with the peak positions of the optical luminosity (bottom-left
 panel) 
and the X-ray surface brightness (bottom-right panel), 
within a smoothing scale ($1\farcm 25$ FWHM).
No significant mass substructure is found in A1750C.
The observed geometry of mass, 
galaxies,
and X-ray emission in A1750C does
not show a signature of recent strong mergers
as seen in A754.
However, we cannot entirely rule out the possibility 
of a minor merger of small mass clumps below the weak lensing sensitivity.
The averaged X-ray temperature of A1750C 
is higher than the temperature predicted by 
the best-fitting SIS model to the tangential shear profile, as shown
in  Figure \ref{fig:tmp} 
(see also \S \ref{subsec:TvsTsis}). 
This is consistent with the lack of cool gas and the excess entropy in
the central region (Belsole et al. 2004).

A1750N has two 
BCGs whose positions coincide with that 
of the X-ray peak of A1750N (top-right panel). 
The mass peak ($5.9\sigma$) of A1750N is 
slightly, but not significantly, offset from the X-ray and
galaxy-luminosity peaks, 
and the offset is within the smoothing scale of ${\rm FWHM}=1\farcm 25$.
The temperature $T_{\rm SIS}$ estimated from weak lensing
coincides with the X-ray temperature $T_X$ 
for A1750N (see \S \ref{subsec:TvsTsis}).

As seen from the bottom-right panel of Figure \ref{fig:a1750},
the reconstructed mass distribution traces an extended X-ray
substructure (XM in Figure \ref{fig:a1750}) between A1750N and A1750C, which is around 
 (RA,DEC)$=(202.77^{\circ}, -1.765^{\circ}$).
Overall, the ICM and galaxy distributions in the binary cluster A1750 and its
components, A1750C and A1750N, are in good agreement with the mass
distribution.

\subsection{A1758} \label{subsec:a1758}

A1758 is a binary cluster at $z=0.2790$. The ROSAT observation
by Rizza et al. (1998)
 showed
that this cluster consists of two clusters (A1758N and A1758S) 
separated by $\sim 8'$ on the sky,
corresponding to the projected 
physical separation of 
$\sim 2{\rm Mpc}$.
David \& Kempner (2004) found from the XMM-Newton observation that
the line-of-sight velocity difference
between
A1758N and A1758S is less than $\Delta V=2100~{\rm km s}^{-1}$.
David \& Kempner (2004) argued that their physical proximity to each
other
and small velocity difference are consistent with those of a
gravitationally interacting system.
The Chandra and XMM-Newton observations showed
that there is no X-ray signature of the interaction between 
A1758N and A1758S (David \& Kempner 2004), contrary to the case of the
binary cluster A1750 which has a high temperature region between
the two components (see \S \ref{subsec:a1750}).
This system is most likely in an earlier merger stage than A1750,
and therefore serves as an ideal target for studying the cluster mass
distribution in the initial stage of the merging process.
As shown in Figure \ref{fig:a1758}, the XMM-Newton X-ray image shows a bridge
connecting between the two cluster components. On the other hand, the
reconstructed mass map shows no significant mass structure corresponding
to the bridge in the X-ray image. Note that the mass structure located
on the X-ray bridge is a local minimum 
(M in Figure \ref{fig:a1758}).
It is also shown in Figure \ref{fig:a1758} that there is no significant
offset between 
the mass structures and the X-ray/optical structures 
of A1758N and A1758S
along the north-south direction connecting between A1758N and A1758S, which is 
consistent with results for A1750 (\S \ref{subsec:a1750}).

The system components, A1758N and A1758S, are both undergoing mergers
(David \& Kempner 2004).  
David \& Kempner (2004) found from XMM-Newton data
average gas temperatures of A1758N and A1758S 
to be $8.2\pm0.4$ keV and $6.4^{+0.3}_{-0.4}$ keV, respectively,
at the 90\% confidence level.
Using Chandra data, David \& Kempner (2004) obtained an average gas
temperature of A1758N to be  $9.0^{-0.6}_{+0.9}$ keV
at the 90\% confidence level.
For A1758N, the XMM-Newton and Chandra measurements of the average
gas temperature are in good agreement with each other.

A1758N has a complex X-ray morphology as seen
in the top right panel of Figure \ref{fig:a1758}: 
there are northwest and southeast X-ray subclumps in the cluster central
region. 
The Chandra X-ray image
suggests that the northwest subclump 
(XC in Figure \ref{fig:a1758})
is currently moving towards the north and that 
the southeast subclump 
(XSE in Figure \ref{fig:a1758})
is currently moving towards the southeast (David \& Kempner 2004).  
The optical 
luminosity distribution (bottom-left panel) reveals two luminous
subclumps of cluster-sequence galaxies in A1758N.
The X-ray peak position of the northwest subclump
coincides with that of the luminous BCG (top-right panel).
The southeast gas subclump is, on the other hand, 
offset 
by $\approx 290$ kpc
to the northwest of the other galaxy subclump (top-right panel).

Figure \ref{fig:a1758} shows that 
mass and light are similarly distributed in A1758N.
The mass map shows double peaks: the first peak 
with a significance of $13.5\sigma$ (SE in Figure \ref{fig:a1758})
corresponding to the southeast luminous galaxy clump, 
and the second peak 
with a significance of $11.0\sigma$ (C in Figure \ref{fig:a1758})
corresponding to the
northwest luminous galaxy clump (bottom-left panel).
A visual inspection also reveals several gravitational arc candidates
around the two mass peaks in A1758N,
supporting the bimodal mass distribution in A1758N.
Figure \ref{fig:a1758N_zoom}
displays zoom in views of tangential arc candidates in the $R_{\rm
c}$-band image.
The tangential arc candidates A, B, and C are bluer than
the cluster red sequence in $g'-R_{\rm c}$ color. 
The tangential arc candidates D and E are associated with
cluster galaxy concentrations.
This bimodal feature of A1758N in the weak lensing mass map was
previously reported by Dahle et al. (2002) based on weak lensing data
taken with ALFOSC on Nordic Optical Telescope.
We note that 
Dahle et al. (2002) also found an arc-like image in A1758N.

The northwest galaxy clump has a higher luminosity than the southeast one.
The angular extent of the northwest galaxy clump is larger than that of
the southeast one.
These optical features would 
suggest that the northwest mass structure (C)
is the primary component of A1758N and the southwest clump (SE) is 
the merging substructure.
$N$-body simulations of $\Lambda$CDM models
show that
weak lensing measurements of cluster mass peaks
have a large scatter 
($\sim 100\%$ for less massive systems with
 $M_{200}\sim 10^{14}M_{\odot} h^{-1}$) 
due to line-of-sight projection effects of intervening mass structures 
as well as due to intrinsic ellipticities of background galaxies
(White, van Waerbeke \&  Mackey 2002; Wu et al. 2006).
Therefore the relative peak heights of the mass peaks SE and C
($13.5\sigma$ and $11.0\sigma$ significance, respectively)
could have been substantially affected by such effects.
The southeast structure in the mass/galaxy map 
is located in front of the X-ray southeast subclump moving
towards the southeast.
This geometry regarding the southeast substructure
is consistent with the results from the joint optical photometric,
X-ray, and weak lensing analysis of the merging cluster 1E0657-56
(Clowe, Gonzale, \& Markevitch 2004).
On the other hand, the northwest structure, which is likely to be the
main component of A1758N, shows no significant offset among the 
galaxy, ICM, and mass distributions,
which is 
different from the results of 1E0657-56.

A1758S has an elongated X-ray emitting core 
located close to the first peak in the galaxy luminosity distribution
(right panels).
David \& Kempner (2004) pointed out that the X-ray 
core has two bow-like shaped edges
which are curved up toward north and south, respectively.
They suggested that these features are 
a signature of the merger
along a line between the northwest and southeast sides  
with a small impact parameter.
No optical counterpart has been detected in the Subaru image (left
panels), whereas two luminous galaxy concentrations are found along the
northeast-southwest direction, which is perpendicular to the
hypothetical merging direction.
The second optical peak associated with A1758S is offset 
from the X-ray core to the southwest by $\sim 480$ kpc.
The high density region with $\kappa > 9\sigma$ significance is elongated along the 
northeast-southwest direction, aligned with the two luminous galaxy
concentrations. 
The mass peak associated with A1758S is located in the middle of the two
optical clumps, and has a peak height of $9.9\sigma$.
The angular separation between the two luminous subclumps is comparable
to the smoothing scale ${\rm FWHM}=1\farcm 25$ for the weak lensing
reconstruction.
At this smoothing scale,
we cannot therefore 
resolve possible mass substructures originally associated with
the optical subclumps.
Nonetheless,
the elongated mass and X-ray structures 
and the existence of two luminous galaxy concentrations
indicate that A1758S is not yet in dynamical equilibrium
and is an on-going merger.
The merger geometry of A1758S is puzzling, 
because the curved-up directions of
the X-ray core are almost
 perpendicular to both directions of 
the elongations of the mass and light distributions.
If the observed direction of the elongation in mass and light 
is regarded as the original collision axis, 
then our results would indicate that 
not all arc-like cores
in the X-ray emission can be used to trace the path of the moving core.

\subsection{A1914} \label{subsec:a1914}

The cluster  A1914 is an on-going merger at $z=0.1712$
with averaged temperature of $10.9\pm 0.7$ keV 
at 90\% confidence (Govoni et al. 2004). 
A1914 has two X-ray cores embedded in a round X-ray halo:
As shown in the top-right panel of Figure \ref{fig:a1914} (see also
 Govoni et al. 2004), the larger X-ray core 
(XE in Figure \ref{fig:a1914})
is elongated in the east-west direction, 
 and the smaller core 
(XW in Figure \ref{fig:a1914})
is located to the  west of the larger core.
Based on Chandra X-ray data,
Govoni et al. (2004) found a filamentary hot region 
along the northeast-southwest direction connecting between
the two small cores, and argued that
the filamentary hot region is a signature of the shock-heated gas
induced by the elongated small X-ray core moving towards the west.

We have obtained for A1914 a high resolution mass map with a Gaussian 
FWHM of $0\farcm 75$ as shown in Figure \ref{fig:a1914} (left panels).
The cluster mass distribution is highly irregular,
whereas the Chandra X-ray map shows a nearly circular morphology except
the central core region (top-right panel).
We found seven mass concentrations 
above a detection threshold of $4\sigma$ significance
in the $\kappa$ map covering a $16'\times 15\farcm 5$ field around
A1914.
There are two prominent mass peaks with $\kappa\simgt 0.3$:
The first mass peak 
(C1 in Figure \ref{fig:a1914})
 has a peak height of  $9.1\sigma$, and its position 
is in good agreement with the second brightest
cluster galaxy (BCG2 in Figure \ref{fig:a1914})
which is close to the elongated X-ray core.
The position of the second mass peak 
with a significance of $7.8\sigma$ (C2 in Figure \ref{fig:a1914}), 
on the other hand,
coincides well with that of the brightest cluster galaxy 
(BCG1 in Figure \ref{fig:a1914}) sitting in the optical center.
The observed geometry of the galaxy and mass distributions 
would suggest that the second mass peak (C2) associated with the
optical cluster center (BCG1) is the primary cluster center,
and the first mass peak (C1) is the merging substructure.
By visual inspection 
we found an arc-like image tangentially-oriented 
with respect to BCG2/C1, as pointed out
by Dahle et al. (2002) and Sand et al. (2005). 
The curvature of this tangential arc candidate is 
consistent with the existence of the high density peak (C1)
around BCG2 (see Figure \ref{fig:a1914_zoom}).

However, it remains puzzling that the projected mass distribution in
A1914 is so irregular contrary to the quasi circular X-ray distribution
in the outer region. 
A detailed analysis will be presented in a forthcoming paper 
(Umetsu \& Okabe in preparation).

\subsection{A2034} \label{subsec:a2034}

A2034 is a cold front cluster at $z=0.1130$ well studied with
ROSAT, ASCA and Chandra X-ray satellites 
(David, Forman, \& Jones  1999;  White 2000; Kempner \& Sarazin 2003).
Based on the Chandra observation, Kempner \& Sarazin (2003)
found a sharp 
discontinuity in the X-ray surface brightness 
$\sim 3'$ to the north of the cluster center,
 similar to the surface brightness jump associated with
cold fronts. 
The X-ray morphology suggests that this dense core is 
moving toward the north direction.
Kempner \& Sarazin (2003) showed that 
the temperature is fairly constant to a
radius of at least about $5'$ 
from the center of the main cluster in A2034, and 
hence 
no significant temperature jump, associated with the dense core,
is found.
They also argued that
the excess emission from the southern part of the dense core 
is likely to be a background
structure. 
Kempner \& Sarazin (2001) found evidence for the existence of a radio
relic near the cold front.

The resulting mass distribution in A2034 is highly irregular and
quite different from the X-ray surface brightness distribution of the
XMM-Newton data as shown in Figure \ref{fig:a2034}.
In the central region of this cluster 
($\sim 18.'5\times 18.'5$), 
there are six significant 
mass concentrations with peak heights greater than 
$4\sigma$ significance (left panels).
The main mass peak 
with a significance of $6.4\sigma$ (C in Figure \ref{fig:a2034})
coincides with the first peak in the X-ray surface brightness
distribution (see the bottom-right panel of Figure \ref{fig:a2034}).
There are two BCGs close to the first X-ray and mass peaks,
corresponding to the main cluster of A2034.

The second mass peak with a significance of $6.3\sigma$ 
(S in Figure \ref{fig:a2034})
is located to the south of the main mass peak, 
and coincides with the south excess X-ray emission
discovered by Kempner \& Sarazin (2003). 
No spectroscopic data are available in this region.
The reconstructed mass map reveals another 
three mass structures (W1, W2 and W3 in Figure \ref{fig:a2034})
associated without X-ray counterparts
in the west of the second mass peak.
We confirmed that they are background structures at
$z \sim0.116-0.118$ from SDSS spectroscopic data. 
The velocity difference
between the main cluster and the west structures is $c\delta z \sim
900-1500~{\rm km s^{-1}}$, 
and the comoving radial separation is ${\mathcal O}(10{\rm Mpc})$ 
by neglecting their proper motions.
Their apparent proximity suggests that the cluster A2034 and the west
mass structures could be part of a filamentary structure.

The primary result of our joint analysis on A2034 is the detection
of the northern mass subclump 
with a significance of $5.4\sigma$ (N in Figure \ref{fig:a2034})
located ahead of the north cold front shown in the top-right
panel. The second brightest member galaxy of A2034 is located close to the
center of the north mass subclump (left panels).
This merger geometry is consistent with the 
results of other cold front clusters, 
A2142 (\S \ref{subsec:a2142}), 
 A1758N (\S \ref{subsec:a1758}),
and 1E0657-56 (Clowe, Gonzale, \& Markevitch 2004).

\subsection{A2142} \label{subsec:a2142}

A2142 is a cold front cluster at $z=0.0909$. 
Cold fronts were discovered for the first time 
in this cluster by the Chandra X-ray observation
(Markevitch et al.  2000). 
There are two cold fronts in the central region of A2142, 
as indicated in the top-right panel of Figure \ref{fig:a2142}:
One is at the northwest edge 
(Cold Front NW) of a
large core, and the other at the south edge 
(Cold Front S) of a small core in the X-ray
surface brightness distribution.
The small core is apparently 
contained within the large core in the sky plane.
Those edges indicate that the large and small cores are moving
toward the northwest and the south directions, respectively.
The temperatures change abruptly across the cold fronts by a factor
of two. 
There are two BCGs (BCG1 and BCG2) in the cluster core region,
as indicated in the top-right panel of Figure \ref{fig:a2142}: 
BCG1 is located inside the small core, and BCG2 is apparently located in
the large core.
The two BCGs have a large line-of-sight velocity difference
of $1840$ ${\rm km s^{-1}}$ (Oegerle, Hill, \& Fitchett 1995).

Okabe, Umetsu \& Hattori (2008) have obtained a mass map with
low angular resolution of ${\rm FWHM}=3\farcm 3$ 
using a different set of Subaru weak lensing data
taken under bad weather conditions.
Okabe et al. (2008) found that
the mass distribution is similar to the luminosity distribution
of spectroscopically selected bright cluster member galaxies,
but is quite different from the ICM distribution. 
Okabe et al. (2008), based on the joint X-ray/optical/weak-lensing
analysis, found a marginal detection ($2.1\sigma$) 
of a northwest mass substructure, which is located ahead of
the northwest cold front. The projected
separation between the northwest cold front and mass substructure is
$\sim 540$ kpc.

In the present study we have derived a high resolution mass map with 
${\rm FWHM}=1\farcm 00$, as shown in  the left panels 
of Figure \ref{fig:a2142}, using a new Subaru data set with long exposure.

The number density of background galaxies usable for a weak lensing
analysis is increased by a factor of $\sim 3$ as compared with the
previous data set (Okabe et al. 2008).

Using the new high-resolution mass map,
this northwest mass structure 
(NW in the top-left panel of Figure \ref{fig:a2142}) 
has been detected at a significance of $\kappa=3.0 \sigma$, 
which is located in the same region as the previous detection.
This improved detection using an independent data set confirms our
previous tentative detection, and supports the existence of the
northwest mass substructure located ahead of the northwest cold front.
In the bottom-left panel of Figure \ref{fig:a2142}, we see a slight
excess luminosity of cluster sequence galaxies 
associated with the northwest mass
substructure, as found earlier by Okabe et al. (2008) using the 
spectroscopically selected sample of bright cluster galaxies.
The Chandra X-ray image shows no X-ray counterpart to the northwest mass substructure
(right panels of Figure \ref{fig:a2142}).
XMM-Newton observations also show no substructure 
in the X-ray emission associated with the northwest mass substructure 
(Okabe et al. 2008).
The merger geometry of A2142, that is the existence of mass
substructures in front of cold fronts, is similar to other cold front
clusters, such as A2034 (\S \ref{subsec:a2034}) and  1E0657-56
(Clowe, Gonzale, \& Markevitch 2004).

We have also discovered a mass concentration associated with a luminous
galaxy concentration located in front of the south cold front.
This south mass substructure (S in Figure \ref{fig:a2142})
has a peak height of 
$4.3 \sigma$ in the $\kappa$ map, and is located around  
$({\rm RA, DEC} )= (239.54^\circ, 27.18^\circ)$.
The line connecting between BCG1
and this south substructure is roughly parallel 
to the moving direction of the south cold front.

The new high resolution mass map has revealed  more complex and detailed 
structures in A2142
than the low-resolution mass map by Okabe et al. (2008).
The significance of the first mass peak is $10.9\sigma$ in the $\kappa$
map, and the peak position is close to BCG1.
We note the presence of two tangential arc candidates
around BCG1 as shown in panels A and B of Figure \ref{fig:a2142_zoom}.

The line-of-sight velocity (LOSV) of BCG1
is close to the central value (mode) of the LOSV distribution  of
cluster galaxies (Oegerle, Hill, \& Fitchett 1995). 
The X-ray, optical, and mass peak positions of the main cluster are in
good agreement with each other within 
${\rm FWHM}=1\farcm 00$ of the weak lensing reconstruction.
On the other hand, 
no significant clumpy mass structure 
is seen around BCG2.
Further we did not find any tangential arc candidate around
BCG2. We show in panel C of Figure \ref{fig:a2142_zoom} a zoom in view
of the $0\farcm 8 \times 0\farcm 8$ region around BCG2.
The projected separation between BCG1 and BCG2 is
larger than the angular resolution of the mass map,
${\rm FWHM}=1\farcm 00$.
BCG2 has a large peculiar velocity, and is 
in the high-velocity tail of 
the LOSV distribution in the cluster rest-frame (Okabe et al. 2008). 

There are few member galaxies associated with BCG2
in the LOSV distribution.
The mass distribution around BCG2 is
peculiar and different from all other BCGs in our target
clusters, which are sitting at high density regions with large
$\kappa$-values.
If BCG2 was originally associated
with the northwest mass substructure, then the results may suggest 
the possibility that BCG and the dark-matter halo underwent
different dynamical processes, which then resulted in a large offset
between the BCG and the dark-matter halo after the merger.

\subsection{A520} \label{subsec:a520}

Markevitch et al. (2005) discovered with Chandra X-ray observations
a bow-shaped shock close to a dense gas clump in the cluster A520 at $z=0.199$.
They derived a Mach number of ${\it M}=2.1^{+0.4}_{-0.3}$ with the
Rankine-Hugoniot condition. 
The post and pre-shock temperatures are found to be 
$11.5^{+6.7}_{-3.1}$ keV and $4.8^{+1.2}_{-0.8}$ keV, respectively.
The X-ray feature is very similar to the bullet cluster
1E0657-56 (Markevitch et al. 2005), in that a shock detached from a
dense core is prominently observed.  
Therefore this cluster is a good target for a weak lensing study of
'bullet'-type merging clusters.

Two compact gas clumps (XSW1 and XSW2 in Figure \ref{fig:a520}) are
located in the southwest of the cluster center.  
The gas clump XSW2 associated with the bow shock is close
to but slightly offset from a luminous BCG. 
The extended tail gas from XSW2 is also seen 
in Figure \ref{fig:a520}.
Markevitch et al. (2005) argued that XSW2 is 
moving along the northeast-southwest direction. 
Govoni et al. (2004) found 
that the temperature is higher along this direction based on Chandra data.
On the other hand,
the gas clump XSW1 is elongated in the east-west direction,
and its X-ray contours are compressed towards the northwest direction
(see the right panels of Figure \ref{fig:a520}),
suggesting that XSW1 is currently moving towards the northwest direction.
The presence of two gas clumps might suggest that
A520 is a three-component merger.

Our weak lensing analysis of A520 is based on archival $i'$-band
data retrieved from SMOKA.
We have three $i'$ imaging data sets
used for our weak lensing analysis
as described in \S \ref{subsubsec:subarudata}.
Here we shall focus on the analysis results using the deepest 
data set (see Table \ref{tab:subaru_data}), 
which is based on the same data used 
by Mahdavi et al. (2007). A comparison of mass maps
with the three different data sets is devoted 
in Appendix \ref{app:a520comp}.

We show in the left panels of Figure \ref{fig:a520} 
contour plots of the reconstructed mass distribution in A520. 
The smoothing scale for the weak lensing mass reconstruction 
is ${\rm FWHM} = 1\farcm 25$.
The first mass peak (C1 in Figure \ref{fig:a520}) 
is located in the southeast of the optical cluster center, 
and the detection signal-to-noise ratio is $6.1\sigma$.
The mass peak C1 is located $\sim0\farcm 5$ east of 
the central X-ray surface brightness peak 
(XC in Figure \ref{fig:a520}). 
Recently Mahdavi et al. (2007) reported the presence of  
a massive {\it dark core} (the mass peak 3 of Mahdavi et al. 2007)
located $\sim 0\farcm 7$ south of the central X-ray peak, XC;
the offset between the dark core and XC is smaller than
the angular resolution of Mahdavi et al.'s mass map (${\rm FWHM}=1'$).   
The first peak C1 in our mass map and the dark core of Mahdavi et al. (2007)
coincide with XC within Gaussian FWHMs of our and their mass maps,
and the projected distance between C1 and the dark core 
is about $1\farcm 0$.
In the central region of the $\kappa$ map,
there are 
two more significant local maxima (C2 and C3 in Figure \ref{fig:a520}):
C2 with a significance of $5.8\sigma$ is located in the north of C1,
and C3 with a significance of $5.4\sigma$ in the east of C1.
The locations of C2 and C3 agree well with 
those of the main and a local maxima of the optical
luminosity density distribution 
(see the bottom-left panel of Figure \ref{fig:a520}), 
respectively. 
In the central high-density region of A520, the shape of mass contours
with $\kappa \simgt 4\sigma$ significance is similar to that of the
cluster galaxy luminosity distribution.

We note that although Mahdavi et al. (2007) used 
the same Subaru/Suprime-Cam $i'$ data as in the present paper,
their analysis method is different from ours in the following way:
Our weak lensing analysis is based on three sets of
stacked Subaru/Suprime $i'$ data (see Appendix \ref{app:a520comp}),
while Mahdavi et al. (2007) performed shape measurements 
separately for each of seven Subaru exposures and combined individual
weak lensing catalogs to improve the accuracy for the shear estimates.
We have confirmed the existence of the dark core (C1) reported by 
Mahdavi et al. (2007) in two of the imaging data sets; however,
the peak C1 is not particularly pronounced in the mass map based solely
on $i'$ images taken without AG.
We have detected C3 at high significance levels
($\simgt 4.5 \sigma$)
from all of the three imaging data sets (see Appendix
\ref{app:a520comp}), while
Mahdavi et al. (2007) found no significant mass structure 
around C3,

The second mass peak (SW2 in Figure \ref{fig:a520}) 
with a significance of $6.0\sigma$
is associated with the southwest substructure.
In the southwest region, mass and light are similarly distributed, and
the peak positions in the mass and luminosity maps coincide well with
each other (see the bottom-left panel of Figure \ref{fig:a520}).
On the other hand,
the X-ray surface brightness peak associated with the southwest X-ray
core (XSW2 in Figure \ref{fig:a520})  
is slightly offset from the corresponding galaxy-luminosity and mass peaks; 
the separation between the X-ray and mass peaks is smaller than the
Gaussian smoothing scale of ${\rm FWHM} = 1 \farcm 25$ 
used for the weak lensing mass reconstruction.
The southwest compact gas, XSW2,
seems to be within the potential well of the
southwest mass substructure, SW2.
We also find a local mass peak 
with a significance of $3.5\sigma$ (SW1 in Figure \ref{fig:a520}) 
located ahead of the northwest of the gas clump XSW1.
The mass peak SW1 is associated with a concentration of
cluster member galaxies (see the bottom-left panel of Figure
\ref{fig:a520}). 
The projected separation between SW1 and the northwest edge of XSW1 is
about $1\farcm 6$.

We note that, in addition to these mass peaks, 
the reconstructed $\kappa$ map shows three more 
local maxima above a significance of $4\sigma$, which are indicated
as N, NE1, and NE2 in Figure \ref{fig:a520}: 
the mass clumps NE2 and N are associated with slight concentrations
of cluster member galaxies.
On the other hand, there is no apparent concentration of galaxies
having similar $(V-i')$ colors around the location of NE1,
although this mass structure is seen all of the three data sets
(see Appendix \ref{app:a520comp}).

Finally, 
the results from 
our joint optical, X-ray, and weak-lensing analysis show
that although the X-ray features of A520 are very similar to 
those of the bullet cluster 1E0657-56, 
the relative positions of the galaxy, X-ray,
and mass substructures are different, 
in that there is no apparent offset between
the mass (SW2) and the gas (XSW2) clumps associated with 
the bow shock.

\section{GLOBAL CLUSTER PROPERTIES} \label{sec:glob}

We present and compare global cluster properties,
such as the total mass, 
         optical luminosity, 
    and X-ray temperature, 
for our sample of merging clusters based on the joint
optical-photometric/X-ray/weak-lensing analysis.

\subsection{Cluster Mass}

We show in Figure \ref{fig:all_zeta} radial profiles of 
the projected mass 
$M_{\zeta}(\theta)$ for our sample of seven merging clusters
together with the best-fitting NFW and SIS profiles.
Note that the error bars are correlated;
in the model fitting, 
we take into account the full error covariance matrix
for the $M_{\zeta}$ measurements.
The projected mass $M_{\zeta}(\theta)$ is measured using 
aperture densitometry, $\zeta_c$, by
$M_{\zeta}(<\theta)\equiv
\pi(D_d\theta)^2\Sigma_{\rm cr}
\zeta_c(\theta; \theta_{\rm inn},\theta_{\rm out})$
(see equation [\ref{eq:zetac}] and \S \ref{subsubsec:fitting}).
Table \ref{tab:zetamodel} summarizes 
the best-fitting SIS and NFW
model parameters derived for our sample of merging clusters.
For each target and its cluster component (if exists) we chose the position of
the BCG as the cluster center
for measuring global cluster properties.
The $\theta_{\rm inn}$ and $\theta_{\rm out}$ define the inner and
outer boundaries of the background annulus for the $\zeta_c$-statistic
measurement. 
The values of $(\theta_{\rm inn},\theta_{\rm out})$ are 
chosen so as to avoid  
cluster binary components and/or significant background substructures, 
and they are summarized in Table \ref{tab:global}.
The radial range used for the fit is different for each cluster: 
the upper limit, $\theta_{\rm max}$, 
is set to the maximum radius which will avoid
contamination by neighboring substructures and foreground/background
mass structures; the lower limit, $\theta_{\rm min}$, 
is set so as to avoid the
strong-lensing regime (see Table \ref{tab:zetamodel}).
As we have seen in \S \ref{sec:maps},
the mass distribution in merging clusters is highly irregular
and complex.
Indeed, when fitting NFW/SIS models to the $\zeta_c$-statistic 
measurements,
we found for some clusters
large $\chi^2$-values relative to the degrees-of-freedom (dof).
For all of the cluster targets, 
the virial mass $M_{\rm vir}$ of the best-fitting 
NFW model is consistent with 
that of the SIS model (see \S \ref{subsubsec:fitting})
within $1\sigma$ statistical error.
On the other hand, 
the NFW concentration parameter $c_{\rm vir}$
was poorly constrained, except for a marginal detection
of A2142 ($\approx 2.5\sigma$),
because of contamination by neighboring substructures and
clusters in projection space.

As defined by equation (\ref{eq:zetac}), 
$M_{\zeta}(<\theta)$ yields 
a lower bound to the true enclosed mass $M(<\theta)$
due to the subtraction of the mean $\kappa$ over the background annulus,
$\bar{\kappa}(\theta_{\rm inn}<\theta<\theta_{\rm out})$. 
In Table \ref{tab:global} we summarize the results of weak-lensing
total mass estimates,
$M_{\zeta}(<\theta_{M})$,
for our sample of merging clusters. 
The aperture radius $\theta_{M}$  is
set to the maximum radius available for each target cluster,
i.e. $\theta_{M} \approx \theta_{\rm max}$.

\subsection{Mass-to-light Ratio}

We measure for individual targets
the cluster luminosity from color-selected member galaxies
(see \S \ref{subsec:gal}). 
The cluster luminosity is measured in $R_{\rm c}$-band for all
clusters, except in $i'$-band for A520.
Here
the aperture and the background annulus 
for each target
are taken to be the same as in aperture densitometry.
Table \ref{tab:global} lists 
the mass-to-light ($M/L$) ratio 
derived for our sample of merging clusters,
where the errors include a systematic uncertainty in the
background luminosity density estimation as well as 
$1\sigma$ random uncertainties in the cluster luminosity
and mass measurements; 
the lower limit on the cluster $M/L$ ratio is based on the 
cluster luminosity without background subtraction
and $1\sigma$ random uncertainties for the $M/L$ ratio.
The clusters A754 and A1750 have considerably high $M/L$ ratios of 
$M/L \simgt 600h (M/L)_{\odot}$, which is simply due to the fact that
cluster galaxies are still dominating in the background annular region
and hence the background luminosity density is overestimated.

Girardi et al. (2000) studied optical properties of 
a sample of 105 nearby Abell-ACO clusters ($z<0.15$)
for which virial mass estimates are available,
and found typical values for $M/L_{B_j}$ ratios to be $230-250
h (M/L_{B_j})_\odot$.
Sanderson \& Ponman (2003) conducted a joint optical/X-ray
study of the mass
composite of a sample of 66 relaxed groups/clusters.
Based on the X-ray mass measurements,
they obtained a logarithmic mean value of 
$M/L_{B_j}\approx 350 h (M/L)_\odot$, 
which is scaled to 
$M/L_{R_{\rm c}}\approx 190h (M/L_{R_{\rm c}})_\odot$ 
assuming 
$(L/L_\odot)_{B_j}=0.547 (L/L_\odot)_{R_{\rm c}}$
from a typical color 
$(B_j-R)=1.702$
of early-type galaxies (equation [4] of Sanderson \& Ponman 2003).

Clowe et al. (2006) performed a weak lensing analysis of
20 high-$z$ clusters ($z>0.4$) selected from the ESO Distant Cluster
Survey
(EDisCS)
using deep 3-color optical images taken with the VLT/FORS2. 
They derived for an individual cluster
$M/L$ ratios, in $I$ and $B$ passbands,
within $500$ kpc from photo-$z$ selected cluster galaxies.
They found for their optically-selected EDisCS clusters
that the clusters tend to have a lower $M/L$ ratio at higher redshifts,
but found no change in their $M/L$ ratios with cluster mass. Clowe et
al. (2006) also showed that the clusters having additional structures
in projection space
have a higher measured $M/L$ ratio, which is 
probably due to the projection effects.

\subsection{X-ray vs. SIS Temperatures}\label{subsec:TvsTsis}

We compare  for individual cluster targets
the X-ray temperature, $T_X$, and the best-fitting SIS
temperature, $T_{\rm SIS}$, obtained from the aperture densitometry
measurements.
In Table \ref{tab:global} we summarize the best-fitting SIS 
velocity dispersion, $\sigma_v$, from the weak lensing analysis.
The X-ray temperatures of the clusters 
are taken from the literature (see Table \ref{tab:sample}).
We note there is a good agreement between 
the Chandra and XMM-Newton temperatures of A1758N (David \& Kempner
2004),
although Chandra and XMM-Newton have different instrument
responses as well as sky and instrumental background levels.
The X-ray temperature measurement is performed within apertures of
$\theta \sim 10'$, 
while the weak lensing measurement of aperture densitometry
within apertures of $\theta_M \sim 5'-15'$ (see Table \ref{tab:global}), 
depending on the availability  of the {\it clean} 
background aperture without any significant mass structures.
We emphasize that the SIS temperature $T_{\rm SIS}$ based
on gravitational lensing does not depend on the dynamical and physical
state of the cluster system.

Figure \ref{fig:tmp} compares the X-ray temperature and the SIS
temperature for our sample of merging clusters.
It is shown in Figure \ref{fig:tmp} that
there is an overall trend that the ICM temperatures
from the X-ray observations are higher than
the SIS temperature from the weak lensing analysis.
Figure \ref{fig:tratio} shows explicitly the X-ray to SIS temperature
ratio, $T_X/T_{SIS}$, as a function of $T_{\rm SIS}$.
The sample mean for $T_X/T_{\rm SIS}$ is estimated as 
$\langle T_X/T_{\rm SIS}\rangle =2.05 \pm 0.11$.

For A1750N, where mass, light, and X-ray emission are similarly
distributed (see \S \ref{subsec:a1750}), 
the temperature ratio $T_X/T_{\rm SIS}$ is consistent with unity
within the $1\sigma$ error bar.
The ICM temperature of A1750C, which shows an 
excess entropy in the central region,
is higher than $T_{\rm SIS}$ (\S \ref{subsec:a1750}),
but is consistent with $T_{\rm SIS}$ within $2\sigma$:
$T_X/T_{\rm SIS} = 1.72 \pm 0.53$ ($1\sigma$).
The averaged X-ray temperature of A520,
$T_X=7.1\pm 0.54$ keV ($1\sigma$),
is slightly higher than the SIS temperature,
$T_{\rm SIS}=5.9\pm 1.1$ keV ($1\sigma$),
but is marginally consistent within $1\sigma$
with $T_{\rm SIS}$: 
$T_X/T_{\rm SIS}=1.18\pm 0.25$ ($1\sigma$);
However, 
we note that the post-shock temperature of A520,
$T_X=11.5^{+6.7}_{-3.1}$ keV ($90\%$CL),
 is significantly higher than 
$T_{\rm SIS}$,
while 
its pre-shock temperature, $T_X=4.8^{+1.2}_{-0.8}$ keV ($90\%$CL),
is marginally consistent with $T_{\rm SIS}$.

For the rest of the sample
(A754, A1758N, A1758S, A1914, A2034, A2142)
consisting of on-going mergers and cold-front clusters,
the ratios $T_X/T_{\rm SIS}$ are in the range of $1.4-3.1$,
with the sample mean of 
$\langle T_X/T_{\rm SIS}\rangle =2.41 \pm  0.16$. 
Hence, presumably the most promising scenario is that
the ICM temperature is 
increased by heating processes triggered by cluster mergers.
On the other hand,
the temperature ratios for the cold front clusters,
A2034 and A2142, are  in a moderate range of
$1.94 \pm 0.38$ and $1.41\pm 0.19$, 
respectively.

Although the averaged temperatures $T_X$ 
of the cold-front clusters were measured
including the cold, dense cores, 
the X-ray temperatures outside the
dense cores  agree with
the global temperatures (including the cold cores) within errors 
(Markevitch et al. 2000; Kempner \& Sarazin 2003). 
Therefore, the observed variations and trends in 
the $T_X/T_{\rm SIS}$ ratios
would be naturally explained by the 
different energy scales injected into the ICM
or different merger states of the cluster mergers.
Figure \ref{fig:tratio} shows no clear correlation between
the temperature ratio $T_X/T_{\rm SIS}$ and the SIS temperature 
$T_{\rm SIS}$. This might suggest
that
the enhancement of the ICM temperature does not
depend strongly on the mass of the main cluster but on merging stages and conditions.

\section{DISCUSSION AND CONCLUSIONS}\label{sec:dis}

We presented and compared
weak lensing mass, optical light, and X-ray emission maps for a sample
of seven merging clusters of galaxies.
We have selected 
for the present study
seven nearby Abell clusters ($0.0542 \le z \le 0.279$)
of different merging stages and properties,
and conducted systematic, deep imaging observations of the seven target
clusters with Suprime-Cam on Subaru telescope.
Our seven target clusters, representing various merging stages and conditions,
allow us to investigate in details the 
physical interplay
between dark matter, ICM, and galaxies
associated with hierarchical structure formation.
The clusters A1750 and A1758 are binary systems,
each of which consists of two cluster-sized components and has not yet
experienced the first impact. A520, A754, A1758N, A1758S, and A1914, 
on the other hand, are classified as on-going merging clusters. 
A2034 and A2142 are ``cold front'' clusters.

In the binary cluster A1750,
the projected mass, optical light, and X-ray distributions all 
revealed clear binary structures associated with the two cluster components
(A1750N and A1750C),
and are overall similar and regular without significant substructures 
(\S \ref{subsec:a1750}).
A1758 also exhibits binary structures associated with the two system
components
(A1758N and A1758S) 
in the projected mass, optical light, and X-ray
distributions (\S \ref{subsec:a1758}).
Along the line between A1758S and A1758N,
which is parallel to the dynamical motion of the binary components,
no significant offset is found between the 
mass, X-ray, and optical peaks in the binary components.
On the other hand, the ICM distribution in individual components,
each of which is an undergoing merger,
appears to be different from the distributions of mass (dark matter)
and light (galaxies).
Our results on the binary clusters
indicate that, in an early merging phase,
the distributions of mass and baryons are overall similar 
and the main mass peak of the cluster coincides well with the 
corresponding X-ray peak and galaxy concentration.

The mass distributions of on-going merging clusters, 
A520, A754, A1758N, A1758S, and A1914, are found to be highly irregular.
Overall, 
the mass distribution appears to be similar to the galaxy
luminosity distribution, whereas their distributions are totally
different from the ICM distribution.
Discrepancies between mass and ICM distributions are different in a
variety of ways.
Our results on the on-going mergers can be formally classified 
into three types depending on 
the moving direction and relative positions 
of mass and X-ray substructures.
Here we speculate on the moving direction of cool, 
dense cores from the direction of compression and elongation of 
X-ray cores and trails, and summarize the results as follows:

\begin{itemize}
\item[(1)] 
{No significant offset between X-ray and mass peaks}
\end{itemize}

The mass peak position coincides with the peak position of 
X-ray surface brightness within Gaussian smoothing FWHM of 
the weak lensing mass reconstruction.
This apparent feature would imply that these dense cores are 
trapped in the gravitational potential well of the mass concentration.
This is found in 
the southwest clump (SW2 and XSW2) of A520, 
the northwest clump (C and XC) of A1758N,
and the first mass peak and the elongated gas core 
(C1 and XE) of A1914.
The position of the BCG in A1758N coincides with the peak position
of the dense core in the ICM (XC).
For the case of A520 and A1914, the associated luminous galaxy 
is slightly offset 
perpendicular to the moving direction of the gas core.

\begin{itemize}
\item[(2)] {Mass clump in front of the dense core}
\end{itemize}

The second type is the case where the mass clump is located 
ahead of the moving gas core, as seen in the southwest structure 
(SE and XSE) in A1758N.
The bullet cluster 1E0657-56 is categorized into this type.
We note that this configuration is also found in our cold front
clusters, A2034  (N and the cold front)
and A2142 (NW and S), 
as we will discuss below.
The following scenario could explain this type of merger configurations:
a dense gas core, which is originally bound in a merging substructure,
is stripped away from the mass and galaxy components of the substructure
by the ram pressure of the ICM.

\begin{itemize}
\item[(3)] {Mass clump behind the dense core}
\end{itemize}

The third type is where 
the mass clump is located behind the dense core, 
which is opposite to the second case.
This is found in the east substructure (E and XE) of A754.
A possible explanation is that
this cluster is in a merging phase right after
reaching its apocenter of the merger orbit (see \S \ref{subsec:a754}),
although no direct evidence from currently available data.

Cluster mergers depend on a certain set of 
merging parameters,
such as the mass ratio of the main to sub cluster,
initial velocity and angular momentum of the merging substructure,
and the shape of the gravitational potential of the main cluster.
Furthermore, clusters are in various phases of the
merging process.
Understanding and constraining
the dynamical process that causes offsets between the mass/ICM/galaxy
components therefore require a more detailed, quantitative study of 
these merged substructures,
which will be presented elsewhere.
For example, 
line-of-sight 
kinematic information of member galaxies 
from spectroscopic observations, measurements of the density,
temperature, and entropy of the pre/post-shock regions will be
valuable.

The mass maps of the cold front clusters A2034 and A2142 reveal
irregular mass structures, which are quite different from the ICM
distribution.
Mass structures with low gas mass fractions have been detected in front
of three cold fronts, which is found earlier in the bullet cluster
1E0657-56 (Clowe et al. 2004).
Such merger configurations are classified as the second type of on-going
mergers as discussed above.
To date four cold fronts have been studied via joint X-ray/weak-lensing
analysis including 1E0657-56, and all of the cold fronts show 
such a configuration of mass/cold-front substructures.
This might be thus a common feature of cold front clusters
associated with the formation of cold fronts.
Current results from the joint analysis of cold front clusters suggest
a possible scenario for the formation of cold fronts that
a substructure falls into and passes through
the primary cluster halo, while the hot gas originally bound in the
substructure is stripped away by ram pressure stripping
(see Clowe et al. 2004).
A more detailed, statistical study of the relationship between
cold fronts and associated mass structures is required to draw more
definitive conclusions on the formation of cold fronts.

The observed positions of dense cores in A2034 and A2142 are 
apparently located in the primary cluster halo in projection space.
This is different from the case of the bullet cluster 1E0657-56,
which shows a significant offset between the mass and X-ray halos
of the primary cluster as well as between those of the subcluster
(Clowe et al. 2004).
However, since no line-of-sight information of 
substructures is available from the current data sets, 
the merger geometry in three-dimensional space has not yet been fully
constrained. 
Such line-of-sight information 
of the primary and sub clusters
will be necessary 
to place quantitative constraints on the 
evolution of the gas temperature and pressure
of the ICM substructure 
against the gravitational potential of the primary cluster.

Our quantitative comparison of projected distributions of
mass, galaxies, and ICM in merging clusters shows that
the dark matter and ICM components exhibit different behaviors during
the cluster merger process.
Similar features have been reported based on cosmological
$N$-body/hydrodynamical simulations. 
Tormen, Moscardini, \& Yoshida (2004) have
shown 
by their numerical simulations
that dark matter halos of merged substructures
freely move and oscillate in the primary cluster halo, while 
gas halos initially bound in the substructure are stripped 
by various physical processes,
such as ram pressure stripping and dynamical friction.
These behaviors depend on the mass ratio between the primary and sub cluster.
It is easier for dark matter of small substructures to survive longer,
while for the ICM of small substructures to be destroyed or digested by
the primary cluster.
They also showed that the separation between the dark matter and gas
centers for a merged substructure, whose mass is more than $0.01$ times
of the primary cluster mass, is very small in the early phase of $\sim
1$ Gyr.
These simulation results are consistent with our results of the
binary cluster A1750 (\S \ref{subsec:a1750}).
Takizawa (2006) studied the X-ray and mass distributions of
1E0657-56 based on $N$-body/hydrodynamical simulations, 
and reproduced
a clear offset between X-ray and mass peak as found from
the joint X-ray/weak-lensing analysis by Clowe et al. (2004).
The mass ratio of the sub to main halo is assumed to be $1/16$,
which is based on the weak lensing mass estimate by Clowe et al. (2004).
Takizawa (2006) also derived an analytic expression for the ram pressure
stripping conditions of the substructure in mergers of two NFW dark
matter halos, which is useful to understand the  merging conditions of
the first and second types defined above.
Ricker \&  Sarazin (2001) and 
Mathis, Lavaux, Diego, \&  Silk (2005) studied cluster mergers 
of nearly-equal-mass with
numerical simulations and showed that 
the hot gas of a merged substructure escapes from the local
gravitational potential well when it reaches the apocenter in merger
orbits, because the inward motion of the gas is delayed compared with 
the associated dark matter clump. 
Their simulation results would be a clue to understand the third type of
on-going mergers as found in A754.
Nagai \& Kravtsov (2003) searched for cold fronts and their counterparts
in their high-resolution cluster simulations, and found cold fronts
appear to be common in major and minor cluster mergers in hierarchical
models. Their results indicate cold fronts are non-equilibrium transient
phenomena during cluster mergers, although no significant offset between
dark matter and gas distributions is shown in their figure.
Ascasibar \& Markevitch (2006) used high-resolution simulations of
cluster mergers and showed that cold fronts in relaxed clusters are 
due to sloshing of the cool gas in the gravitational potential of the
primary cluster, and can be easily triggered by minor mergers and
persist for gigayears.
Based on three-dimensional hydrodynamic simulations
Takizawa (2005) investigated the dynamical evolution of the ICM
driven by a radially-moving substructure and its observational
implications, 
and found that the subcluster's cold gas is pushed out of its
potential well around the turn around, resulting in that
the cold gas clump appears to be in front of the main X-ray peak
of the subcluster (see Figure 6 of Takizawa 2005).

A comparison of the ICM and SIS temperatures of merging clusters
from X-ray
and weak-lensing analyses, respectively, provides us with 
interesting pieces of evolutionary information of the ICM
during the cluster merger.
Our results show that the ICM temperature in on-going and cold-front
clusters is significantly higher than the SIS temperature that is
a measure of the virial temperature of the cluster
(Figure \ref{fig:tmp}).
This would suggest that the ICM temperature is increased 
by the cluster merger.
Similar results were found from numerical simulations.
Randall, Sarazin, \& Ricker (2002) and 
Rowley, Thomas, \& Kay (2004) 
studied the effects of merger boosts on the ICM properties
and showed that cluster mergers can boost 
for a duration of the sound crossing time
the X-ray luminosity and
temperature of the merged cluster above the virial 
equilibrium values for the merged system.
Our results show no significant correlation between
the SIS temperature, or the cluster mass, and 
the X-ray to SIS temperature ratio, $T_X/T_{\rm SIS}$;
this might indicate that 
the energy release from cluster mergers is determined
not only by the cluster mass but also by other merging parameters.
Ricker \&  Sarazin (2001) studied offset mergers between clusters
based on $N$-body/hydrodynamic simulations as a function of 
impact parameter and mass ratio of the colliding clusters,
and show that
the variations in global properties of clusters, namely total luminosity
and average temperature,
indeed depend on the impact parameter and the mass ratio,
as well as the observed epoch of mergers.

Based on visual inspection, we have identified 
gravitational arc candidates in three of our cluster targets:
A1758, A1914, and A2142 (see Figures \ref{fig:a1758N_zoom}, 
\ref{fig:a1914_zoom}, and \ref{fig:a2142_zoom}).
A further examination and confirmation of 
the lensing hypothesis of these gravitational arc candidates
will require follow up spectroscopy and higher-resolution imaging,
which will provide strong lensing constraints
needed for 
a detailed modeling of the observed arc-cluster systems.

In summary we compared 
projected distributions 
of mass and baryons 
in a sample of seven merging clusters based on the joint
optical-photometric/X-ray/weak-lensing analysis.
The global cluster parameters, such as the cluster mass, 
the cluster mass-to-light ratio, and the ICM temperature, 
are derived for individual targets of our cluster sample.

A joint analysis of optical photometric, X-ray, and weak lensing data
provides us with a 
comprehensive picture of cluster mergers,
which will help 
to better understand the complex physical interplay between
dark matter, ICM, and galaxies in the cluster formation process.
For merging clusters 
an X-ray analysis alone cannot constrain the cluster mass distribution
since the ICM is not in hydrostatic equilibrium.
We have demonstrated that weak lensing is indeed a unique and powerful
method to map the mass distribution in merging clusters
that are in the process of active evolution and formation,
since it does not require any assumption of the physical/dynamical state
of the system.
Furthermore, thanks to the Subaru/Suprime-Cam
with excellent image quality and wide-field imaging capability,
we were able to achieve high-resolution and wide-field imaging in weak
lensing mass reconstructions. 
In particular, it is worth noting that the Subaru/Suprime-Cam makes it
possible to reveal a mass distribution of a very nearby cluster at
$z\sim 0.05$ (A754), where the lensing signal is very low due to its
proximity to the observer.
A detailed study of individual targets using the joint data sets
is the next step for 
understanding 
the merger dynamics of clusters
and associated physical processes of the ICM.
To date galaxy clusters have been studied extensively and systematically
in X-ray and optical photometric observations.
On the other hand, 
the weak-lensing database for clusters is not sufficiently large
as compared with X-ray and optical,
although much progress has been made in the recent years
(e.g., Dahle et al. 2002; Cypriano et al. 2004; 
      Bardeau et al. 2005, 2007; Clowe et al. 2006; Hoekstra 2007).

Therefore a systematic weak-lensing study of a larger sample of clusters
needs to be conducted for further understanding of clusters and their
formation/evolution process.

\section*{Acknowledgments}

This work is in part supported by 
Grants-in-Aid for the 21st Century COE Program 
``Exploring New Science by Bridging Particle-Matter Hierarchy''
in Tohoku University
and 
``Towards a New Basic Science; Depth and Synthesis'' 
in Osaka University, 
funded by the Ministry of Education, Science, Sports and Culture of
Japan.
The work is partially supported by the National Science Council of Taiwan
under the grant NSC95-2112-M-001-074-MY2.
This work is in part supported by a Grant-in-Aid for Science Research in a Priority Area "Probing the Dark Energy through an Extremely Wide and Deep Survey with Subaru Telescope" (18072001) from the Ministry of Education, Culture, Sports, Science, and Technology of Japan .

We gratefully thank H. Furusawa, the Subaru Support Astronomer 
of the Suprime-Cam.
We acknowledge fruitful and intensive discussions with
M. Takada, 
H. Hoekstra,
T. Broadhurst, 
T.~T. Takeuchi,
M. Takizawa, 
M. Hattori,
N. Ota,
S. S. Sasaki,
T. Nagao,
and Y. Fujita. 
We are also grateful to N. Kaiser for developing the IMCAT
package publicly available.

\appendix
\section{Lensing Mass Models}\label{app:halo}

The SIS density profile is given by
\begin{eqnarray}
 \rho_{\rm SIS}(r)=\frac{\sigma_v^2}{2\pi G r^2} \label{eq:sis}
\end{eqnarray}
where $\sigma_v$ is the one-dimensional velocity dispersion of
the SIS halo.
The $\sigma_v$ is related with the 
virialization epoch $z_{\rm vir}$
and the virial mass $M_{\rm vir}$ of the SIS halo
as
\begin{eqnarray}
\sigma_v(M_{\rm vir},z_{\rm vir}) &=& \frac{1}{2} r_{\rm vir} H_0 
\sqrt{\Omega_{m0} \Delta_{\rm vir} (1+z_{\rm vir})^3},\\
M_{\rm vir} &=& \frac{4\pi}{3}\bar{\rho}(z_{\rm vir})\Delta_{\rm
vir}r_{\rm vir}^3, 
\end{eqnarray}
where $r_{\rm vir}=r_{\rm vir}(M_{\rm vir},z_{\rm vir})$ 
is the virial radius, and $\Delta_{\rm vir}$
is the mean overdensity with respect to the mean cosmic density
$\bar{\rho}(z_{\rm vir}) = \Omega_m(z_{\rm vir})\rho_{\rm cr}(z_{\rm vir})$ 
at the virialization epoch,
predicted by the dissipationless spherical
tophat collapse model (Peeebles 1980; Eke, Cole, \& Frenk 1996; Bullock
et al. 2001). We assume the cluster redshift $z$ is
 equal to the cluster virial redshift $z_{\rm vir}$.
We use the following fitting formula in a flat 3-space with cosmological
constant (see Oguri, Taruya, Suto 2001):
\begin{eqnarray}
\Delta_{\rm vir} &=& 18 \pi^2 (1+0.4093 \omega_{\rm vir}^{0.9052}), 
\end{eqnarray}
where $\omega_{\rm vir}\equiv 1/\Omega_m(z_{\rm vir})-1$.
 
We thus have $\sigma_{v}\propto M_{\rm vir}^{1/3}$.
In the case of an SIS lens, the lensing convergence is obtained as
\begin{equation}
\kappa_{\rm SIS}(\theta)=\frac{\theta_E}{2\theta},
\end{equation}
where 
$\theta_E = 4\pi (\sigma_v/c)^2 (D_{ds}/D_s)$ is the 
Einstein radius. Then the averaged convergence within $\theta$ is 
\begin{equation}
\bar{\kappa}_{\rm SIS}(<\theta) = \frac{\theta_E}{\theta} = 2\kappa_{\rm
 SIS}(\theta).
\end{equation}

The NFW universal density profile 
has a two-parameter functional form as
\begin{eqnarray}
 \rho_{\rm NFW}(r)= \frac{\rho_s}{(r/r_s)(1+r/r_s)^2} \label{eq:nfw}
\end{eqnarray}
where $\rho_s$ is a characteristic inner density, and $r_s$ is a
characteristic inner radius.
In stead of using $r_s$, we introduce the concentration parameter,
$c_{\rm vir}\equiv r_{\rm vir}/r_s$.
The inner density $\rho_s$ can be expressed in terms of other virial
properties of the NFW halo:
\begin{equation}
\rho_s = \bar{\rho}(z_{\rm vir}) 
\frac{\Delta_{\rm vir}}{3}\frac{c_{\rm vir}^3}{\ln(1+c_{\rm vir})-c_{\rm vir}/(1+c_{\rm vir})}
\end{equation}
$\rho_{\rm cr}$ is the
critical density of the universe at the cluster redshift $z_d$,
and $r_s$ is the NFW scaling radius.
Hence, for a given cosmological model and a halo virial redshift
($z_{\rm vir}$), we can specify the NFW model with the halo virial mass
$M_{\rm vir}$ and the halo concentration parameter $c_{\rm vir}$.

For an NFW profile, it is useful to decompose the convergence 
$\kappa(\theta)$ and the
averaged convergence $\bar{\kappa}(<\theta)$ as
\begin{eqnarray}
\kappa_{\rm NFW}(x) &=& \frac{b}{2}f(x),\\
\bar{\kappa}_{\rm NFW}(<x) &=& \frac{b}{x}g(x),
\end{eqnarray}
where $b=4\rho_s r_s/\Sigma_{\rm cr}$ is the dimensionless scaling
 convergence, 
 $x=\theta/(r_s/D_d)$ is the dimensionless angular radius,
and $f(x)$ and $g(x)$ are dimensionless functions. 
We have analytic expressions for $f(x)$ and $g(x)$
as (Bartelmann 1996):
\begin{eqnarray}
f(x) &=&
\left\{
  \begin{array}{ll}
  \frac{1}{1-x^2}\left[-1+ \frac{2}{\sqrt{1-x^2}} {\rm arctanh}\sqrt{\frac{1-x}{1+x}}\right]  & (x<1)
   \\
  \frac{1}{3}~~~~~~~~~~~~~~~~~~~~~~~~~~(x=1) &
  \\
  \frac{1}{x^2-1}\left[+1- \frac{2}{\sqrt{x^2-1}} {\rm arctan}\sqrt{\frac{x-1}{x+1}}  \right]  & (x>1).
  \end{array}
\right. .\\
g(x) &=&
\left\{
  \begin{array}{ll}
  \frac{2}{x\sqrt{1-x^2}} {\rm arctanh}\sqrt{\frac{1-x}{1+x}} + \frac{1}{x}\ln\frac{x}{2}  & (x<1)
   \\
   1-\ln 2   &(x=1)
   \\
  \frac{2}{x\sqrt{x^2-1}} {\rm arctan}\sqrt{\frac{x-1}{x+1}} + \frac{1}{x}\ln\frac{x}{2}  & (x>1)
  \end{array}
\right. .
\end{eqnarray}

Finally, we use the following identity to calculate the theoretical
$\zeta_c$-statistic for a given set of the aperture parameter
$(\theta_{\rm inn},\theta_{\rm out})$:
\begin{eqnarray}
\label{eq:zeta_c}
\zeta_c(\theta) &\equiv& \bar{\kappa}(<\theta) -
 \bar{\kappa}(\theta_{\rm inn}<\theta<\theta_{\rm out}) \\
&=& \bar{\kappa}(<\theta) -\bar{\kappa}(<\theta_{\rm inn})\\
 &&+\frac{1}{1-(\theta_{\rm inn}/\theta_{\rm out})^2}\left( 
 \bar{\kappa}(<\theta_{\rm inn}) -
 \bar{\kappa}(<\theta_{\rm out})\right).
\end{eqnarray}
In practice, we use a discretized estimator for 
equation (\ref{eq:zetac}). 
We note that, unlike tangential shear measurements,
measurement errors for $\zeta_c(\theta)$ are correlated between
different annular bins.

\section{Comparison of Weak Lensing Mass Reconstructions 
between Three Different Data Sets for A520}\label{app:a520comp}

We retrieved from the Subaru Archival Website, SMOKA,
a total of seven $i'$-band images of A520 
taken under good seeing conditions ($\sim 0\farcs 6$).
The archival $i'$ data of A520 were taken with and without
AG (acquisition and guide)
probe on 17th November 2001 and 19th October 2001, respectively.
The use of AG probe is required for accurate telescope pointing,
which is crucial for weak lensing shape measurements of faint
background galaxies.
We performed a weak lensing analysis separately on the following three
imaging data sets:
(A) all of seven $i'$ images 
taken both with and without guide probe ($7\times 240$s exposure)
used in the weak lensing analysis of Mahdavi et al. (2007),
(B) four $i'$ images taken without guide probe ($4\times 240$s exposures), 
and 
(C) three $i'$ images taken with guide probe ($3\times 240$s exposure). 
Therefore, 
(B) and (C) are independent and taken under different
observing modes, while (A) is a combination of (B) and (C).
We note that
the A520 data were taken with large dithering offsets
of $\approx 2\farcm 3$, 
whereas the other cluster data were taken with a dithering offset of $1'$.

Our weak lensing analysis is based on 
co-added mosaics of multiple exposures (A), (B), and (C),
whereas Mahdavi et al. (2007) performed shape measurements 
separately for each of seven Subaru exposures and combined individual
weak lensing catalogs to improve the accuracy for the shear estimates.
In the present paper,
the deepest data set (A) is used for our main analysis of A520,
and the resulting mass map is compared with 
X-ray and optical-luminosity distributions 
of the cluster in Figure \ref{fig:a520}.
Figure \ref{fig:a520B}, on the other hand,
compares the mass reconstruction based on (B) 
with the X-ray and optical-luminosity distributions.

We show in Figure \ref{fig:a520_3maps} the mass maps of A520
reconstructed from the above three imaging data sets.
The left panel in Figure \ref{fig:a520_3maps} shows the mas map
from the deepest data set (A), which is used for the main analysis of
the present paper (see also Figure \ref{fig:a520});
the middle panel shows the mass map from (B) taken without guide probe;
the right panel shows the mass map from (C) taken with guide probe.
For all of the reconstruction, the Gaussian smoothing FWHM is taken to
be $1\farcm 25$.  
The contours are spaced in units of $1\sigma$ reconstruction error
of each data set.

The primary mass peak C1 found from the data set (A) is also detected
at a significance of $5.6\sigma$ from the data set
(C) taken with guide probe. 
The angular position of C1 agrees with
that of the {\it dark core} discovered by Mahdavi et al. (2007) 
within Gaussian FWHM used for the mapmaking.
However, this mass clump is not particularly pronounced in the mass map
obtained from (B).

The mass clump C3 of (A)
is associated with a local concentration of cluster member
galaxies (see Figure \ref{fig:a520}). The mass clump C3 
is found to be the first peak in the reconstructed mass map
from (B).
The mass structure around C3 is not strongly pronounced 
in the mass map based on (C), 
but the reconstruction shows
a clear feature of the extending mass structure towards the 
southeast direction, which is seen in the optical luminosity map
of the cluster (see the bottom-left panel of Figure \ref{fig:a520}).
However, we note Mahdavi et al. (2007) found no significant mass
structure around C3 from
their multi-telescope, multi-bandpass weak lensing analysis 
based on Subaru/Suprime-Cam $R_{\rm c}$ and $i'$ and CFHT/MegaCam $r'$ data.

The mass clump C2 of (A) corresponds to the primary peak 
in the optical luminosity distribution of the cluster.
The mass reconstruction based on (B) shows a local maximum around C2,
whereas no local maximum is found around C2 in the mass map from (C).

The mass structures SW2, NE1, and NE2
are seen in all of the three reconstructions
above a significance level of $4\sigma$.
The mass clump SW1 is detected at a significance level of
$3.5\sigma$ from  (A), 
and is marginally detected at $\approx 3\sigma$ significance
from (B),
but is not seen in the mass reconstruction from (C).

Details of mass structures are different 
in a qualitative and quantitative way
between the results from (B) and (C)
taken without and with guide probe, respectively.
This difference in detailed mass structures might be due to 
the different pointing modes and exposures between the data sets
(B) and (C), which could affect the shape measurements of 
faint background galaxies. 
Furthermore, the large dithering offsets of $2\farcm 3$ 
could also make it difficult to correct for the PSF anisotropy
in stacked images, as discussed by Mahdavi et al. (2007).
Careful follow-up imaging observations with guide probe and
small dithering offsets ($\sim 1'$) will make it possible to
clarify these discrepancies in detailed mass structures in A520.

\newpage


\begin{table*}
  \caption{
Target Clusters and {\it Subaru}/X-ray Observations}
\label{tab:subaru_data}
  \begin{center}
    \begin{tabular}{cccccc}
\hline
\hline\\
 Cluster         &~~~  Subaru Bands  
                                         &   \& Exposure Time~~~~                   & Seeing 
                & X-ray data  &   Obs. ID                  \\
                &             &  (sec)    & (arcsec) & & \\
 (1)            &    (2)      &           & (3)   & (4)  &  (5) \\
\hline
 A754          ~~& $R_{\rm c}=360{~\rm  s}\times8$ 
                  & $g'=180{~\rm s}\times 4$
                  & 0.75 
                  & Chandra 
                  & 577 \\
 A1750          ~~& $R_{\rm c}=360{~\rm  s}\times8$ 
                  & $g'=180{~\rm s}\times 4$
                  & 0.67 
                  & XMM 
                  & 0112240301 \\
 A1758          ~~& $R_{\rm c}=360{~\rm  s}\times8$ 
                  & $g'=180{~\rm s}\times 4$  
                  & 0.69 
                  & XMM
                  & 0111160101 \& 0142860201 \\
 A1914          ~~& $R_{\rm c}=360{~\rm  s}\times8$ 
                  & $g'=180{~\rm s}\times 4$  
                  & 0.61
                  & Chandra
                  & 3593 \\
 A2034          ~~& $R_{\rm c}=360{~\rm  s}\times8$ 
                  & $g'=180{~\rm s}\times 4$ 
                  & 0.63
                  & XMM
                  & 0149880101 \\
 A2142          ~~& $R_{\rm c}=240{~\rm  s}\times4+360{~\rm  s}\times3$ 
                  & $g'=120{~\rm s}\times 4$
                  & 0.55
                  & Chandra
                  & 5005   \\
A520$^{\rm a}$ ~~ & $i'~=240{~\rm s}\times 7$
                  & $V=450{~\rm s}\times 4+120{~\rm s}\times 3$
                  & 0.65
                  & Chandra
                  & 4215 \\
\hline
\end{tabular}
\end{center}
\textrm{ Note $\singlebond$ 
Col. (1): Cluster name. 
Col. (2): Band name of Subaru/Suprime-Cam
and total exposure time in units of s.  
The former passband is used for the weak lensing analysis.  
Col. (3): Seeing FWHM in
  $R_{\rm c}$ or $i'$ band in units of arcsec. 
Col. (4): X-ray satellite name of archival data.
Col. (5): X-ray observation ID number.}\\
\textrm{~~$^{\rm a}$  
 Data were collected at Subaru telescope and
 obtained from the SMOKA science archive. 
 Four $i'$-band images were taken without guide probe on 19th October 2001,
 while the other three $i'$ images were taken without guide probe on
 17th November 2001.
}
\end{table*}

\newpage

\begin{table*}
  \caption{Cluster X-ray Properties} \label{tab:sample}
  \begin{center}
    \begin{tabular}{lccccc}
\hline
\hline
\\
 Cluster        &  $z$      &  Type   & 1 arcmin  & Components & $T_{\rm ave}$ \\
                &           &         & (${\rm kpc}~h_{70}^{-1}$)    &             &  (keV)   \\
(1)             &  (2)      & (3)     & (4)       & (5)         &  (6) \\  
\hline
 A754           ~~& 0.0542
                  & On-going
                  & 63.1 
                  & 
                  & $10.0\pm0.3^{\rm a}$ \\
 A1750          ~~& 0.0860   
                  & Binary    
                  & 96.7 
                  & A1750C 
                  & $~3.87\pm0.10 ~^{\rm b}$ \\
                ~~&
                  & 
                  & 
                  & A1750N 
                  & $~2.84\pm0.12~^{\rm b}$ \\
 A1758          ~~& 0.2790
                  & Binary 
                  & 254.0
                  & A1758N 
                  & $~8.2\pm0.4~^{\rm c}$ \\
                ~~& 
                  & 
                  &
                  & A1758S  
                  & $~6.4^{+0.3}_{-0.4}~^{\rm c}$ \\ 
A1914           ~~& 0.1712  
                  & On-going
                  & 174.9 
                  &
                  & $10.9\pm0.7~^{\rm a}$\\
 A2034          ~~& 0.1130  
                  & Cold Front 
                  & 123.2 
                  &
                  & $~7.9\pm0.4~^{\rm d}$\\
 A2142          ~~& 0.0909  
                  & Cold Front 
                  & 101.7 
                  &
                  & $~8.1\pm0.4~^{\rm e}$\\
 A520           ~~& 0.1990
                  & On-going  
                  & 197.2
                  &
                  & $~7.1\pm0.9~^{\rm a}$ \\
\hline
    \end{tabular}
  \end{center}
\textrm{ Note $\singlebond$ 
Col. (1): Cluster name. 
Col. (2): Cluster redshift.
Col. (3): Cluster merger type, classified by their X-ray features. 
'Binary' represents the system consisting of two clusters 
separated by the order of Mpc scales.
A 'Cold Front' cluster shows the ICM with the presence of a contact 
discontinuity.
'On-going' is a cluster whose X-ray surface brightness
 and temperature maps show irregular morphologies.
Col. (4): Physical scale in ${\rm kpc}/h_{70}$ unit corresponding to
 $1'$ at the cluster redshift.
Col. (5): Name of a component of binary cluster.
Col. (6): Spatially averaged temperature from 
                $^{\rm a}$ Govoni et al. (2004),
                $^{\rm b}$ Belsole et al. (2004), 
                $^{\rm c}$ David \& Kempner (2004),  
                $^{\rm d}$ Kempner \& Sarazin (2003) 
            and $^{\rm e}$ Markevitch et al. (2000). 
}
\end{table*}

\newpage
\begin{table}
  \caption{Color-Magnitude relation} \label{tab:cmr}
  \begin{center}
    \begin{tabular}{lccccc}
\hline
\hline
\\
 Cluster        & a   &  b  &  c & $L_{\rm tot}/L_{\rm obs}$ \\
(1)             & (2) &     &    & (3) \\  
\hline
 A754           ~~& -0.00999
                  &  1.15662
                  &  0.16
                  &  1.00103 \\
 A1750          ~~& -0.02849
                  &  1.65197
                  &  0.12
                  &  1.00268 \\
 A1758          ~~& -0.11661
                  &  3.93112
                  &  0.16
                  &  1.04095 \\
 A1914          ~~& -0.04522
                  &  2.13056
                  &  0.10
                  &  1.01238 \\
 A2034          ~~& -0.02424
                  &  1.49218
                  &  0.18
                  &  1.00483 \\
 A2142          ~~& -0.02019
                  &  1.51493
                  &  0.11
                  &  1.00302 \\
 A520           ~~& -0.0236 
                  & 1.33718 
                  & 0.12 
                  & 1.00531  \\
\hline
    \end{tabular}
  \end{center}
\textrm{ Note $\singlebond$ 
Col. (1): Cluster name. 
Col. (2): Color-magnitude relation parameters of the form: 
$|{\rm color}-(a\times{\rm mag}+b)|<c$, 
where  ${\rm color}=g'-R_{\rm c}$ and ${\rm mag}=R_{\rm c}$  for all targets,
except ${\rm color}=V-i'$ and ${\rm mag}=i'$ for A520.
Col. (3): Correction factor of galaxy luminosities due to the
 luminosity cutoff, $L_{\rm tot}/L_{\rm obs}=\Gamma(2-p)/\Gamma(2-p,
 L_{\rm lim}/L^*)$.
}\\
\end{table}

\newpage

\begin{table*}
  \caption{Weak Lensing Stellar Sample} \label{tab:estar}
  \begin{center}
    \begin{tabular}{lrrrrrrrr}
\hline
\hline
\\
 Cluster        &  Uncorrected $e^*$     & &          
                &  Residual $\delta e^*$ & & 
                &  $N^*$
                &  $\overline{r_h}^*$\\         

                &  $\bar{e}_1^*\times 10^2$  & $\bar{e}_2^* \times 10^2$  
                   & $\sigma(e_*)$
                &  $\overline{\delta e}^*_1 \times 10^{4}$ &
     $\overline{\delta e}^*_2 \times 10^{4}$   
                   & $\sigma(\delta e^*)$
                &
                &  arcsec\\
(1)             &  (2)   &            &   &  (3)  &       & & (4)   &    (5)             \\  
\hline

 A754           ~~& $-2.10$
                  & $-0.52$
                  & $1.77 \times 10^{-2}$
                  & $+0.63 \pm 1.30$
                  & $+0.10 \pm 0.88$
                  &  $4.25 \times 10^{-3}$
                  & 733
                  & 0.440\\
 A1750          ~~& $+0.65$ 
                  & $-0.05$ 
                  & $1.73\times 10^{-2}$
                  & $-0.46\pm 1.54$
                  & $-0.11\pm 0.92$
                  & $4.60\times 10^{-3}$
                  & 651
                  & 0.384\\
 A1758          ~~& $+0.10$ 
                  & $+0.14$ 
                  & $1.77 \times 10^{-2}$
                  & $0.00\pm 2.44$
                  & $+0.06\pm 1.00$
                  & $4.20 \times 10^{-3}$
                  & 255
                  & 0.380\\
 A1914          ~~& $-0.85$ 
                  & $+0.50$ 
                  & $2.25 \times 10^{-2}$
                  & $+0.42\pm 2.67$
                  & $-0.18\pm 1.68$
                  & $6.11 \times 10^{-3}$
                  & 375
                  & 0.345\\
 A2034          ~~& $-1.03$
                  & $+2.13$
                  & $2.51 \times 10^{-2}$
                  & $ +0.02 \pm  2.82$
                  & $ -2.70 \pm  2.10$
                  & $7.52 \times 10^{-3}$
                  & 458
                  & 0.342\\
 A2142          ~~& $-1.23$
                  & $-2.77$
                  & $2.86 \times 10^{-2}$
                  & $+1.01 \pm 2.64$
                  & $+4.24 \pm 1.82$
                  & $8.39 \times 10^{-3}$
                  & 684 
                  & 0.294\\
 A520           ~~& -2.40   
                  & -2.43   
                  & $2.98\times 10^{-2}$
                  & $+0.19\pm 1.55$     
                  & $+2.10\pm 1.29$     
                  & $7.92 \times 10^{-3}$
                  & 1537                 
                  & $0.374$\\            
\hline
    \end{tabular}
  \end{center}
\textrm{ Note $\singlebond$ 
 Col. (1): Cluster name. 
 Col. (2): Mean and standard deviation of stellar ellipticities before PSF correction.
 Col. (3): Mean and standard deviation of stellar ellipticities after
 PSF correction.
 Col. (4): Number of stellar objects.
 Col. (5): Median stellar half-light radius.
}\\
\end{table*}

\newpage

\begin{table*}
  \caption{Background Galaxy Sample and Weak Lensing Mass
 Reconstruction}\label{tab:fbg}
  \begin{center}
    \begin{tabular}{lccccc}
\hline
\hline
\\
 Cluster        &  Magnitude range
                &  $n_g$            
                &  $\bar{\sigma}_g$
                &  FWHM  
                &  $\sigma_{\kappa}$  \\

                & ABmag
                & ${\rm arcmin}^{-2}$
                &
                & arcmin
                & \\

   (1)            &  (2)     &  (3)   & (4)   & (5)  & (6)\\  

\hline

 A754           ~~&   $19\simlt R_{\rm c}\simlt 27$
                  & 37.3
                  & 0.376
                  & 1.67
                  & 0.01739\\
 A1750          ~~&  $19\simlt R_{\rm c}\simlt 27$
                  & 56.4
                  & 0.405
                  & 1.25
                  & 0.02032\\
 A1758          ~~&  $19\simlt R_{\rm c}\simlt 27$
                  & 46.4
                  & 0.408
                  & 1.25
                  & 0.02253\\
 A1914          ~~& $19\simlt R_{\rm c}\simlt 27$
                  & 47.7 
                  & 0.411
                  & 0.75
                  & 0.03737\\
 A2034          ~~& $19\simlt R_{\rm c}\simlt 27$
                  & 52.4
                  & 0.413
                  & 1.17
                  & 0.02301\\
 A2142          ~~& $19\simlt R_{\rm c}\simlt 27$
                  & 72.1 
                  & 0.395
                  & 1.00
                  & 0.02190\\
 A520           ~~& $ 20\simlt i' \simlt 25.5$ 
                  & 37.6 
                  & 0.424
                  & 1.25
                  & 0.02605\\ 
\hline
    \end{tabular}
  \end{center}
\textrm{ Note $\singlebond$ 
Col. (1): Cluster name. 
Col, (2): Magnitude range of the background galaxy sample.
Col. (3): Surface number density of background galaxies for map-making.
Col. (4): RMS error for the shear estimate per galaxy.
Col. (5): Gaussian FWHM in map-making.
Col. (6): RMS noise in the reconstructed mass map.
}\\
\end{table*}

\newpage

\begin{table*}
  \caption{Red Background Galaxy Sample}\label{tab:red}
  \begin{center}
    \begin{tabular}{lccc}
\hline
\hline
\\
 Cluster        &  Magnitude range
                &  Color
                &  $n_g$\\            

                & ABmag
                &
                & ${\rm arcmin}^{-2}$\\

   (1)            &  (2)     &  (3)   & (4)\\

\hline

 A754           ~~&   $19\simlt R_{\rm c}\simlt 27$
                  &  $1.14 < g'-R_{\rm c} < 15.0 $ 
                  &   6.61 \\
 A1750          ~~&  $19\simlt R_{\rm c}\simlt 27$
                  &   $1.23 < g'-R_{\rm c} < 15.0 $
                  &  8.55\\
 A1758          ~~&  $19\simlt R_{\rm c}\simlt 27$
                  &   $1.20 < g'-R_{\rm c} < 15.0 $
                  &  7.90\\
 A1914          ~~& $19\simlt R_{\rm c}\simlt 27$
                  &  $1.36< g'-R_{\rm c} < 15.0 $
                  &  5.11\\
 A2034          ~~& $19\simlt R_{\rm c}\simlt 27$
                  &  $1.20< g'-R_{\rm c} < 15.0 $ 
                  &  7.31\\
 A2142          ~~& $19\simlt R_{\rm c}\simlt 27$
                  &  $1.29< g'-R_{\rm c} < 15.0 $ 
                  &  7.55\\   
  A520          ~~& $20\simlt i' \simlt 25.5$            
                  & $0.90<V-i'<15.0$
                  & 9.40 \\         
\hline
    \end{tabular}
  \end{center}
\textrm{ Note $\singlebond$ 
Col. (1): Cluster name. 
Col, (2): Magnitude range of the red background galaxy sample.
Col. (3): Color range of the red background galaxy sample.
Col. (4): Mean surface number density of the red background galaxies.
}\\
\end{table*}

\newpage

\newpage

\begin{table*}
  \caption{Best-Fitting Mass Models} \label{tab:zetamodel}
  \begin{center}
    \begin{tabular}{lccccccc}
\hline
\hline
 Cluster        &  SIS                     &             &
                &  NFW                     &             &
                &                          \\
                &  $\sigma_v$
                &  $\chi^2/{\rm dof}$ (dof)
                &  ($\theta_{\rm min}, \theta_{\rm max}$) 
                &  $M_{\rm vir}$
                &  $c_{\rm vir}$ 
                &  $\chi^2/{\rm dof}$ (dof)
                &  ($\theta_{\rm min}, \theta_{\rm max}$)   \\
                &  (${\rm km~s^{-1}}$)             &   &  (arcmin)       
                &  ($h^{-1}10^{14}M_{\odot}$) &
                &     &  (arcmin)  \\
(1)             &    (2)     &  (3)     &  (4)  &  (5)    &    (6)  &(7) &(8) \\  
\hline
 A754           ~~& $703.1 \pm 139.5$ 
                  & $0.32(3)$
                  & ($6.7, 15.0)$
                  & $2.86 \pm 3.77$
                  & $4.97 \pm 5.26$
                  & $2.57(4)$
                  & ($1.5, 5.4$)  \\

 A1750N         ~~& $633.9\pm 80.2$
                  & $1.30(7)$
                  & ($1.0,7.0$)
                  & $2.90\pm 2.08$
                  & $5.21\pm 3.42$
                  & $1.11(6)$  
                  & ($1.0,7.0$) \\

 A1750C         ~~& $588.9\pm 90.6$
                  & $0.91(7)$
                  & ($1.0,6.7$)
                  & $1.88 \pm 1.63$
                  & $5.52 \pm 4.89$
                  & $0.99(6)$
                  & ($1.0,6.7$) \\
 A1758N          ~& $672.5\pm 113.3$
                  & $1.23(4)$
                  & ($1.7, 5.0$)
                  & $5.26\pm 5.70$
                  & $0.24\pm 0.95$
                  & $1.01(5)$
                  & ($1.0, 5.0$) \\
 A1758S          ~& $594.5\pm 94.0$
                  & $3.51(7)$
                  & ($0.5, 4.6$)
                  & $1.20\pm 1.15$
                  & $3.91\pm 6.51$
                  & $1.04(6)$
                  & ($0.5, 4.6$) \\
 A1914          ~~& $845.7\pm 87.4$
                  & $1.08(7)$
                  & ($0.5, 5.4$)
                  & $6.14 \pm 3.19$
                  & $4.13 \pm 2.79$
                  & $0.75(6)$
                  & ($1.0, 7.0$)\\
 A2034          ~~& $792.8 \pm 78.2$
                  & $1.46(10)$
                  & ($1.0, 12.0$)
                  & $7.17 \pm 4.30$
                  & $3.74 \pm 2.25$
                  & $0.84(9)$
                  & ($1.0, 12.0$) \\
 A2142          ~~& $940.4 \pm 62.6$
                  & $2.03(10)$
                  & ($1.0, 15.0$)
                  & $11.95 \pm 5.24$
                  & $4.32\pm 1.70$
                  & $0.83(9)$
                  & ($1.0, 15.0$)\\
 A520           ~~& $960.4 \pm 94.0$
                  & $0.70(4)$
                  & ($5.1,15.0$)
                  & $6.14 \pm 2.38$
                  & $2.90 \pm 1.82$
                  & $0.68(7)$
                  & ($1.7,15.0$) \\
\hline
    \end{tabular}
  \end{center}
\textrm{ Note $\singlebond$ 
Col. (1): Cluster name. 
Col. (2): Best-fitting parameter value for the SIS 1D velocity dispersion.
Col. (3): Reduced $\chi^2$ for the best-fitting SIS model, and the
 degrees-of-freedom in parenthesis.
Col. (4): Radial range of the $\zeta_c$-statistic measurements 
used for the SIS model fitting.
Col. (5): Best-fitting parameter value for the NFW virial mass.
Col. (6): Best-fitting parameter value for the NFW concentration parameter.
Col. (7): Reduced $\chi^2$ for the best-fitting NFW model, and the
 degrees-of-freedom in parenthesis.
Col. (8):  Radial range of the $\zeta_c$-statistic measurements 
used for the NFW model fitting.
}\\
\end{table*}

\newpage

\begin{table*}
  \caption{Global Cluster Properties} \label{tab:global}
  \begin{center}
    \begin{tabular}{lccccc}
\hline
\hline
 Cluster        &  $M_\zeta(<\theta_{M}) $   
                &  $L(<\theta_{M})$
                &  $M_\zeta/L(<\theta_{M})$
                &  $\theta_{M}$ 
                &  $(\theta_{\rm inn},\theta_{\rm out})$     \\
                &  ($h^{-1}10^{14}M_{\odot}$) 
                &  ($h^{-2}10^{12}L_{\odot}$) 
                &  ($hM_{\odot}/L_{\odot}$) 
                &  (arcmin) 
                &  (arcmin)   \\
(1)             &  (2)       &  (3)   &  (4)  &  (5)    &    (6)  \\  
\hline
 A754           ~~& $2.73 \pm 1.70$
                  & $0.29^{+0.22}_{-0.03}$
                  & $932^{+635}_{-771}$ 
                  & $15.0$  
                  & ($15.0,20.0$)\\
 A1750N         ~~& $1.49 \pm 0.65$ 
                  & $0.22^{+0.07}_{-0.01}$
                  & $660^{+290}_{-346}$
                  & $6.96$ 
                  & ($16.0,21.0$)\\
 A1750C         ~~& $1.26\pm0.60$
                  & $0.17^{+0.06}_{-0.01}$
                  & $760^{+366}_{-439}$
                  & $6.65$
                  & ($15.0,20.0$)\\
 A1758N         ~~& $4.89\pm1.54$
                  & $2.17^{+0.13}_{-0.02}$
                  & $225^{+71}_{-77}$
                  & $5.08$
                  & ($15.0,20.0$)\\
 A1758S         ~~& $6.06\pm1.10$
                  & $1.20^{+0.18}_{-0.02}$
                  & $506^{+92}_{-136}$
                  & $4.63$
                  &($12.0,17.0$)\\
 A1914          ~~& $4.10\pm1.55$
                  & $1.08^{+0.22}_{-0.02}$
                  & $380^{+144}_{-176}$
                  & $6.96$
                  &($15.0,20.0$)\\
 A2034          ~~& $3.90\pm1.81$   
                  & $1.02^{+0.44}_{-0.05}$
                  & $382^{+178}_{-229}$
                  & $12.0$
                  & ($12.0,17.0$) \\
 A2142          ~~& $3.97\pm2.04$
                  & $1.09^{+1.37}_{-0.18}$
                  & $365^{+197}_{-274}$
                  & $15.0$
                  & ($15.0,20.0$)\\
 A520           ~~& $5.18\pm4.19$       
                  & $3.40^{+1.35}_{-0.16}$
                  & $152^{+123}_{-130}$   
                  & $15.0$
                  & ($15.0,20.0$)\\
\hline
    \end{tabular}
  \end{center}
\textrm{ Note $\singlebond$ 
 Col. (1): Cluster name. 
 Col. (2): Cluster mass $M_{\zeta}(<\theta_{M})$ enclosed within $\theta_{M}$
  from aperture densitometry.
 Col. (3): Cluster luminosity enclosed within $\theta_{M}$
           in $R_{\rm c}$-band for all targets,
           except in $i'$-band for A520.
 Col. (5): Cluster mass-to-light ratio within $\theta_{M}$.  
 Col. (6): Aperture radius in units of arcmin. 
 Col. (8): Inner and outer boundaries of the background annulus
for the cluster mass and luminosity measurements,
given in units of arcmin.
}\\
\end{table*}

\newpage

\newpage
\begin{figure*}
\begin{center}
\FigureFile(90mm,90mm){Fig2.ps}
\end{center}
\caption{
The lower panel shows 
the inverse of the critical surface mass density of gravitational lensing,
$\Sigma_{\rm cr}^{-1} (z_d, z_s)$, 
as a function of lens redshift $z_d$
for three different source redshifts, $z_s=0.8, 1.0, 1.2$
({\it dashed}, {\it  solid}, and {\it dotted-dashed}, respectively),
demonstrating the geometric scaling of gravitational lensing signal.
The top panel shows the relative lensing strength
$\Sigma_{\rm cr}^{-1} (z_d,  z_s)/\Sigma_{\rm cr}^{-1} (z_d, z_s=1.0)$
as a function of lens redshift $z_d$
normalized with respect to the source at $z_s=1$. 
For lensing clusters at low redshifts $z_d$,
$\Sigma_{\rm cr}$ depends very weakly on the background redshift
$z_s$, 
so that the uncertainty in $z_s$ of background galaxies is less
 important in the lensing-based cluster mass determination. 
}\label{fig:lensS}
\end{figure*}

\newpage
\begin{figure*}
\begin{center}
\FigureFile(150mm,150mm){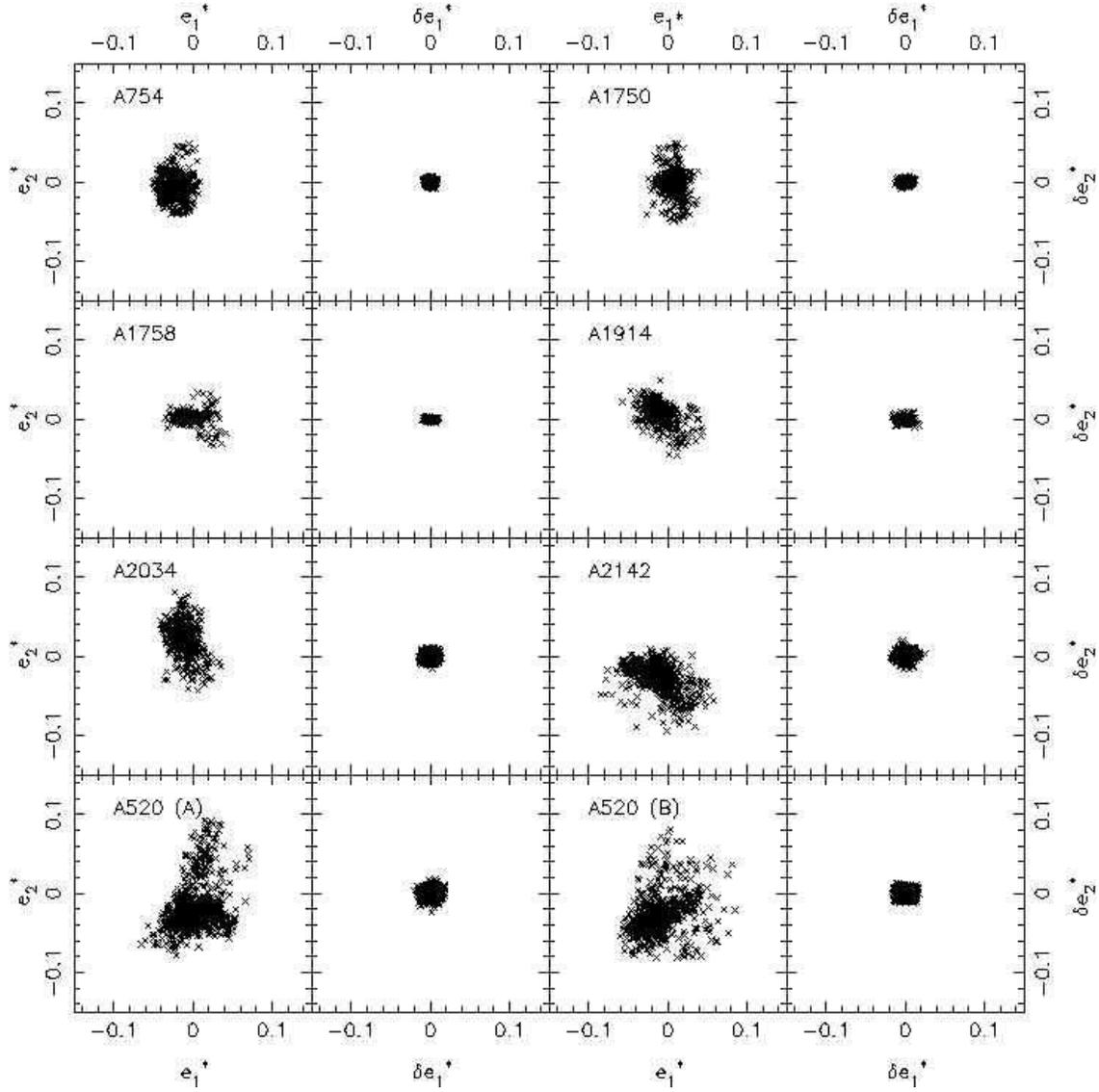}
\end{center}
\caption{
Stellar ellipticity distributions before and after the PSF anisotropy
 correction for individual cluster targets.
For each cluster the left panel shows the raw ellipticity
 components $(e^*_1,e^*_2)$ of stellar objects, 
and the right panel shows the residual ellipticity components
$(\delta e^*_1, \delta e^*_2)$ after the PSF anisotropy correction.
For A520 two different imaging data sets are shown:
A520 (A) is based on a co-added mosaic of seven $i'$ images
($7\times 240$s) taken both with and without guide probe.
A520 (B) is based on a co-added mosaic of four $i'$ images ($4\times  240$s)
taken without guide probe.
}
\label{fig:e+eres}
\end{figure*}

\newpage

\begin{figure*}
\begin{center}
\FigureFile(170mm,170mm){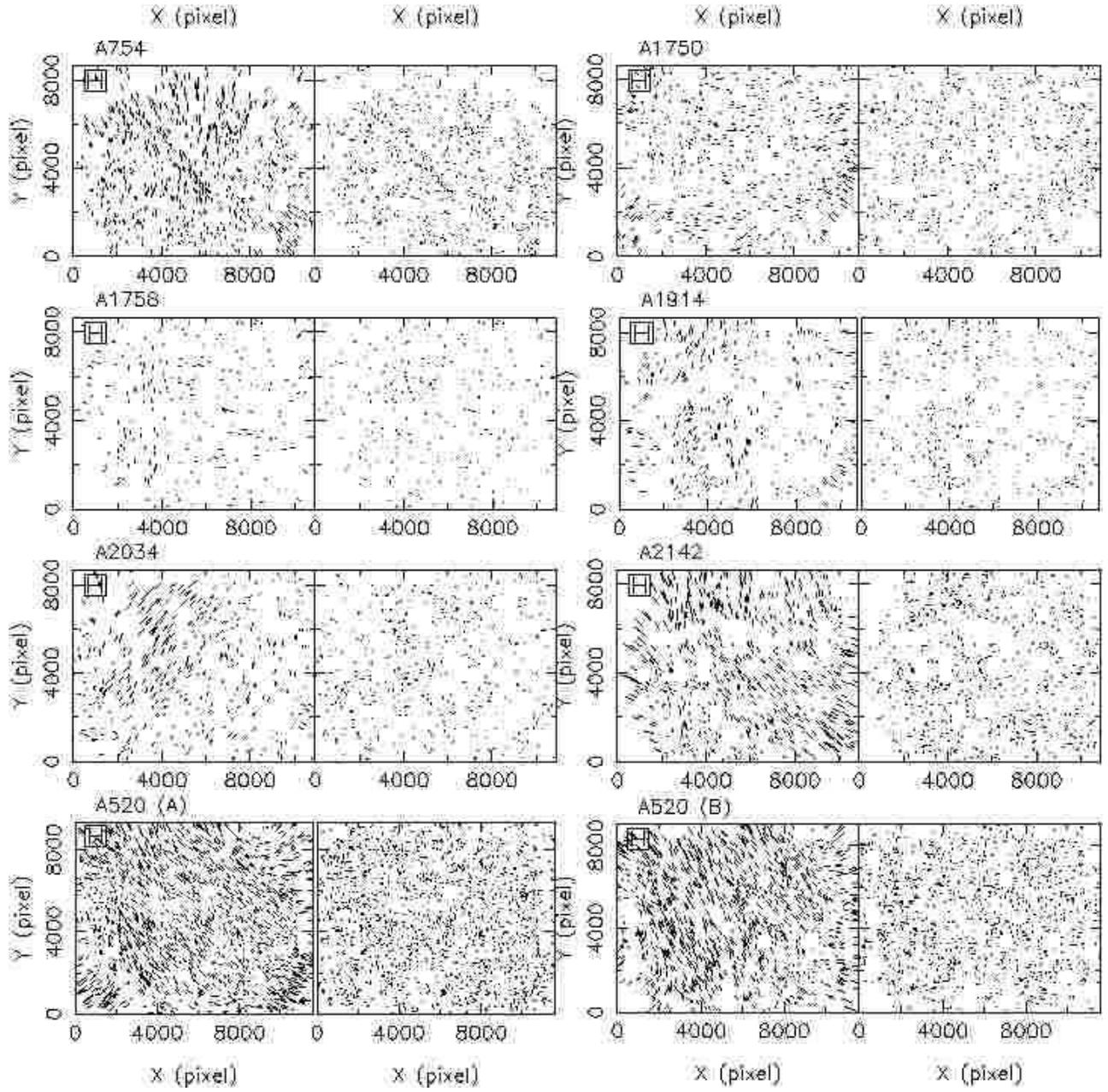}
\end{center}
\caption{
The distortion field of stellar ellipticities before and after the PSF
 anisotropy correction for individual cluster targets.
For each cluster the left panel shows the raw ellipticity field
of stellar objects, and the right panel shows the residual ellipticity
 field after the PSF anisotropy correction.
The orientation of the sticks indicates the position angle of the major
 axis of stellar ellipticity, whereas
the length is proportional to the modulus of stellar ellipticity.
A stick with the length of $10\%$ ellipticity is indicated in the top
 left of the left panel for each target.
For A520 two different imaging data sets are shown:
A520 (A) is based on a co-added mosaic of seven $i'$ images
($7\times 240$s) taken both with and without guide probe.
A520 (B) is based on a co-added mosaic of four $i'$ images ($4\times  240$s)
taken without guide probe.
}\label{fig:emap}
\end{figure*}

\newpage

\begin{figure*}
\begin{center}
\FigureFile(170mm,90mm){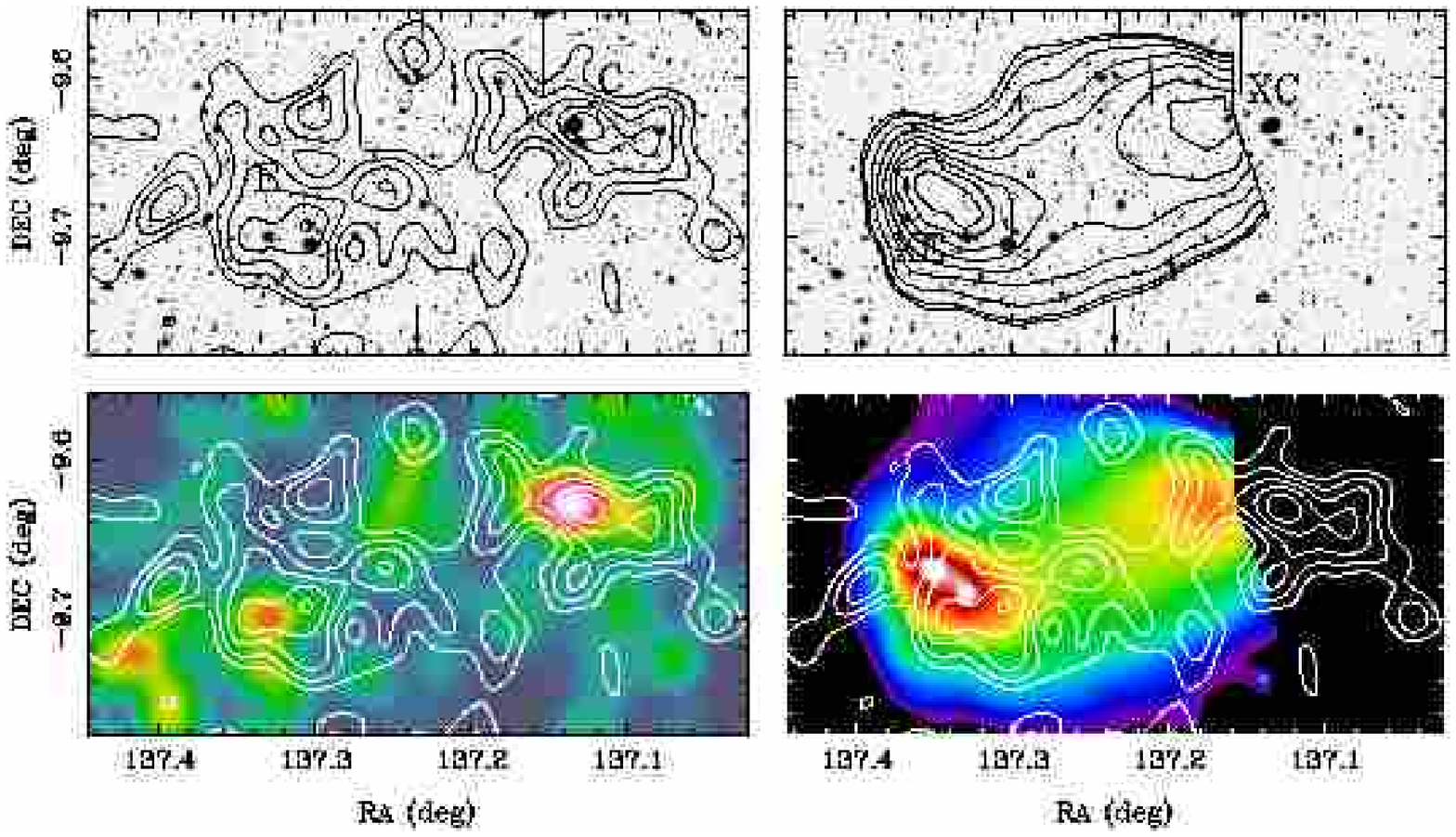} 
\end{center}
\caption{A754 \singlebond  
Top-left:
Subaru $R_{\rm c}$-band image of the central $\sim 25'\times 13'$ 
cluster region. Overlayed are contours of the lensing $\kappa$-field
reconstructed from weak shear data.
The contours are spaced in units of $1\sigma$ reconstruction error
(see Table \ref{tab:fbg}).
The Gaussian FWHM 
used for the mass reconstruction is $1\farcm 67$.
Top-right: adaptively-smoothed Chandra X-ray contours ($0.7-7.0$ keV)
overlaid on the same $R_{\rm c}$-band image.
Bottom-left: 
Cluster luminosity density distribution 
in $R_{\rm c}$-band smoothed to the same
angular resolution of the mass map.
Overlayed are the same mass contours as in the top-left panel.
Bottom-right: 
The same mass contours overlayed on the adaptively-smoothed Chandra
 X-ray
image ($0.7-7.0$ keV).
 }\label{fig:a754} 
\end{figure*}

\newpage

\begin{figure*}
\begin{center}
\FigureFile(170mm,140mm){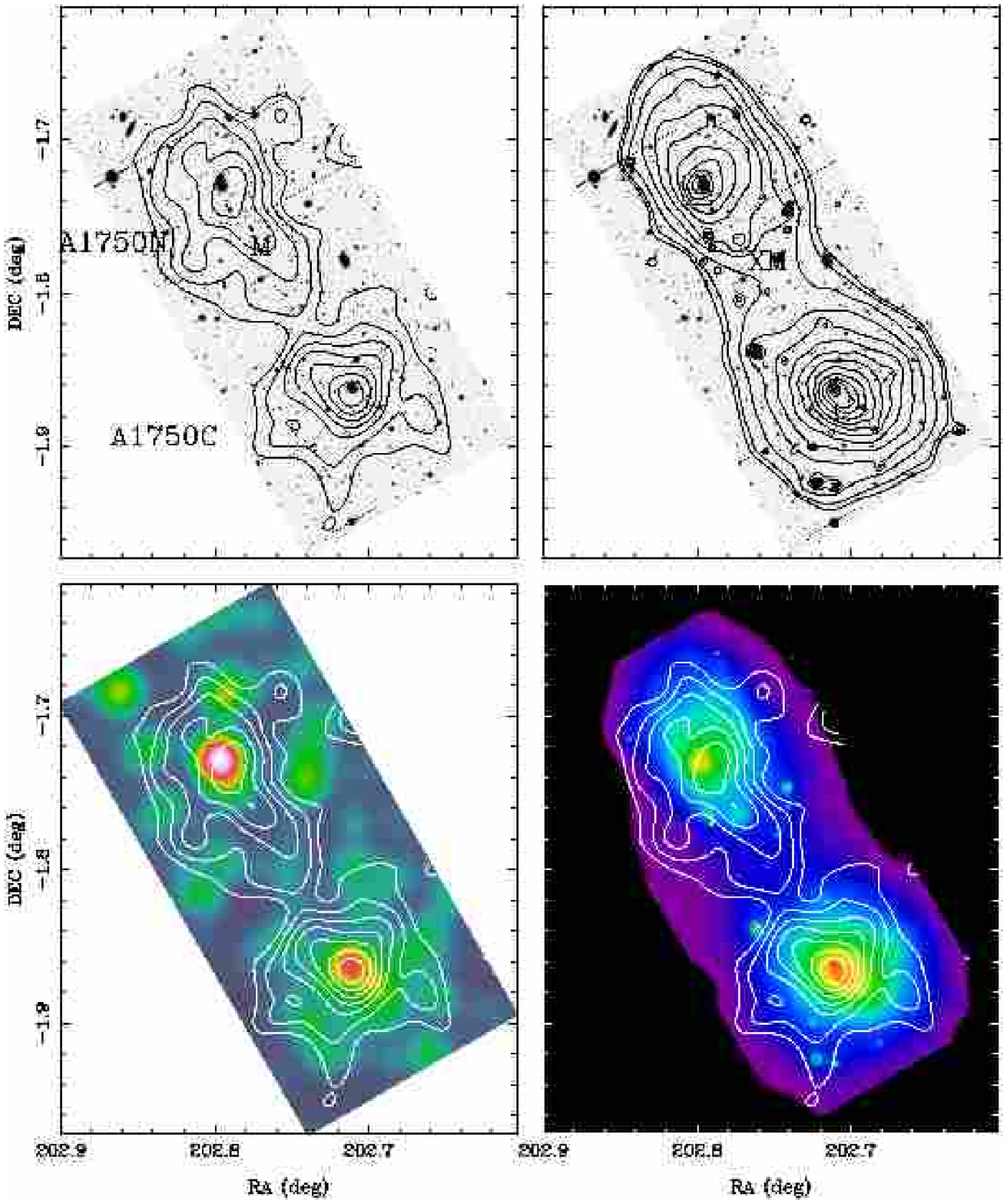}
\end{center}
\caption{A1750 \singlebond  
Top-left:
Subaru $R_{\rm c}$-band image of the central $\sim 21',\times 18'$ 
cluster region. Overlayed are contours of the lensing $\kappa$-field
reconstructed from weak shear data.
The contours are spaced in units of $1\sigma$ reconstruction error
(see Table \ref{tab:fbg}).
The Gaussian FWHM 
used for the mass reconstruction is $1\farcm 25$.
Top-right: adaptively-smoothed XMM-Newton X-ray contours ($0.5-7.0$ keV)
overlaid on the same $R_{\rm c}$-band image.
Bottom-left: 
Cluster luminosity density distribution 
in $R_{\rm c}$-band smoothed to the same
angular resolution of the mass map.
Overlayed are the same mass contours as in the top-left panel.
Bottom-right: 
The same mass contours overlayed on the adaptively-smoothed XMM-Newton
 X-ray
image ($0.5-7.0$ keV).
}\label{fig:a1750} 
\end{figure*}

\newpage

\begin{figure*}
\begin{center}
\FigureFile(170mm,140mm){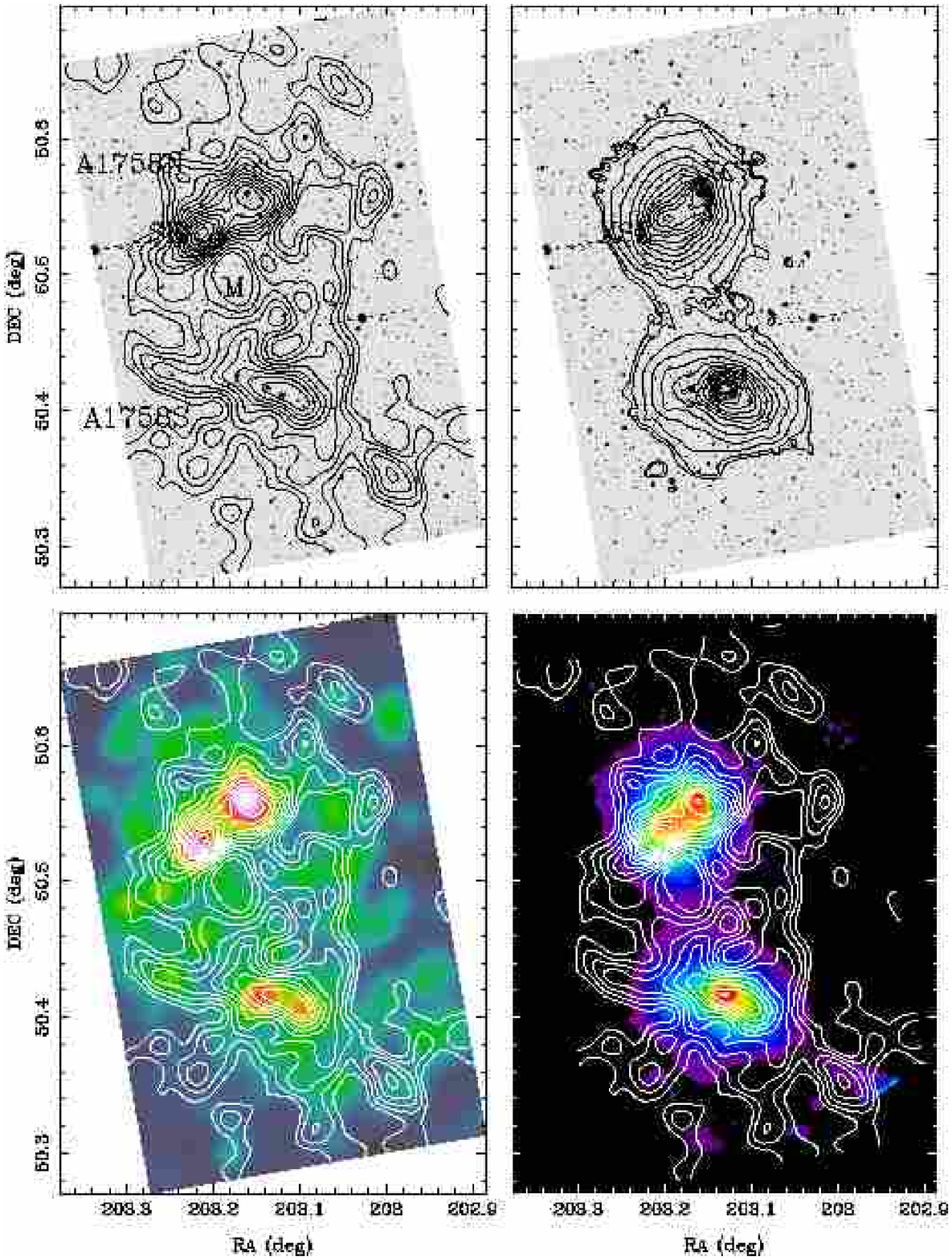}
\end{center}
\caption{A1758 \singlebond  
Top-left:
Subaru $R_{\rm c}$-band image of the central $\sim 26',\times 19'$ 
cluster region. Overlayed are contours of the lensing $\kappa$-field
reconstructed from weak shear data.
The contours are spaced in units of $1\sigma$ reconstruction error
(see Table \ref{tab:fbg}).
The Gaussian FWHM 
used for the mass reconstruction is $1\farcm 25$.
Top-right: adaptively-smoothed XMM-Newton X-ray contours ($0.5-7.0$ keV)
overlaid on the same $R_{\rm c}$-band image.
Bottom-left: 
Cluster luminosity density distribution 
in $R_{\rm c}$-band smoothed to the same
angular resolution of the mass map.
Overlayed are the same mass contours as in the top-left panel.
Bottom-right: 
The same mass contours overlayed on the adaptively-smoothed XMM-Newton
 X-ray
image ($0.5-7.0$ keV).
}\label{fig:a1758} 
\end{figure*}

\newpage

\begin{figure*}
\begin{center}
\FigureFile(170mm,140mm){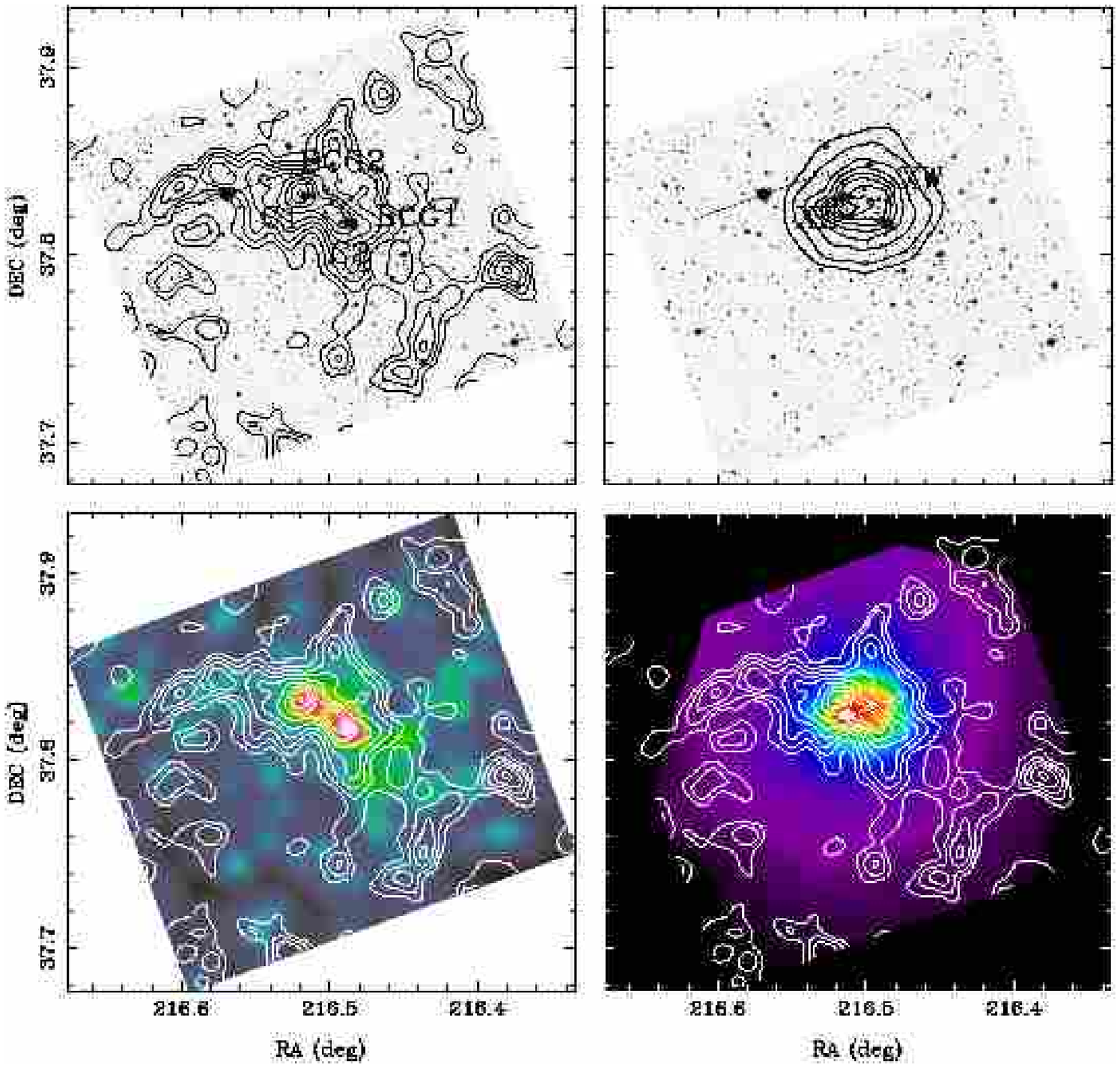}
\end{center}
\caption{A1914 \singlebond  
Top-left:
Subaru $R_{\rm c}$-band image of the central $\sim 16',\times 15\farcm 5$ 
cluster region. Overlayed are contours of the lensing $\kappa$-field
reconstructed from weak shear data.
The contours are spaced in units of $1\sigma$ reconstruction error
(see Table \ref{tab:fbg}).
The Gaussian FWHM 
used for the mass reconstruction is $0\farcm 75$.
Top-right: adaptively-smoothed Chandra X-ray contours ($0.7-7.0$ keV)
overlaid on the same $R_{\rm c}$-band image.
Bottom-left: 
Cluster luminosity density distribution 
in $R_{\rm c}$-band smoothed to the same
angular resolution of the mass map.
Overlayed are the same mass contours as in the top-left panel.
Bottom-right: 
The same mass contours overlayed on the adaptively-smoothed Chandra
 X-ray
image ($0.7-7.0$ keV).
}\label{fig:a1914} 
\end{figure*}

\newpage

\begin{figure*}
\begin{center}
\begin{tabular}{cc}
\FigureFile(170mm,140mm){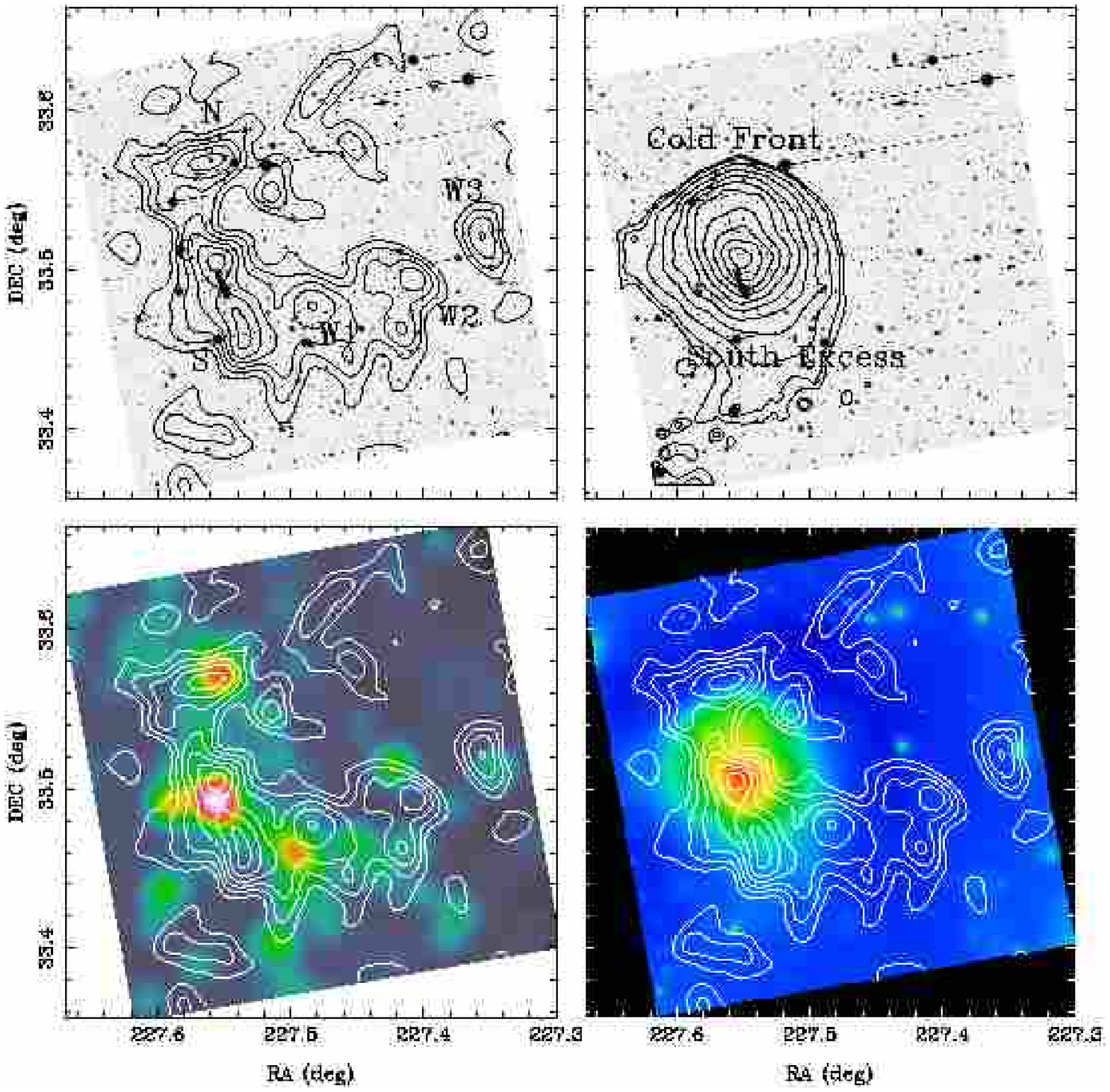}
\end{tabular}
\end{center}
\caption{A2034 \singlebond  
Top-left:
Subaru $R_{\rm c}$-band image of the central 
$\sim 18\farcm 5,\times 18\farcm 5$ 
cluster region. Overlayed are contours of the lensing $\kappa$-field
reconstructed from weak shear data.
The contours are spaced in units of $1\sigma$ reconstruction error
(see Table \ref{tab:fbg}).
The Gaussian FWHM 
used for the mass reconstruction is $1\farcm 17$.
Top-right: adaptively-smoothed XMM-Newton X-ray contours ($0.5-7.0$ keV)
overlaid on the same $R_{\rm c}$-band image.
Bottom-left: 
Cluster luminosity density distribution 
in $R_{\rm c}$-band smoothed to the same
angular resolution of the mass map.
Overlayed are the same mass contours as in the top-left panel.
Bottom-right: 
The same mass contours overlayed on the adaptively-smoothed XMM-Newton
 X-ray
image ($0.5-7.0$ keV).
}\label{fig:a2034} 
\end{figure*}

\newpage

\begin{figure*}
\begin{center}
\begin{tabular}{cc}
\FigureFile(170mm,140mm){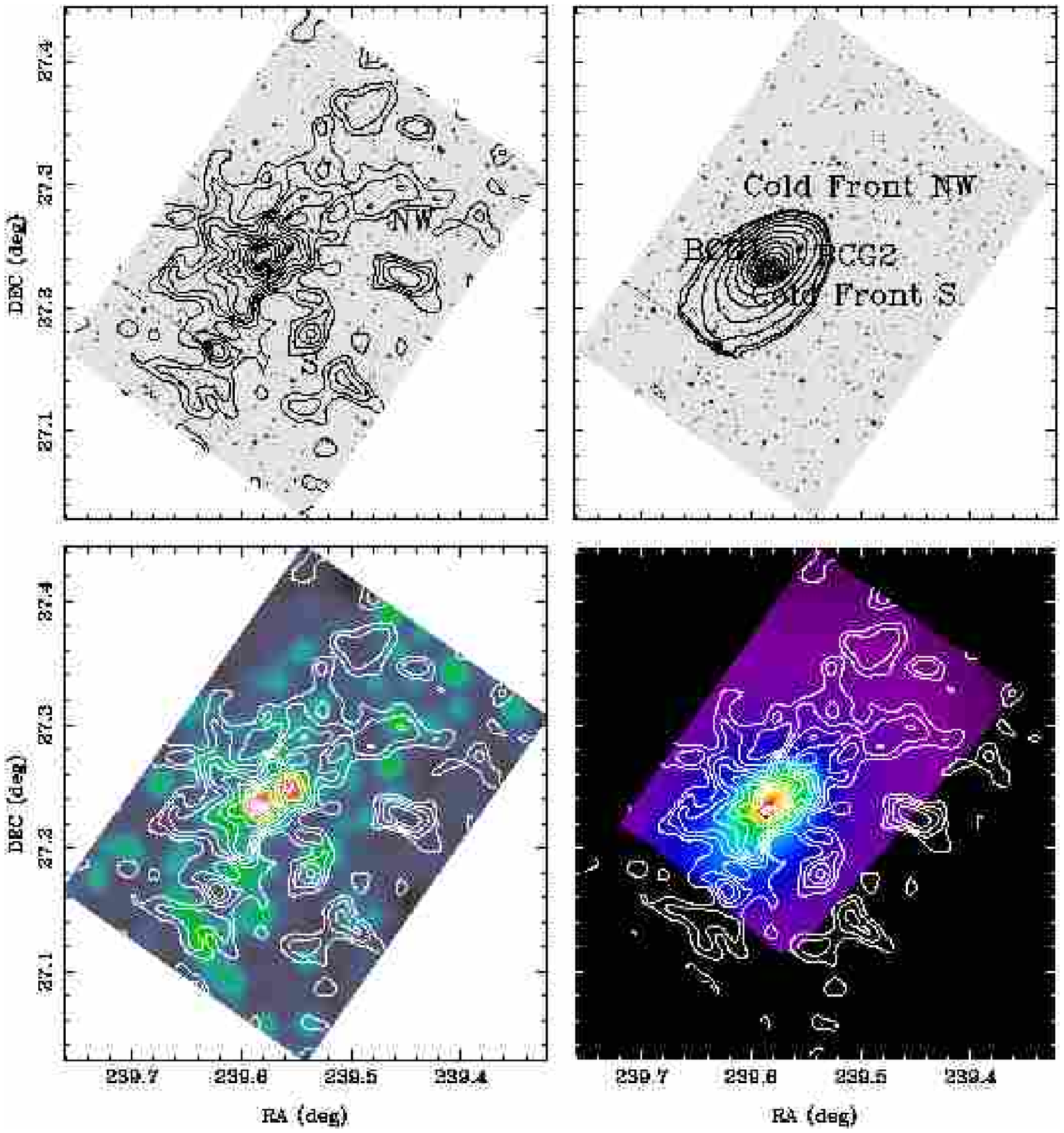}
\end{tabular}
\end{center}
\caption{A2142 \singlebond  

Top-left:
Subaru $R_{\rm c}$-band image of the central $\sim 25',\times 23\farcm 5$ 
cluster region. Overlayed are contours of the lensing $\kappa$-field
reconstructed from weak shear data.
The contours are spaced in units of $1\sigma$ reconstruction error
(see Table \ref{tab:fbg}).
The Gaussian FWHM 
used for the mass reconstruction is $1\farcm 00$.
Top-right: adaptively-smoothed Chandra X-ray contours ($0.7-7.0$ keV)
overlaid on the same $R_{\rm c}$-band image.
Bottom-left: 
Cluster luminosity density distribution 
in $R_{\rm c}$-band smoothed to the same
angular resolution of the mass map.
Overlayed are the same mass contours as in the top-left panel.
Bottom-right: 
The same mass contours overlayed on the adaptively-smoothed Chandra
 X-ray
image ($0.7-7.0$ keV).
}\label{fig:a2142} 
\end{figure*}

\newpage

\begin{figure*}
\begin{center}
\FigureFile(170mm,140mm){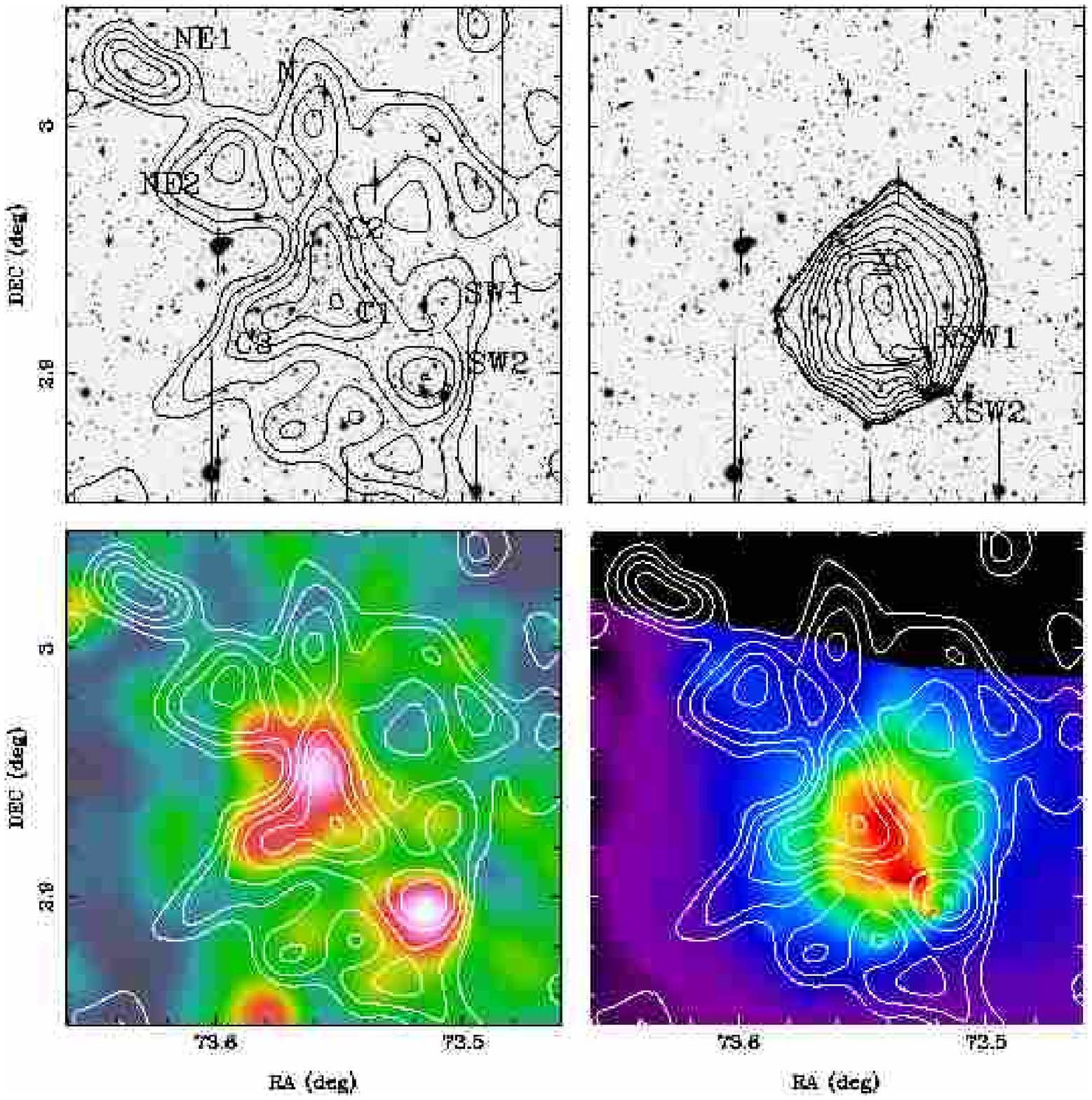}
\end{center}
\caption{A520 \singlebond  
Top-left:
Subaru $i'$-band image of the central $\sim 12',\times 12'$ 
cluster region. Overlayed are contours of the lensing $\kappa$-field
reconstructed from weak shear data.
The contours are spaced in units of $1\sigma$ reconstruction error
(see Table \ref{tab:fbg}).
The Gaussian FWHM 
used for the mass reconstruction is $1\farcm 25$.
Top-right: adaptively-smoothed Chandra X-ray contours ($0.7-7.0$ keV)
overlaid on the same $i'$-band image.
Bottom-left: 
Cluster luminosity density distribution 
in $i'$-band smoothed to the same
angular resolution of the mass map.
Overlayed are the same mass contours as in the top-left panel.
Bottom-right: 
The same mass contours overlayed on the adaptively-smoothed Chandra
 X-ray
image ($0.7-7.0$ keV).
}\label{fig:a520} 
\end{figure*}

\newpage

\begin{figure*}
\begin{center}
\FigureFile(140mm,140mm){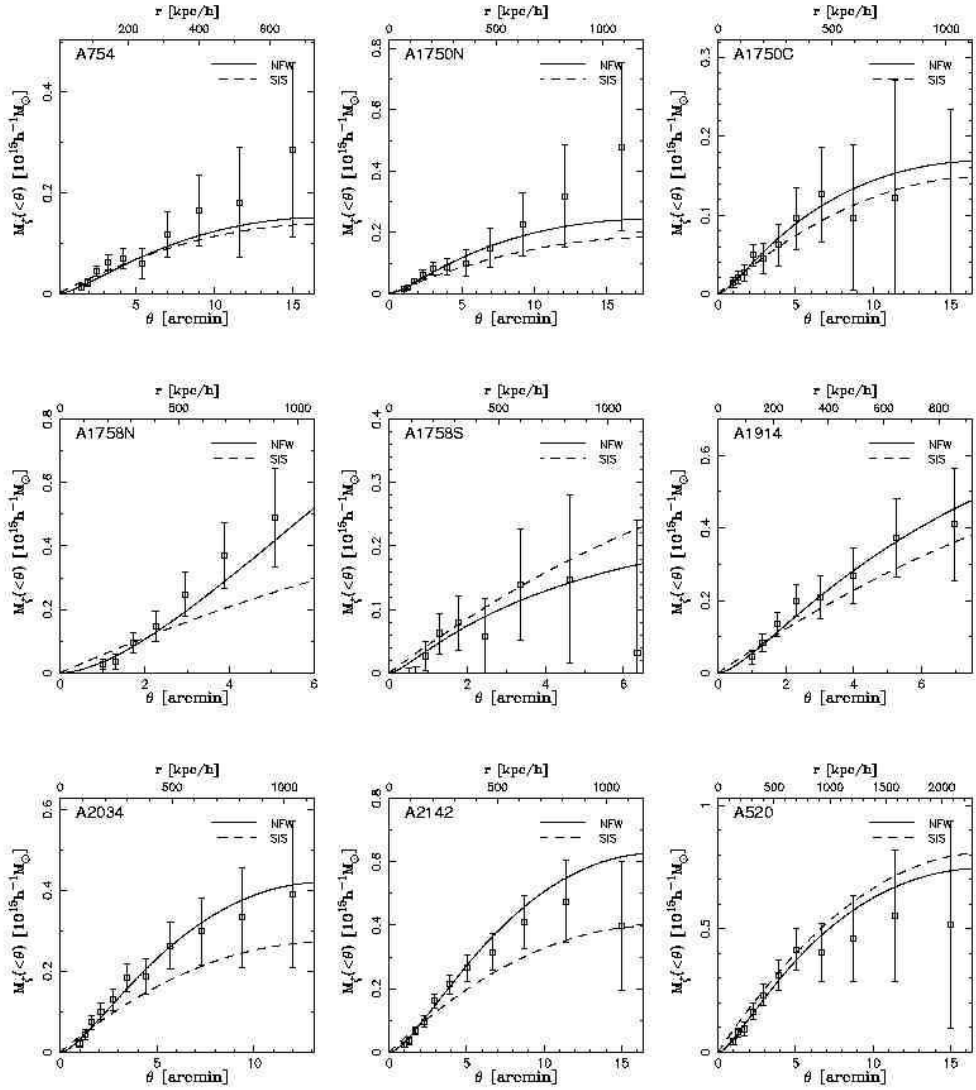} 
\end{center}
\caption{
Radial profiles of the projected mass
$M_{\zeta}(\theta)$ for a sample of seven merging clusters
of galaxies as measured by the weak lensing $\zeta$-statistic.
The error bars are correlated.
Also plotted are the best-fitting NFW and SIS profiles.
}
\label{fig:all_zeta} 
\end{figure*}

\begin{figure*}
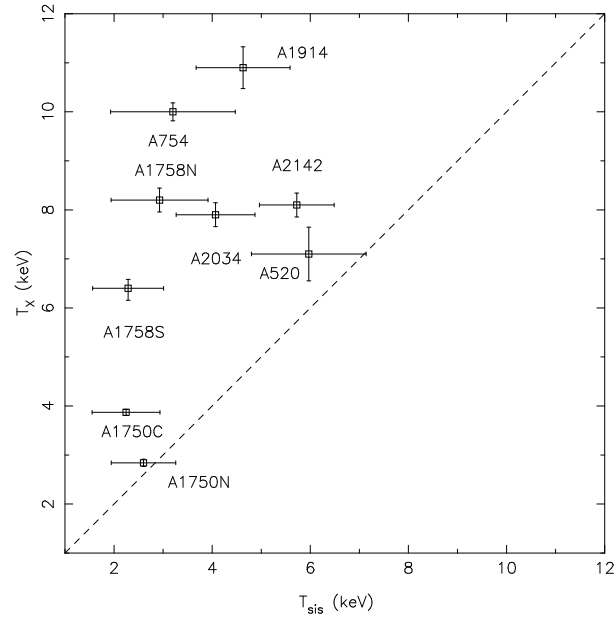

\begin{center}
\FigureFile(80mm,80mm){Fig13.ps} 
\end{center}
\caption{
Comparison between the best-fitting SIS temperature $T_{\rm SIS}$ 
from the $\zeta_c$-statistic measurement vs. the averaged X-ray
 temperature $T_X$ for a sample of nearby merging clusters.
The dashed line indicates $T_{\rm SIS}=T_X$.
For all clusters except A1750N,
the observed X-ray temperature is significantly higher than 
the expected virial temperature of the cluster mass.
}
\label{fig:tmp} 
\end{figure*}

\begin{figure*}
\begin{center}
\FigureFile(80mm,80mm){Fig14.ps} 
\end{center}
\caption{
The temperature 
ratio $T_X/T_{\rm SIS}$ v.s. the best-fitting SIS temperature
 $T_{\rm SIS}$.
The dashed line indicates $T_{\rm SIS}=T_X$.
}
\label{fig:tratio} 
\end{figure*}

\newpage

\begin{figure*}
\begin{center}
\FigureFile(80mm,80mm){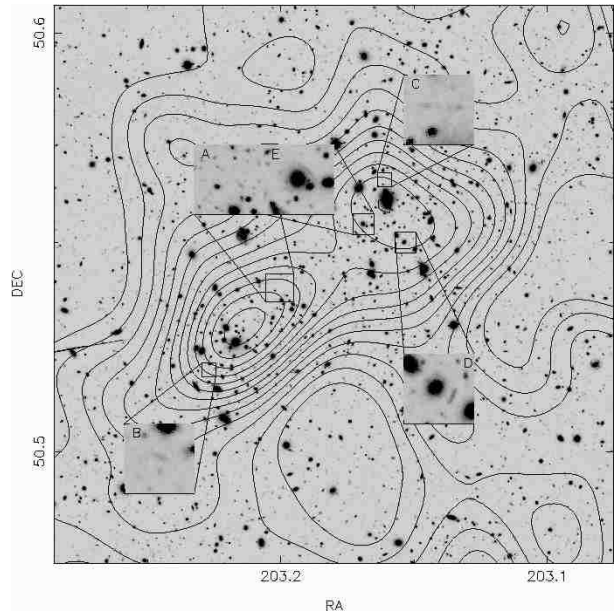} 
\end{center}
\caption{Subaru $R_{\rm c}$-band image of the central 
$8\farcm \times 8\farcm$  cluster region of A1758N. 
Overlayed are contours of the reconstructed
projected mass distribution of the cluster.
The insets zoom in on tangential arc candidates 
identified based on visual inspection.
The panels A ($0\farcm4 \times 0\farcm4$) and B ($0\farcm2 \times 0\farcm2$) 
are zoom in of blue tangential arcs around the SE mass clump. 
The panel C ($0\farcm2 \times 0\farcm2$) is a zoom in view of blue
tangential arcs  around the mass clump C.
The panels D ($0\farcm3 \times 0\farcm3$) 
and E ($0\farcm3 \times 0\farcm3$) are 
zoom in of tangential arcs 
associated with cluster galaxy concentrations.
}
\label{fig:a1758N_zoom} 
\end{figure*}

\begin{figure*}
\begin{center}
\FigureFile(80mm,80mm){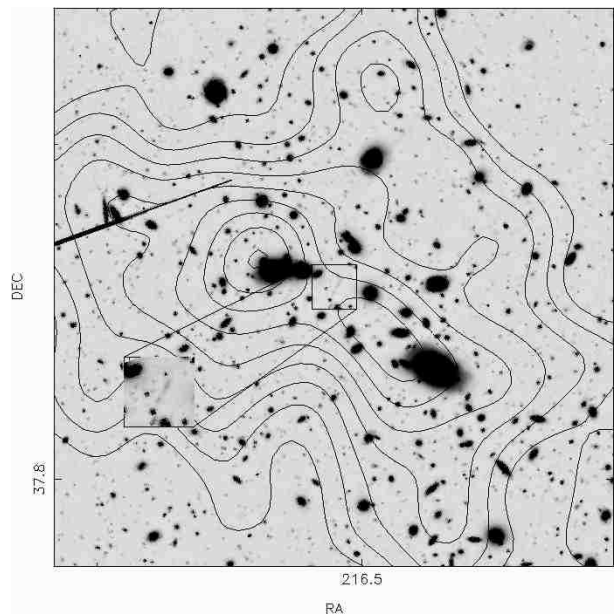} 
\end{center}
\caption{Subaru $R_{\rm c}$-band image of the central 
$5\farcm \times 5\farcm$ cluster region of A1914. 
Overlayed are contours of the reconstructed projected mass distribution
of the cluster.
The inset panel ($0\farcm4 \times 0\farcm4$) shows a zoom in view of
two blue tangential arcs around BCG2.}
\label{fig:a1914_zoom} 
\end{figure*}

\begin{figure*}
\begin{center}
\FigureFile(80mm,80mm){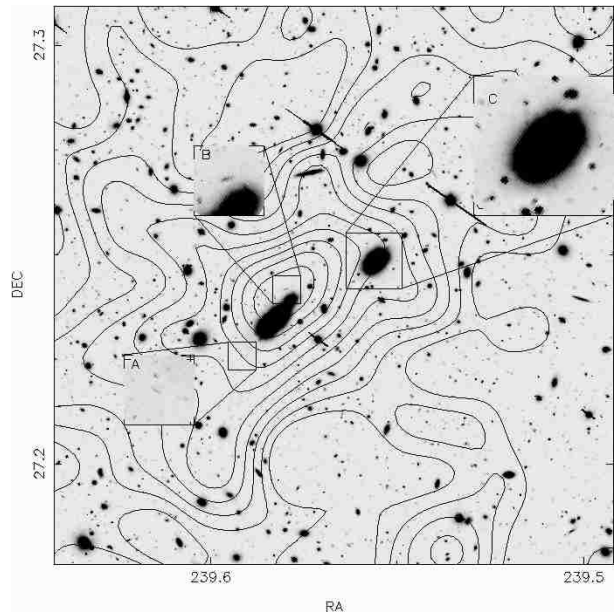} 
\end{center}
\caption{
Subaru $R_{\rm c}$-band image of the central 
$8\farcm \times 8\farcm$  cluster region of A2142. 
Overlayed are contours of the reconstructed projected mass distribution
of the cluster.
The insets zoom in on tangential arc candidates 
identified based on visual inspection.
The panel A shows a zoom in view 
($0\farcm 4 \times 0\farcm 4$) of a tangential arc.
The panel B shows a zoom in view ($0\farcm4 \times 0\farcm4$) 
of a tangential arc.
The panel C sows a zoom in view ($0\farcm8 \times 0\farcm8$) 
around BCG2, where no arc-like image is found by visual inspection.
 }
\label{fig:a2142_zoom} 
\end{figure*}

\newpage

\begin{figure*}
\begin{center}
\FigureFile(170mm,140mm){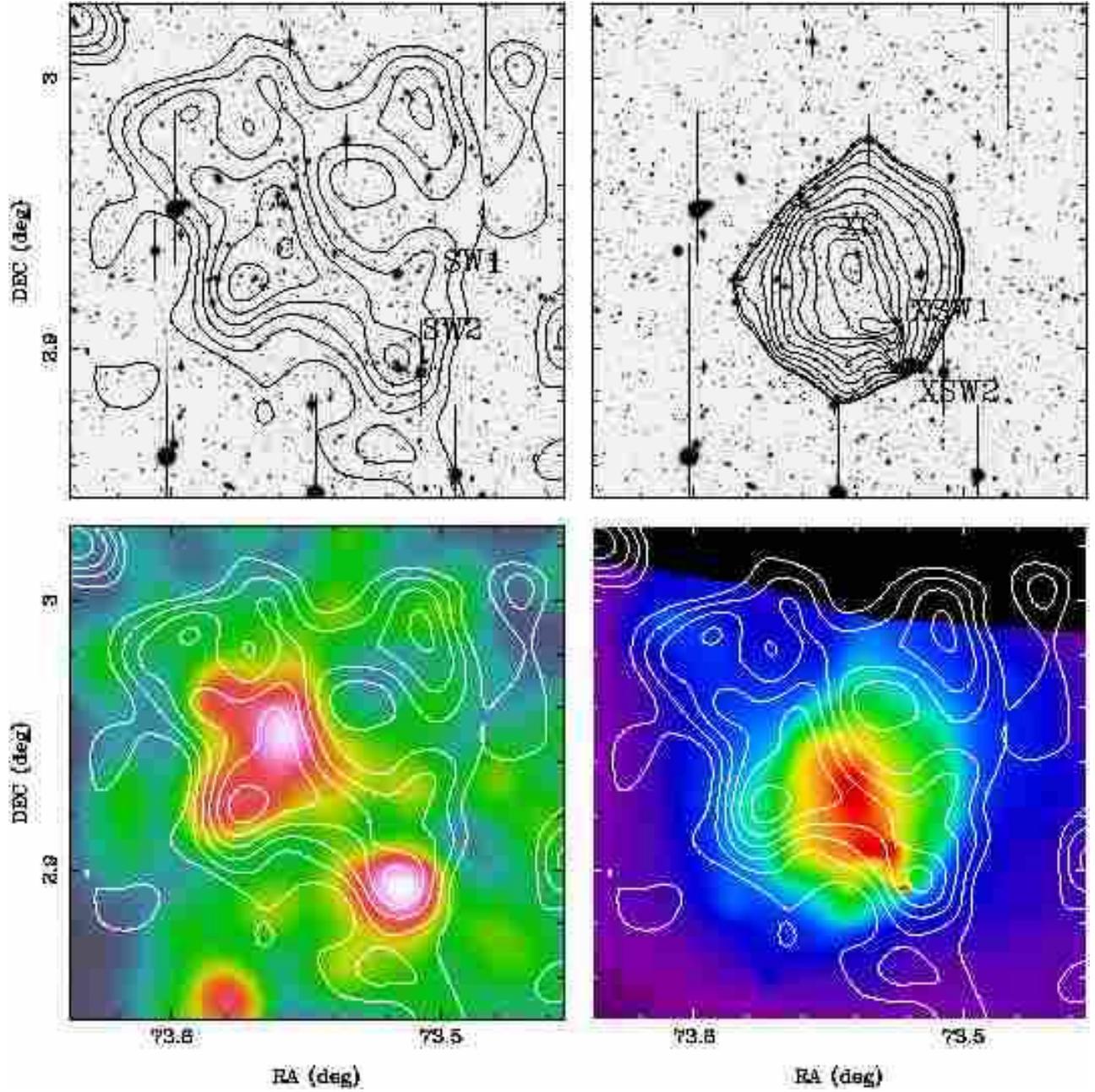}
\end{center}
\caption{A520 based on 
$4\times 240$s $i'$-band images
taken without guide star
\singlebond  
Top-left:
Subaru $i'$-band image of the central $\sim 11',\times 11'$ 
cluster region. Overlayed are contours of the lensing $\kappa$-field
reconstructed from weak shear data.
The contours are spaced in units of $1\sigma$ reconstruction error
(see Table \ref{tab:fbg}).
The Gaussian FWHM  
used for the mass reconstruction is $1\farcm 25$.
Top-right: adaptively-smoothed Chandra X-ray contours ($0.7-7.0$ keV)
overlaid on the same $i'$-band image.
Bottom-left: 
Cluster luminosity density distribution 
in $i'$-band smoothed to the same
angular resolution of the mass map.
Overlayed are the same mass contours as in the top-left panel.
Bottom-right: 
The same mass contours overlayed on the adaptively-smoothed Chandra
 X-ray
image ($0.7-7.0$ keV).
}\label{fig:a520B} 
\end{figure*}

\newpage

\begin{figure*}
\begin{center}
\FigureFile(150mm,150mm){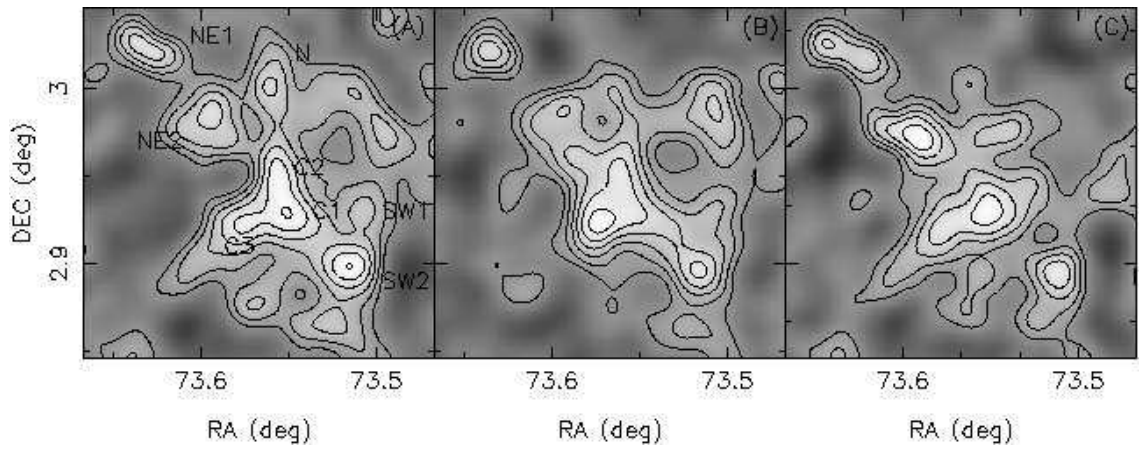}
\end{center}
\caption{Comparisons of A520 weak lensing analyses of
different data sets.
Left (A) : The lensing $\kappa$-field of ($12\farcm\times 12\farcm$),
reconstructed from seven $i'$ images with and without AG (acquisition
 and guide) probe ($7\times 240$s exposure),
which was also used by Mahdavi et al. (2007).
The Gaussian FWHM used for the mass reconstruction is $1\farcm 25$.
The contours are spaced in units of $1\sigma$ reconstruction error.
Middle (B) : The same figure as left panel, reconstructed from
 four $i'$ images taken without guide probe ($4\times 240$s exposure).
Right (C) : The same figure as left panel, reconstructed from
 three $i'$ images taken with guide probe ($3\times 240$s exposure).
}\label{fig:a520_3maps}
\end{figure*}


\begin{thebibliography}{}

\bibitem[Abell, Neyman, \&  Scott 1964]{ab64}
Abell, G. O., Neyman, J., \& Scott, E. L., AJ, 69, 529.

\bibitem[Ascasibar \& Markevitch 2006]{asc06} 
Ascasibar, Y. \& Markevitch, M. 2006, ApJ, 650, 102.


\bibitem[Arnaud et al. 2002]{arn02}
Arnaud, M., et al. 2002, A\&A, 390, 27.

\bibitem[Ashman, Bird \& Zepf 1994]{ash94}
Ashman, K., M., Bird, C., M. \& Zepf, S., E., 	
	1994, AJ, 108, 2348.

	
\bibitem[Bardeau et al. 2005]{bar05}
Bardeau, S., Kneib, J.-P., Czoske, O., Soucail, G., Smail, I., Ebeling,
				       H. \&  Smith, G. P. 2005, A\&A,
				       434, 433.

\bibitem[Bardeau et al. 2007]{bar07}
Bardeau, S., Soucail, G., Kneib, J. P., Czoske, O., Ebeling, H.,
				       Hudelot, P., Smail, I. \& Smith,
				       G. P. 
 2007, A\&A, 470, 449.



\bibitem[Bartelmann96]{bar96}
Bartelmann 1996, A\&A, 313, 697


\bibitem[Bartelmann \& Schneider 2001]{bar01} Bartelmann, M. \&
				      Schneider, P.,  2001, PhR, 340, 291.


\bibitem[Baier \& Ziener 1977]{ba77} Baier, F. W. \& Ziener, R., AN,
			    298, 87.


\bibitem[Belsole et al. 2004]{bel04}
Belsole, E., Pratt, G. W., Sauvageot, J.-L. \& Bourdin, H.
2004, 415, 821. 


\bibitem[Bertin \& Arnouts 1996]{ber96} Bertin, E. \& Arnouts, S. 1996,
                                        A\&AS, 117, 393.





\bibitem[Brada ${\check {\rm c}}$ et al. 2006]{bra06} Brada${\check
				       {\rm c}}$, M. et al. 2006, ApJ,
				       652,  937.
	


\bibitem[Broadhurst et al. 2005]{bro05} 
Broadhurst, T., Takada, M., Umetsu, K.,  Kong, X., Arimoto, N.,  Chiba,
				      M. \& Futamase, T.  2005, ApJ,
				      619, L143.

\bibitem[Bruzual  \& Charlot 2003]{bru03}
Bruzual, G. \& Charlot, S. 2003, MNRAS, 344,1000.

\bibitem[Bullock et al. 2001]{bul01}	
	Bullock, J. S., Kolatt, T. S.,  Sigad, Y., Somerville, R. S.,
				      Kravtsov, A. V., Klypin, A. A.,
				      Primack, J. R. \&  Dekel, A.,
				      2001, MNRAS, 321, 559.


\bibitem[Clowe et al. 2000]{clo00} 
Clowe, D., Luppino, G. A., Kaiser, N., \& Gioia, I. M., 2000, ApJ, 539, 540.


\bibitem[Clowe,Gonzale \& Markevitch 2004]{clo04}
Clowe, D., Gonzalez, A. \& Markevitch, M.  2004, ApJ, 604, 596.


\bibitem[Clowe et al. 2006]{clo06} 
Clowe, D. et al. 2006, A\&A, 451, 395. 

\bibitem[Cypriano]{clp04} Cypriano, E. S., Sodre, L., Jr., Kneib, J.-P. \&
				       Campusano, L. E.  2004, ApJ, 613,
				       95.

\bibitem[Dahle et al. 2002]{dah02}
Dahle, H.,  Kaiser, N., Irgens, R. J., Lilje, P.  B. \& Maddox,
			    S. J. 2002, ApJS, 139, 313.


\bibitem[David, Forman \& Jones 1999]{dav99}
David, L. P., Forman, W. \&  Jones, C.  1999, ApJ, 519, 533.

\bibitem[David \& Kempner 2004]{dav04}
David, L. P. \& Kempner, J.  2004, ApJ, 613, 831.

	


\bibitem[Donnelly et al. 2001]{don01}
Donnelly, R. H., Forman, W., Jones, C., Quintana, H., Ramirez, A.,
			     Churazov, E.\& Gilfanov, M.
 2001, ApJ, 562, 254

\bibitem[Eke, Cole, \& Frenk 1996]{eke96} Eke, V. R., Cole, S. \& Frenk,
				      C. S. 1996, MNRAS, 282, 263.

\bibitem[Erben et al. 2001]{erb01}
Erben, T., van Waerbeke, L., Bertin, E., Mellier, Y., \& Schneider, P. 2001, A\&A, 366, 717

\bibitem[Fahlman et al. 1994]{fah94} Fahlman, G., Kaiser, N., Squires G. \&
                                                     Woods, D. 1994,
                                                      ApJ, 437, 56.	

\bibitem[Finoguenov, B$\ddot{{\rm o}}$hringer \& Zhang]{fin03}	
	Finoguenov, A., B$\ddot{{\rm o}}$hringer, H. \&  Zhang, Y.-Y., 	
	2005, A\&A, 442, 827.

\bibitem[Forman et al. 1981]{for81}
Forman, W., Bechtold, J., Blair, W., Giacconi, R., van Speybroeck, L. \&
			     Jones, C.
 1981, ApJ, 243, L133. 
	

\bibitem[Geller \& Beers 1982]{ge82}
	
	Geller, M. J. \&  Beers, T. C., PASP, 94, 421.

\bibitem[Girardi et al. 2000]{gir00} Girardi, M.,  Borgani, S.,
				      Giuricin, G., Mardirossian, F. \&
				      Mezzetti, M., 2000, ApJ, 530, 62.


\bibitem[Goto et al. 2002]{got02}
        Goto, T. et al. 2002, PASJ,  54, 515.


\bibitem[Govoni et a. 2004]{gov04}
Govoni, F., Markevitch, M.,  Vikhlinin, A., VanSpeybroeck, L., Feretti,
			     L.\&  Giovannini, G. 2004, ApJ, 605, 695.

\bibitem[Hamana et al. 2003]{ham03} Hamana et al., 2003, ApJ,  597, 98.

\bibitem[Hoekstra 2007]{hoe07}
Hoekstra, H. 2007, MNRAS, 379, 317.


\bibitem[Henriksen \& Markevitch 1996]{hen96}
Henriksen, M. J. \& Markevitch, M.
1996,ApJ, 466,L79.

\bibitem[Henry \& Briel 1995]{hen95}
Henry, J. P. \& Briel, U. G.
1995, ApJ,  443. L9.

\bibitem[Henry, Finoguenov \&  Briel 2004]{hn04}
Henry, J. P., Finoguenov, A. \& Briel, U. G.,
 2004, ApJ, 615, 181.


\bibitem[Hudson, Gwyn, Dahle \& Kaiser]{hud98}
Hudson, M. J., Gwyn, S. D. J., Dahle, H. \&  Kaiser, N., 1998, ApJ, 503,
				      531.

\bibitem[Hetterscheidt et al 2007]{het07} Hetterscheidt, M. et al.,
	 		     2007, A\&A, 468, 859  (astro-ph/0606571).

\bibitem[Hoekstra et al. 1998]{hoe98} Hoekstra, H., Franx, M., Kuijken,
				      K., \& Squires, G., 1998, ApJ,
				      504, 636.


\bibitem[Kaiser \& Squires 1993]{kai93} Kaiser, N. \&  Squires, G. 1993, ApJ, 404, 441.


\bibitem[Kaiser 1995]{kai95a} Kaiser, N. 1995, ApJL, 439, L1.

\bibitem[Kaiser\, Squires \& Broadhurst 1995]{kai95b} Kaiser, N.,
					  Squires, G., Broadhurst, T.,
					  1995, ApJ, 449, 460.

\bibitem[Katgert et al. 2004]{kat04} Katgert, P., Biviano, A. \& Mazure,
				      A.  2004, ApJ, 600, 657.


\bibitem[Kempner \& Sarazin 2001]{kem01}
Kempner, J. C. \& Sarazin, C. L. 2001 ApJ, 548, 639.

\bibitem[Kempner \& Sarazin 2003]{kem03}
Kempner, J. C. \& Sarazin, C. L. 2003 ApJ, 593, 291.


\bibitem[Novicki et al. 1998]{nov98}
Novicki, M., Jones, C., \& Donnelly, R. H. 1998, AAS, 193, 3807.

\bibitem[Mahdavi et al. 2007]{mah07} Mahdavi, A., Hoekstra, H., Babul,
				       A., Balam, D.\& Capak, P. 2007,
				       ApJ, 668, 806.


\bibitem[Markevitch et al.  2000]{mar00}
Markevitch, M. et al. 2000, \apj, 541, 542.

\bibitem[Markevitch et al.  2000]{mar02}
Markevitch, M., Gonzalez, A. H.,  David, L.,  Vikhlinin, A.,   Murray,
			    S.,  Forman, W.,   Jones, C., \& Tucker, W.  2002, ApJ, 567, L27.


\bibitem[Markevitch et al. 2003]{mar03}
Markevitch, M. et al. 2003, ApJ, 586, L19. 


\bibitem[Markevitch et al. 2005]{mar05}
Markevitch, M., Govoni, F., Brunetti, G. \&  Jerius, D. 2005, ApJ, 627, 733.

\bibitem[Mathis, Lavaux, Diego, \&  Silk 2005]{mat05}
Mathis, H., Lavaux, G., Diego, J. M. \& Silk, J., 2005, MNRAS, 357, 801.



\bibitem[Medezinski et al. 2007]{med07} Medezinski, E., Broadhurst, T.,
Umetsu, K.  et al., ApJ, 663, 717 

\bibitem[Miyazaki  et al. 2002]{miy02}  Miyazaki, S. et al.  2002,  PASJ, 54, 833,801.


\bibitem[Nagai \& Kravtsov 2003]{nag03}
Nagai, D., \& Kravtsov, A. V.,  2003, ApJ, 587, 514.

\bibitem[Navarro, Frenk, \& White 1996]{nav96} Navarro, J. F.,Frenk, C. S. \& White, S. D. M. 1996, ApJ, 462, 563.

\bibitem[Novicki et al. 1998]{nov98} 
Novicki, M., Jones, C. \&  Donnelly, R. H. 1998, AAS, 193, 3807.


\bibitem[Oegerle,\ Hill \& Fitchett 1995]{oeg95} Oegerle, W. R.,  Hill, J. M. \&
				 Fitchett, M.  J., 1995, AJ, 110, 32.



\bibitem[Oguri, Taruya, \& Suto 2001]{}
Oguri, M., Taruya, A., Suto, Y. 2001, ApJ, 559, 572


\bibitem[Okabe, Umetsu \& Hattori 2008]{oka08}
 Okabe, N., Umetsu, K. \& Hattori, M., in preparation


\bibitem[Ouchi et al. 2004]{ouc04} Ouchi, M., et al. 2004, ApJ, 611, 660.


\bibitem[Peebles 1980]{} 
Peebles, P. J. E., 1980  The large-scale structure of
				      the universe (Princeton University Press) 


\bibitem[Randall, Sarazin \& Ricker 2002]{ran02} 
				      Randall, S. W.,
                                       Sarazin, C. L. \& Ricker,
                                       P. M. 2002, ApJ, 577, 579.

\bibitem[Ricker \&  Sarazin 2001]{ric01} Ricker, P. M. \& Sarazin,
                                       C. L.  2001, ApJ, 561, 621.


\bibitem[Rizza et al. 1998]{riz98}
Rizza, E., Burns, J. O., Ledlow, M. J., Owen, F. N., Voges, W. \&
			    Bliton, M.
1998,MNRAS, 301, 328.


\bibitem[Rowley, Thomas \&  Kay 2004 ]{row04}
        Rowley, D. R., Thomas, P. A. \&  Kay, S. T., 2004, MNRAS, 352, 508.

\bibitem[Sanderson \&  Ponman 2003]{san03} Sanderson, A. J. R. \&
				      Ponman, T. J., 2003, MNRAS, 345, 1241.

\bibitem[Sand et al. 2005]{san05} Sand, D. J., Treu, T., Ellis, R. S. \&
				       Smith, G. P.  2005, ApJ, 627, 32.

\bibitem[Sato et al. 2003]{sat03} Sato, J., Umetsu, K.,  Futamase, T. \&
				      Yamada, T. 2003, ApJ, 582, L67.

\bibitem[Schechter 1976]{sch76} Schechter, P. 1976, ApJ, 203, 297.

\bibitem[Seitz \& Schneider 1995]{sei95} Seitz, S. \& Schneider,
					  P., 1995, A\&A, 297, 287.
\bibitem[Seitz \& Schneider 1996]{sei96} Seitz, S. \& Schneider,
					  P., 1996, A\&A, 305, 383.
\bibitem[Seitz \& Schneider 2001]{sei01} Seitz, S. \& Schneider, 2001,
				      A\&A, 374, 740.



\bibitem[Takizawa 2005]{tak05}
Takizawa, M. 2005, ApJ, 629, 791.


\bibitem[Takizawa 2006]{tak06}
Takizawa, M. 2006,  PASJ, 58, 925 




\bibitem[Umetsu \& Okabe 2007]{ume07}
Umetsu \& Okabe, in preparation

\bibitem[UTF99]{}
Umetsu, K., Tada, M., \& Futamase, T. 1999, Prog. Theor. Phys. Suppl.,
133, 53 

\bibitem[UB07]{}
Umetsu, K. \& Broadhurst, T. 2007, submitted to ApJ
(arXiv:astro-ph/0712.3441) 


\bibitem[Tormen, Moscardini \&  Yoshida 2004]{tor04}	
	Tormen, G., Moscardini, L. \& Yoshida, N.,  2004, MNRAS, 350, 1397.

 
\bibitem[Van Waerbeke  et al. 2000]{wae00}
 Van Waerbeke et al., 2000A\&A, 358, 30.

\bibitem[White 2000]{whi00}
White, D. A, 2000, MNRAS, 312, 663.

\bibitem[White 2002]{whi02}
White, M., Van Waerbeke, L., Mackey, J. 2002, ApJ, 575, 640


\bibitem[Wu et al. 2006]{wu06} 
Wu, J.-M., Umetsu, K., Chien, C.-H., Chiueh, T. 2006, 
submitted to ApJ (astro-ph/0607542).


\bibitem[Yagi et al. 2002]{yag02} 
Yagi, M., Kashikawa, N., Sekiguchi,
			    M., Doi, M., Yasuda, N., Shimasaku, K. \&  Okamura, S., 2002, AJ, 123, 66.


\bibitem[Zabludoff et al. 1993]{zab93}
Zabludoff, A., I., Geller, M. J., Huchra, J., P. \& Vogeley,
				      M., S., 1993, AJ, 106, 1273.


\end{thebibliography}
\end{document}